\ifx\texorpdfstring\undefined\def\texorpdfstring#1#2{#1}\fi
\ifx\pdfsuppresswarningpagegroup\undefined\else\pdfsuppresswarningpagegroup=1\fi
\PassOptionsToPackage{unicode}{hyperref}
\documentclass[aps,twocolumn,showpacs,showkeys,preprintnumbers,floatfix,nofootinbib,superscriptaddress,longbibliography]{revtex4-2}
\usepackage{amsfonts} 
\usepackage{amssymb} 
\usepackage{amsmath} 
\usepackage{graphicx} 
\usepackage{subfigure} 
\usepackage{array} 
\usepackage{dcolumn} 
\usepackage{latexsym} 
\usepackage{longtable} 
\usepackage{bbold}
\usepackage{color}
\usepackage{multirow}
\usepackage{rotating}
\usepackage{comment}

\newcommand\suppcite[2]{#1~\ref{#2}}
\newcommand\suppeqcite[2]{#1~\eqref{#2}}
\let\mainref=\suppcite
\let\maineqref=\suppeqcite

\providecommand \doibase {http://dx.doi.org/}%

\graphicspath{{./figs/}{./figs_gS/}}

\newcommand{\MeV}{\mathop{\rm MeV}\nolimits}
\newcommand{\GeV}{\mathop{\rm GeV}\nolimits}

\newcommand{\kk}{\mathbf{k}}

\newcommand{\nn}{\mathbf{n}}
\newcommand{\mpi}{M_\pi}
\newcommand{\mN}{M_N}
\newcommand{\spiN}{\sigma_{\pi N}}

\newcommand{\changed}[1]{{{#1}}}

\usepackage{textgreek}
\usepackage[unicode]{hyperref} 

\providecommand{\abs}[1]{\lvert#1\rvert}
\providecommand{\matrixe}[3]{\langle#1\lvert#2\rvert#3\rangle}

\definecolor{green}{rgb}{0.1, 0.8, 0.1}

\usepackage{textgreek}
\usepackage[unicode]{hyperref} 

\begin{document}


\title{The \changed{pion--}nucleon sigma term from lattice QCD}

\author{Rajan Gupta}
\email{rajan@lanl.gov}
\affiliation{Los Alamos National Laboratory, Theoretical Division T-2, Los Alamos, NM 87545, USA}

\author{Sungwoo Park}
\email{sungwoo@jlab.org}
\affiliation{Los Alamos National Laboratory, Theoretical Division T-2, Los Alamos, NM 87545, USA}
\affiliation{Center for Nonlinear Studies, Los Alamos National Laboratory, Los Alamos, New Mexico 87545, USA}

\author{Martin Hoferichter}
\email{hoferichter@itp.unibe.ch}
\affiliation{Albert Einstein Center for Fundamental Physics, Institute for Theoretical Physics, University of Bern, Sidlerstrasse 5, 3012 Bern, Switzerland}

\author{Emanuele Mereghetti}
\email{emereghetti@lanl.gov}
\affiliation{Los Alamos National Laboratory, Theoretical Division T-2, Los Alamos, NM 87545, USA}

\author{Boram~Yoon}
\email{boram@lanl.gov}
\affiliation{Los Alamos National Laboratory, Computer, Computational and Statistical Sciences Division CCS-7, Los Alamos, NM 87545, USA}

\author{Tanmoy~Bhattacharya}
\email{tanmoy@lanl.gov}
\affiliation{Los Alamos National Laboratory, Theoretical Division T-2, Los Alamos, NM 87545, USA}

\preprint{LA-UR-21-24759}

%
%
%
%
\begin{abstract}
We present an analysis of the pion--nucleon $\sigma$-term, $\sigma_{\pi N}$, using
six ensembles with 2+1+1-flavor highly improved staggered quark
action generated by the MILC collaboration. The most serious
systematic effect in lattice calculations of nucleon correlation
functions is the contribution of excited states. We estimate these 
using chiral perturbation theory ($\chi$PT), and show that the
leading contribution to the isoscalar scalar charge
comes from $N \pi $ and $N\pi\pi$ states. Therefore, we carry out
two analyses of lattice data to remove excited-state contamination,
the standard one and a new one including $N\pi$ and $N\pi\pi$ states.  We find
that the standard analysis gives $\sigma_{\pi N} =  41.9(4.9)$ MeV, consistent with
previous lattice calculations, while \changed{our preferred} $\chi$PT-motivated analysis
gives $\sigma_{\pi N} = 59.6(7.4)$ MeV, which is
consistent with phenomenological values obtained using $\pi
N$ scattering data.  Our data on one physical pion mass ensemble was
crucial for exposing this difference, therefore, calculations on
additional physical mass ensembles are needed to confirm our result
and resolve the tension between lattice QCD and phenomenology.
\end{abstract}
\maketitle

\section{Introduction}
\label{sec:into}

This Letter presents results for the pion--nucleon $\sigma$-term, 
$\spiN \equiv { 
  m}_{ud}\, g_S^{u+d} \equiv {m}_{ud} \allowbreak\, \langle N({\kk},s)| \bar{u}
u + \bar{d} d | N({\kk},s) \rangle$ calculated in the isospin symmetric limit with 
${m}_{ud} = (m_u + m_d)/2$  the average of the light quark masses. 
It is a fundamental
parameter of QCD that quantifies the amount of the nucleon mass 
generated by  the $u$- and $d$-quarks. The scalar charge $g_S^q$ is determined from 
the forward matrix element of the scalar
density $\bar{q} q$ between the nucleon state:
\begin{align}
g_S^q =  \langle N({\kk}=0,s)| Z_S\ \bar{q}  q | N({\kk}=0,s) \rangle,
\label{eq:gSdef}
\end{align}
where $Z_S$ is the renormalization constant and the nucleon spinor has
unit normalization.
The connection between $g_S^q$ and the rate of variation of the
nucleon mass, $\mN$, with the mass of quark with flavor $q$ is given by the
Feynman--Hellmann (FH)
relation~\cite{Hellmann:1937,Feynman:1939zza,Gasser:1979hf}
\begin{align}
\frac{\partial \mN}{\partial m_q}  =  \langle N({\kk},s)| \bar{q}  q | N({\kk},s) \rangle = g_S^q/Z_S.
\label{eq:FH}
\end{align}
The charge,  $g_S^q$, determines the coupling of the nucleon to the scalar quark
current---an important input quantity in the search for physics beyond the
Standard Model (SM), including in direct-detection searches for dark
matter~\cite{Bottino:1999ei,Bottino:2001dj,Ellis:2008hf,Crivellin:2013ipa,Hoferichter:2017olk},
lepton flavor violation in $\mu\to e$ conversion in
nuclei~\cite{Cirigliano:2009bz,Crivellin:2014cta}, and electric dipole
moments~\cite{Engel:2013lsa,deVries:2015gea,deVries:2016jox,Yamanaka:2017mef}. In particular, 
$\spiN$ is a rare example of a matrix element that,
despite the lack of scalar probes in the SM, can still be extracted
from phenomenology---via the Cheng--Dashen low-energy
theorem~\cite{Cheng:1970mx,Brown:1971pn}---and thus defines an
important benchmark quantity for lattice QCD.

The low-energy theorem establishes a connection between $\spiN$
and a pion--nucleon ($\pi N$) scattering amplitude, albeit evaluated
at unphysical kinematics. Since the one-loop corrections are free of chiral
logarithms~\cite{Bernard:1996nu,Becher:2001hv}, the remaining corrections to the low-energy theorem scale
as $\spiN\mpi^2/\mN^2\approx 1\MeV$, leaving the
challenge of controlling the analytic continuation of the \changed{isoscalar $\pi N$
amplitude $\Sigma_{\pi N}$.} Stabilizing this extrapolation by means of dispersion
relations \changed{(and clarifying the relation between $\spiN$ and $\Sigma_{\pi N}$)}, Refs.~\cite{Gasser:1988jt,Gasser:1990ce,Gasser:1990ap}
found $\spiN\approx 45\MeV$ based on the partial-wave
analyses from Refs.~\cite{Koch:1980ay,Hohler:1984ux}. More recent
partial-wave analyses~\cite{Arndt:2006bf,Workman:2012hx} favor higher
values, e.g., $\spiN=64(8)\MeV$~\cite{Pavan:2001wz}. Similarly,
$\chi$PT analyses depend crucially on the $\pi N$ input, with
$\spiN$ prediction varying
accordingly~\cite{Fettes:2000xg,Alarcon:2011zs}\changed{. Other works that exploit this relation to $\pi N$ scattering include Refs.~\cite{Koch:1982pu,Ericson:1987uf,Hohler:1990tz,Olsson:1999jt,Hite:2005tg,Hadzimehmedovic:2007dsb,Stahov:2012ca,Matsinos:2013era}.}\looseness-1

\begin{figure}[th]
  \subfigure{ \includegraphics[height=0.8in]{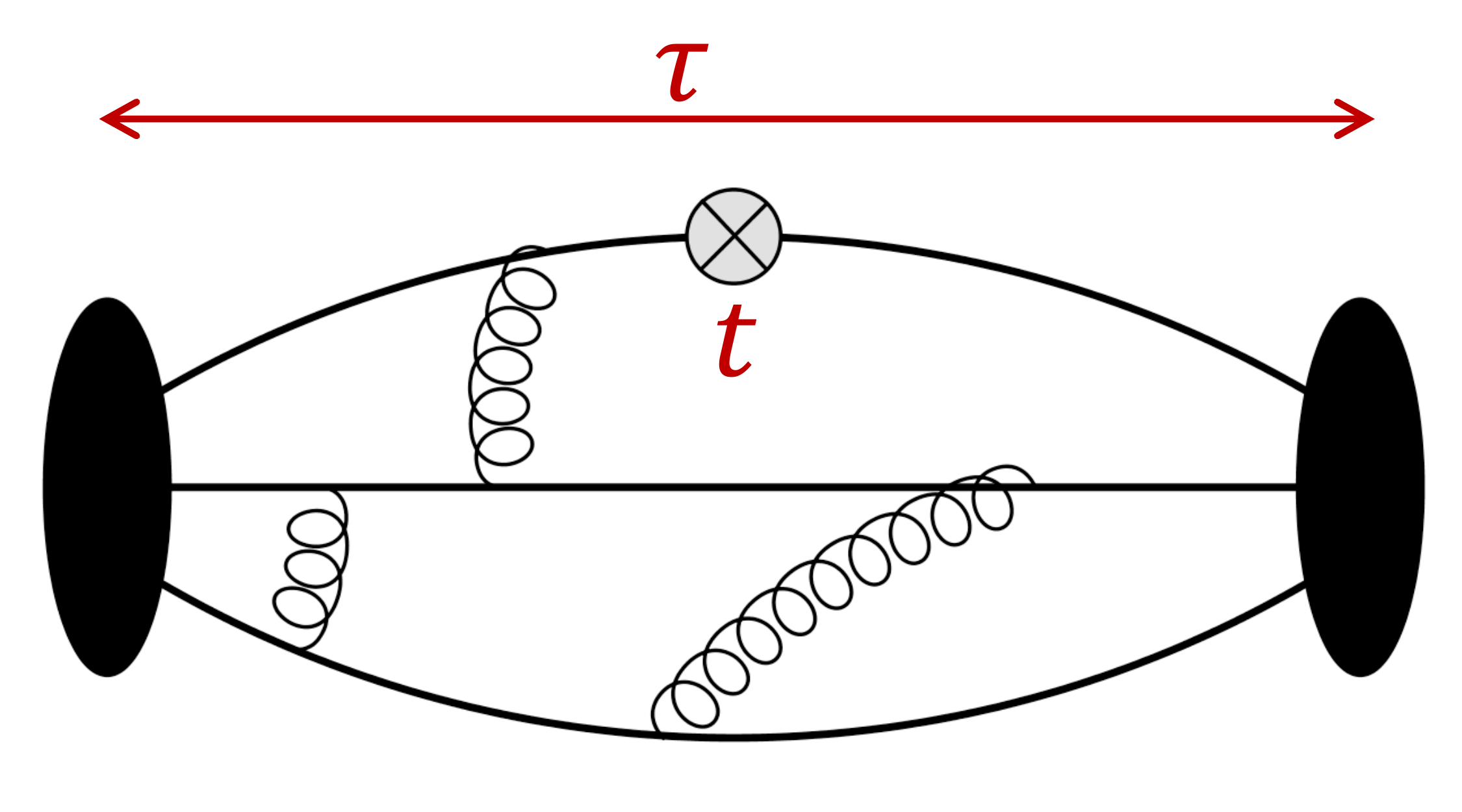}  \hspace{0.1\linewidth}
    \includegraphics[height=1.0in]{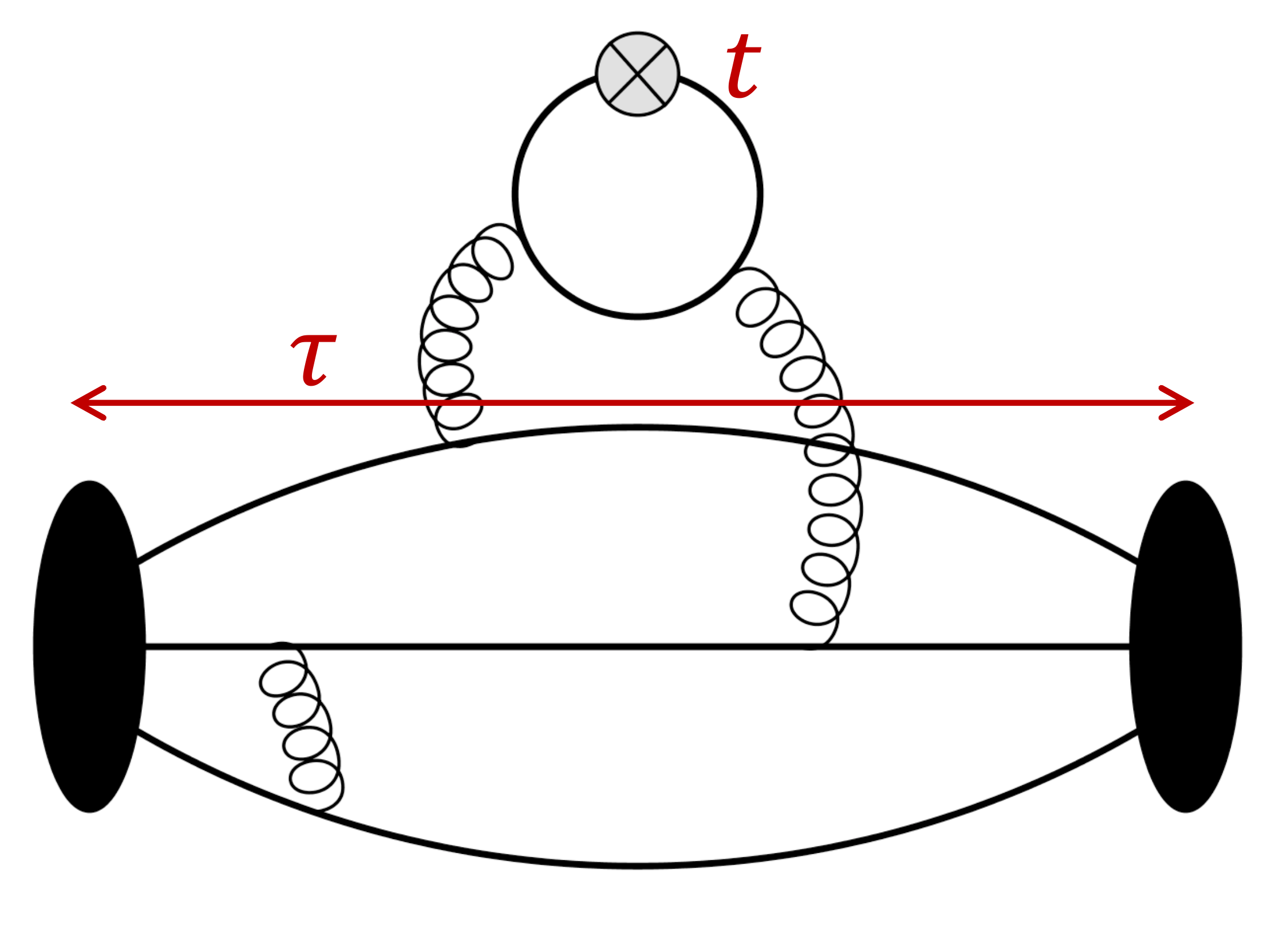}\hspace{0.1\linewidth}
  }
 \changed{\includegraphics[height=0.8in]{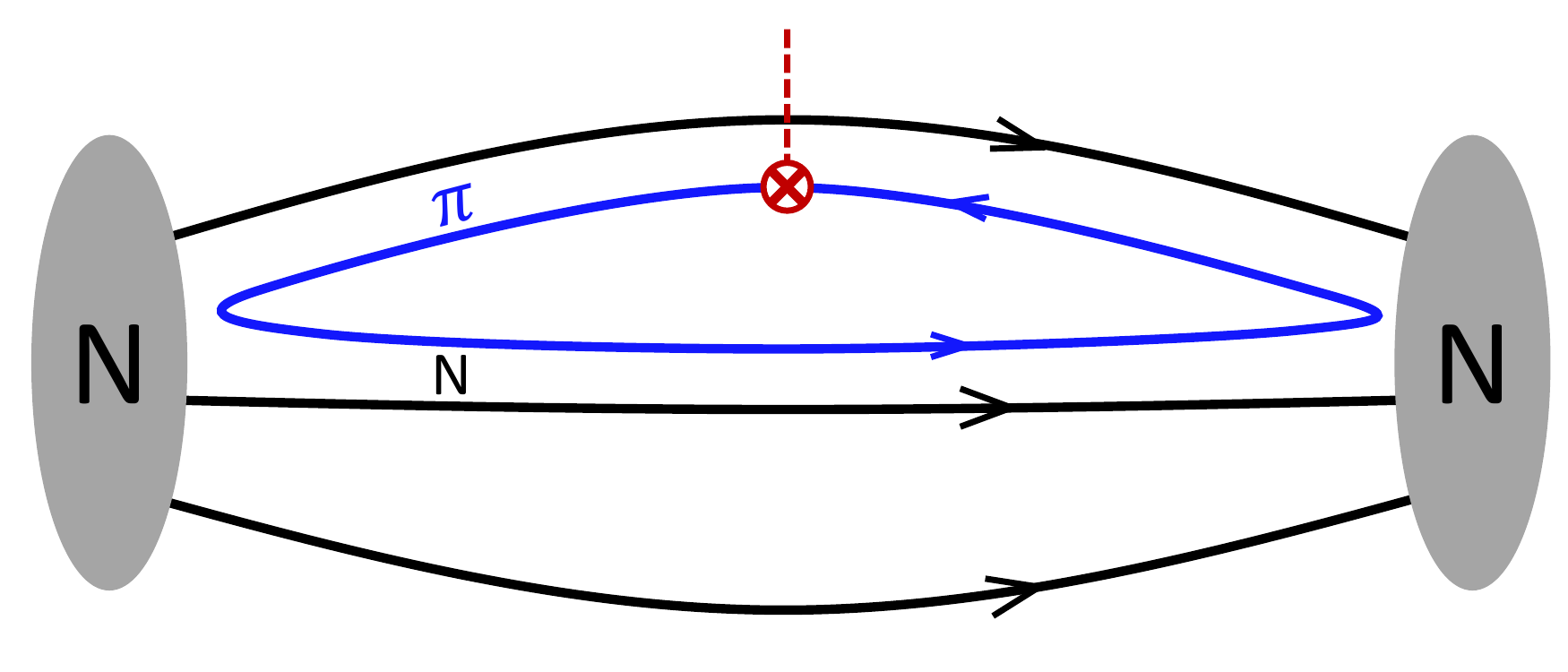}}
\vspace{-0.1in}
\caption{ \changed{The upper connected (left) and disconnected  (right) 
diagrams contribute to the 3-point
  function that determine the matrix element of
  flavor-diagonal scalar operators (shown by the symbol $\otimes$ at time slice $t$) within
  the nucleon state. The black \changed{and gray} blobs denote nucleon source and sink, 
  separated by Euclidean time $\tau$.  The bottom diagram illustrates that the disconnected diagrams include an 
 $N\pi$-intermediate state configuration that can give an enhanced contribution.}\looseness-1
  \label{fig:Npisketch}
  \label{fig:conn_disc}}
\vspace{-0.1in}
\end{figure}

The analytic continuation can be further improved in the framework of
Roy--Steiner
equations~\cite{Ditsche:2012fv,Hoferichter:2012wf,Hoferichter:2015dsa,Hoferichter:2015tha,Hoferichter:2015hva,Hoferichter:2016ocj,Hoferichter:2016duk,Siemens:2016jwj,RuizdeElvira:2017stg},
whose constraints on $\spiN$ become most powerful when combined
with pionic-atom data on threshold $\pi N$
scattering~\cite{Strauch:2010vu,Hennebach:2014lsa,Hirtl:2021zqf,Baru:2010xn,Baru:2011bw}. Slightly
updating the result from
Refs.~\cite{Hoferichter:2015dsa,Hoferichter:2015hva} to account for
the latest data on the pionic hydrogen width~\cite{Hirtl:2021zqf}, one
finds $\spiN=59.0(3.5)\MeV$. In particular, this determination
includes isospin-breaking
corrections~\cite{Gasser:2002am,Hoferichter:2009ez,Hoferichter:2009gn,Hoferichter:2012bz}
to ensure that $\spiN$ coincides with its definition in
lattice QCD calculations~\cite{Hoferichter:2016ocj}. The difference from 
Refs.~\cite{Gasser:1988jt,Gasser:1990ce,Gasser:1990ap} traces back to
the scattering lengths implied by
Refs.~\cite{Koch:1980ay,Hohler:1984ux}, which are incompatible with
the modern pionic-atom data. Independent constraints from experiment
are provided by low-energy $\pi N$ cross-sections, including more
recent data on both the elastic
reactions~\cite{Brack:1989sj,Joram:1995gr,Denz:2005jq} and the charge
exchange~\cite{Frlez:1997qu,Isenhower:1999aj,Jia:2008rt,Mekterovic:2009kw},
and a global analysis of low-energy data in the Roy--Steiner framework
leads to $\spiN=58(5)\MeV$~\cite{RuizdeElvira:2017stg}, in
perfect agreement with the pionic-atom result.  In contrast, so far lattice 
QCD calculations~\cite{Durr:2011mp,Bali:2012qs,Durr:2015dna,Yang:2015uis,Abdel-Rehim:2016won,Bali:2016lvx,Yamanaka:2018uud,Alexandrou:2019brg,Borsanyi:2020bpd}
have favored low values $\spiN\approx 40\MeV$ (with the exception of Ref.~\cite{Alexandrou:2014sha}), and it is this
persistent tension with phenomenology that we aim to address in this
Letter.

There are two ways to calculate $\spiN$ using lattice QCD, which
are called the FH and the direct
methods~\cite{Aoki:2019cca}. In the FH method, the nucleon mass is
obtained as a function of the bare quark mass ${m}_{ud} $
(equivalently $\mpi^2$) from the nucleon 2-point correlation
function, and its numerical derivative multiplied by ${m}_{ud} $
gives $\spiN$. In the direct method, the matrix element of $ \bar{u} u
+ \bar{d} d$ is calculated within the ground-state nucleon. Both
methods have their challenges. In the FH method, one needs to
calculate the derivative about the physical ${ m}_{ud} $, which is
computationally very demanding.  Most calculations extrapolate from
heavier masses or fit the data for $\mN$ versus $\mpi^2$ to an ansatz
motivated by $\chi$PT and evaluate its derivative at
${ m}_{ud} $. \changed{On the other hand, the signal in the matrix element is noisier since
it is obtained from} a 3-point function with the insertion of the scalar
density. In both methods, one has to ensure that all  excited-states contamination (ESC) 
has been removed. \changed{Both methods give
$\spiN \approx 40\MeV$---see Fig.~\ref{fig:FLAG}, review by the Flavour 
Lattice Averaging Group (FLAG) in 2019~\cite{Aoki:2019cca}, and the two subsequent works~\cite{Alexandrou:2019brg,Borsanyi:2020bpd}. }

Here, we present a new direct-method calculation. 
Our main message is that $N\pi$ and $N\pi\pi$ excited states, which have not
been included in previous lattice calculations, can make a significant
contribution. We provide motivation for this effect from heavy-baryon $\chi$PT \cite{Jenkins:1990jv,Bernard:1992qa}, and show that including the excited states in fits
to the spectral decomposition of the 3-point function increases the
result by about 50\%. Such a change brings the lattice result in
agreement with the phenomenological value.

\section{Lattice Methodology and Excited States}
\label{sec:MethodologyES}

The construction of all nucleon 2- and 3-point correlations functions
is carried out using Wilson-clover fermions on six 2+1+1-flavor
ensembles generated using the highly improved staggered quark (HISQ)
action~\cite{Follana:2006rc} by the MILC
collaboration~\cite{Bazavov:2012xda}.  In each of these ensembles, the
$u$- and $d$-quark masses are degenerate, and the $s$- and $c$-quark
masses have been tuned to their physical values. Details of the six ensembles 
at lattice spacings, $a \approx 0.12$, $0.09$, and
$0.06$~fm, and $\mpi \approx 315$, $230$, and
$138\MeV$  are given in Table~\ref{tab:results} and  in 
\suppcite{Table}{tab:ens}, and of the analysis in \suppcite{App.}{sec:Lparams}. To obtain flavor-diagonal charges $g_S^q$,
two kinds of diagrams, called connected and disconnected and  illustrated in Fig.~\ref{fig:conn_disc}, are calculated.  The
details of the methodology for the calculation of the connected
contributions (isovector charges) using this clover-on-HISQ
formulation are given in
Refs.~\cite{Bhattacharya:2015wna,Gupta:2018qil} and of the disconnected ones in
Ref.~\cite{Bhattacharya:2015wna}.

The main focus of the analysis is on controlling the ESC. To this end, we 
estimate $\spiN$ using two possible sets of excited-state
masses, $M_1$ and $M_2$, given in Table~\ref{tab:results}. These $M_i$
are obtained from simultaneous fits to the zero momentum 
nucleon 2-point, $C^{2\text{pt}}$, and 3-point,
$C^{3\text{pt}}$, functions using their spectral
decomposition truncated to four and three states respectively:
\begin{align}
  C^{2\text{pt}}(\tau; \kk) &= \sum_{i=0}^3 \abs{\mathcal{A}_i(\kk)}^2 e^{-M_i \tau},  \nonumber \\
C_\mathcal{S}^{3\text{pt}}(\tau; t) &=
   \sum_{i,j=0}^2 {\mathcal{A}_i} {\mathcal{A}_j^\ast}\matrixe{i}{\mathcal{S}}{j} e^{-M_i t - M_j(\tau-t)\
}. 
\label{eq:2pt3pt}
\end{align}
Here ${\mathcal{A}}_i$ are the amplitudes for the
creation or annihilation of states by the nucleon interpolating operator
used on the lattice, $\mathcal{N} = \epsilon^{abc} \left[ {u^a}^T C
  \gamma_5 (1 + \gamma_4) d^b \right] u^c$, with color indices $\{a,
b, c\}$ and charge conjugation matrix $C$. The nucleon source--sink separation
is labeled by $\tau$ and the operator insertion time by $t$.

\begin{figure}[tb]
  \subfigure{
    \includegraphics[width=0.465\linewidth]{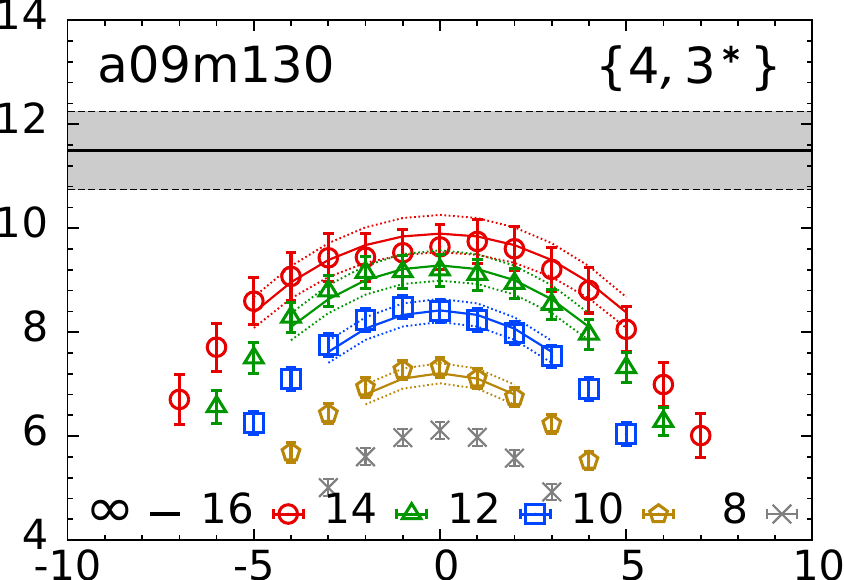}  \hspace{0.001\linewidth}
    \includegraphics[width=0.465\linewidth]{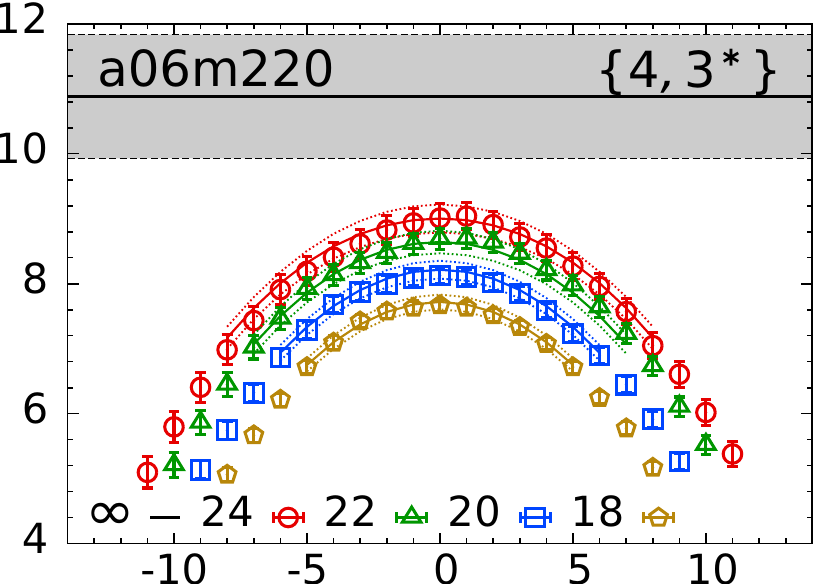}  \hspace{0.001\linewidth}
  }
  \subfigure{
    \includegraphics[width=0.465\linewidth]{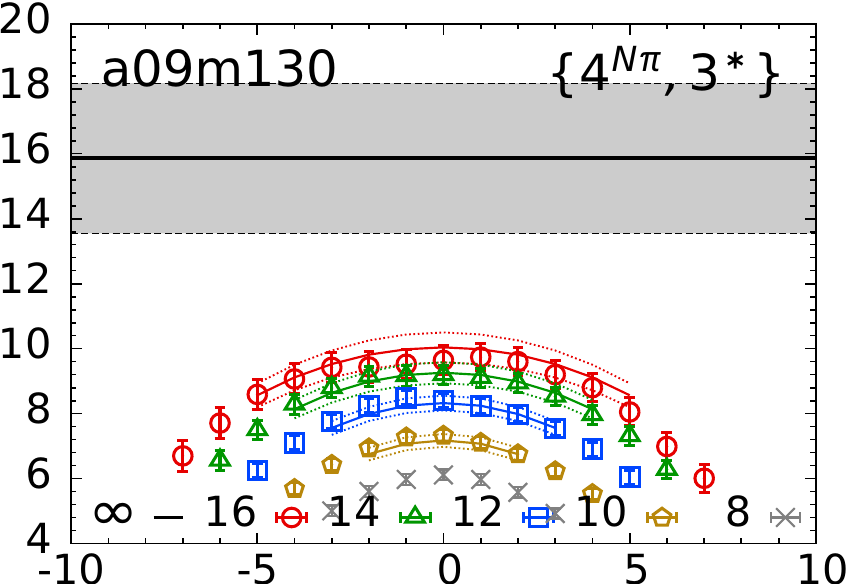}  \hspace{0.001\linewidth}
    \includegraphics[width=0.465\linewidth]{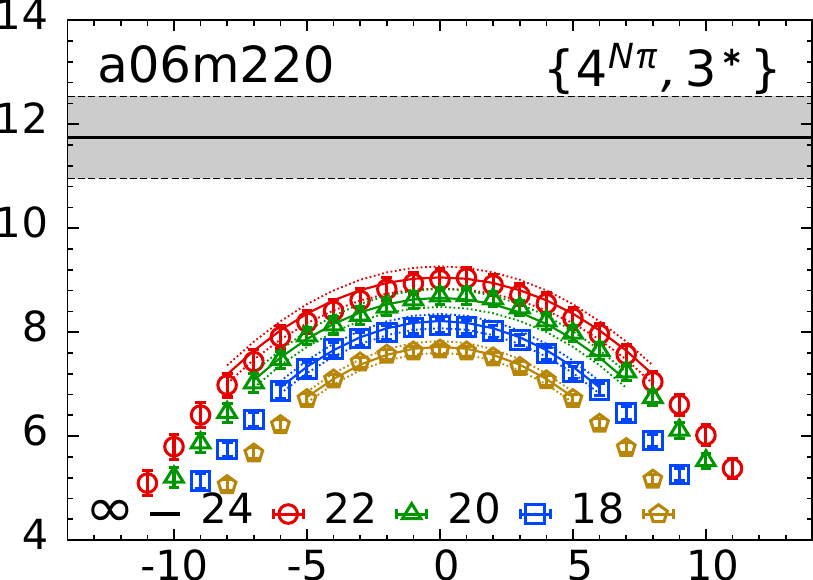}  \hspace{0.001\linewidth}
  }
\vspace{-0.1in}
\caption{Results of the $\{4,3^\ast\}$ (top row) and $\{4^{N\pi},3^\ast\}$ (bottom
  row) fits to the sum of the connected and disconnected data plotted versus $(t - \tau/2)/a$ for ensembles 
  $a09m130$ 
  and
  $a06m220$. Result of the fit is shown by lines of the same color as the data  for various $\tau/a$ listed in the label, and the $\tau=\infty$ value is given 
  by the gray band.\looseness-1
  \label{fig:Sfit}}
\vspace{-0.1in}
\end{figure}

\begin{table*}[htbp]   
\centering
\begin{ruledtabular}
\begin{tabular}{c|c|cccccc|cccccc}
        &    &
            \multicolumn{6}{c|}{$\{4,3^\ast\}$}       & \multicolumn{6}{c}{$\{4^{N\pi},3^\ast\}$}  \\
Ensemble& ${m}_{ud}^{\rm bare}$
        & $ M_0$ & $ M_1$ & $ M_2$ & $\frac{\chi^2}{\text{dof}}$ & $g_S^{u+d,{\rm bare}}$ & $\spiN$
        & $ M_0$ & $ M_1$ & $ M_2$ & $\frac{\chi^2}{\text{dof}}$ & $g_S^{u+d,{\rm bare}}$ & $\spiN$ \\
ID      & (MeV) & (GeV) & (GeV) &  (GeV) &   &   & (MeV) & (GeV) & (GeV) & (GeV) &   &   &  (MeV) \\
\colrule                                                                            a12m310 & 18.7(5)  & 1.09(1) & 1.80(12) & 2.7(1) & $27/28$ &  8.6(0.6) & 160(12)  & 1.09(1) & 1.71(02) & 2.6(1) & $27/28$ &  8.5(0.5) & 160(10) \\
a12m220 &  9.9(5)  & 1.02(1) & 1.76(08) & 3.0(3) & $18/22$ & 10.5(0.5) & 104(07)  & 1.01(1) & 1.50(03) & 2.6(2) & $20/22$ & 11.8(1.0) & 117(11) \\
\colrule
a09m220 &  9.4(1)  & 1.02(1) & 1.66(14) & 2.4(1) & $35/35$ & 10.4(0.8) &  98(07)  & 1.02(1) & 1.47(06) & 2.3(1) & $35/35$ & 11.6(0.9) & 109(09) \\
a09m130 &  3.5(1)  & 0.95(1) & 1.59(09) & 2.8(2) & $47/42$ & 11.5(0.8) &  40(03)  & 0.94(1) & 1.22(01) & 1.8(1) & $51/42$ & 15.9(2.3) &  55(08) \\
\colrule
a06m310 & 17.2(2)  & 1.11(1) & 1.80(11) & 2.9(2) & $56/60$ & 10.4(0.7) & 179(12)  & 1.11(1) & 1.76(06) & 2.8(2) & $56/60$ & 10.6(0.6) & 182(10) \\
a06m220 &  9.1(1)  & 1.02(1) & 1.62(14) & 2.5(2) & $69/81$ & 10.9(1.0) &  98(09)  & 1.02(1) & 1.51(07) & 2.3(1) & $68/81$ & 11.7(0.8) & 106(07) \\
\end{tabular}
\end{ruledtabular}
\caption{The ground- and excited-state masses, $M_0$, $M_1$, and $M_2$\changed{, $\chi^2$} of
  the fit, and the resulting value of the bare isoscalar
  charge and $\spiN$ with the two strategies $\{4,3^\ast\}$ and
  $\{4^{N\pi},3^\ast\}$. The second column gives the bare quark
  mass ${m}_{ud}^\text{bare}$.\looseness-1}
\label{tab:results}
\end{table*}

The issue of ESC arises because $\mathcal{N}$ couples not only to the
ground-state nucleon but to all its excitations including multihadron
states with the same quantum numbers. In the current data, the signal
in $C_\mathcal{S}^{3\text{pt}}$ extends to $\tau \approx 1.5$~fm, at
which source--sink separation the contribution of excited states is
significant as evident from the dependence on $\{\tau,t\}$ in the
ratio ${\cal R}_\mathcal{S}(\tau,t) =
C^{3\text{pt}}_\mathcal{S}(t,\tau)/C^{2\text{pt}}(\tau)$ shown in
Fig.~\ref{fig:Sfit}.  In the limits $t\to \infty$ and $(\tau-t) \to
\infty$, the ratio ${\cal R}_\mathcal{S}(\tau,t) \to g_S$.  
\changed{Fits to $C^{3\text{pt}}$ using Eq.~\eqref{eq:2pt3pt} with the key parameters $M_i$  left as free
parameters have large fluctuations. We, therefore, remove ESC and
extract the ground-state matrix element,
$\matrixe{0}{\mathcal{S}}{0}$, using simultaneous fits to
$C^{2\text{pt}}$ and $C^{3\text{pt}}$ with common $M_i$.  Statistical precision of the data allowed, without overparameterization, four states in $C^{2\text{pt}}$ (labeled $\{4\}$ or 
$\{4^{N \pi}\}$), and three states in $C^{3\text{pt}}$ (labeled $\{3^\ast \}$). 
We also dropped the unresolved $\matrixe{2}{S}{2}$ term in Eq.~\eqref{eq:2pt3pt}. 
Keeping it increases the errors slightly but does not change the values. Using empirical
Bayesian priors for $M_{i}$ and $\mathcal{A}_i$ given in \suppcite{Table}{tab:priors}, 
we calculate $\spiN$ for two plausible but significantly different values 
of $M_1$ and $M_2$ in Table~\ref{tab:results} that give fits with similar $\chi^2$.}  A
similar strategy has been used in the analysis of axial-vector form
factors, where also the $N\pi$ state gives a large contribution as
discussed in Refs.~\cite{Jang:2019vkm,Park:2021ypf}.\looseness-1

Data for $C^{3\text{pt}}$, by Eq.~\eqref{eq:2pt3pt}, should be (i)
symmetric about $\tau/2$, and (ii) converge monotonically in $\tau$
for sufficiently large $\tau$, especially when a single excited state
dominates. These two conditions are, within errors, satisfied by the
data shown in Fig.~\ref{fig:Sfit}. 
In the simultaneous fits, $M_1$ and $M_2$ 
are mainly controlled by the 4-state fits to $C^{2\text{pt}}$,
however, as discussed in Refs.~\cite{Park:2021ypf,Gupta:2018qil},
there is a large region in $M_{i>0}$ in which the augmented $\chi^2$
of fits with different priors for $M_i$ is 
essentially the same, i.e., many
$M_i$ are plausible.  This region covers the towers of positive
parity $N\pi$, $N\pi \pi$, $\ldots$, multihadron states, labeled by
increasing relative momentum $\kk$, that can contribute and whose
energies start below those of radial excitations. To obtain
guidance on which excited states give large contributions to
$C^{3\text{pt}}$, we carried out a $\chi$PT analysis.\looseness-1

We study two well-motivated values of $M_1$ and $M_2$ for the analysis
of $C^{3\text{pt}}$.  The ``standard'' strategy (called the $\{4,3^\ast\}$
fit) imposes wide priors on \changed{$M_{i>0}$}, mostly to stabilize the fits, while the
$\{4^{N\pi},3^\ast\}$ fits use narrow-width priors for $M_1$ centered about
the noninteracting energy of the almost degenerate lowest positive parity multihadron
states, $N(\mathbf{1}) \pi(- \mathbf{1})$ or
$N(\mathbf{0}) \pi(\mathbf{0}) \pi( \mathbf{ 0})$. Thus, the label $N\pi$ 
implies that the contribution of both states is included.  Details of
extracting the $M_i$ from these two four-state fits, $\{4\}$ and
$\{4^{N\pi}\}$ to just $C^{2\text{pt}}$, can be found in
Ref.~\cite{Park:2021ypf}.  For the $a09m130$ ensemble with
$\mpi=138\MeV$, the $\{M_1,M_2\}$ are $\{1.5\changed{9},2.\changed{8}\}$ and
$\{1.22,1.8\}\GeV$ for the two cases, as shown in
Table~\ref{tab:results}.  The fits (see Fig.~\ref{fig:Sfit}) and the
$\chi^2$ with respect to $C^{3\text{pt}}$ data are equally
good, however, the results for the isoscalar charge $g_S^{u+d}$ differ
significantly.

The $\{4,3^\ast\}$ fit leads to a result
consistent with $\spiN \approx 40\MeV$, whereas the $\{4^{N\pi},3^\ast\}$ fit
gives $\approx 60\MeV$.  The major difference comes from
the disconnected quark loop diagram shown in Fig.~\ref{fig:conn_disc},
and is strongly $\mpi$ dependent---the effect of the $N\pi$ states is hard to
resolve in the $\mpi \approx 315\MeV$ data, debatable in the $230\MeV$ data, and
clear in the $\mpi = 138\MeV$ data. 

\begin{figure*}[t]
\subfigure{
    \includegraphics[width=0.24\linewidth]{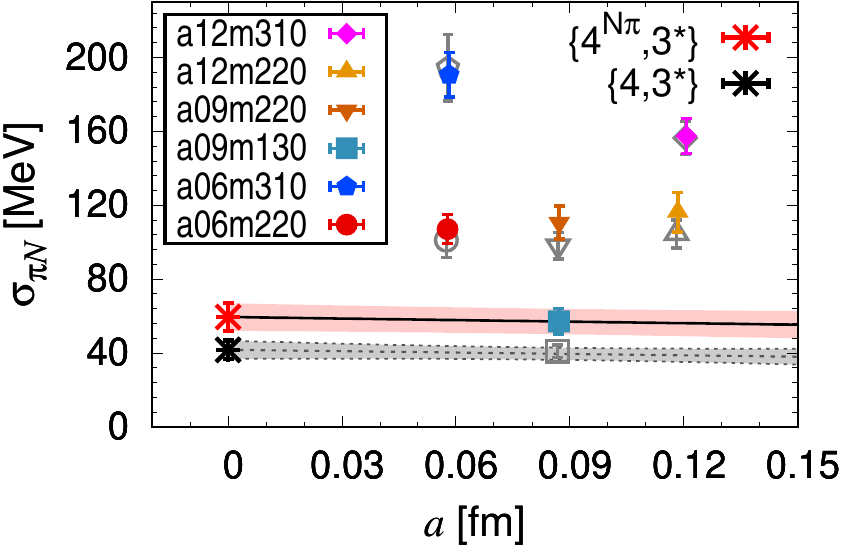}
    \includegraphics[width=0.24\linewidth]{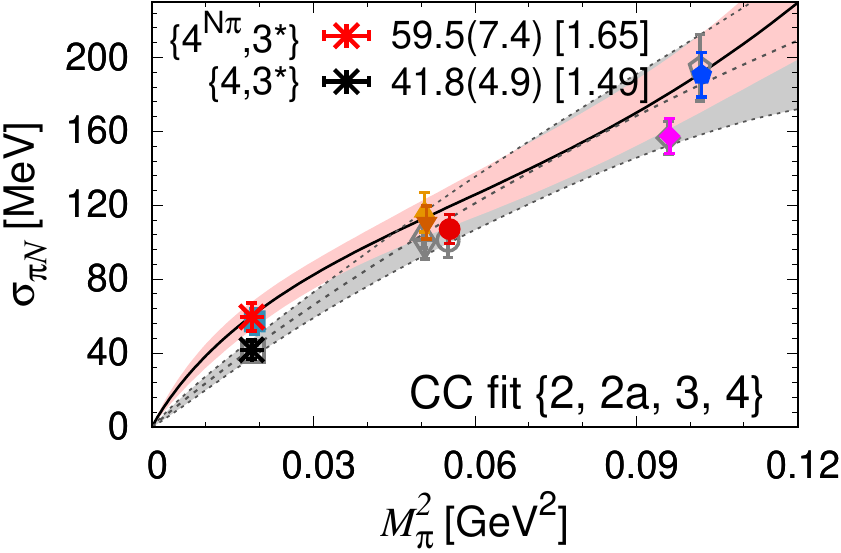}
    \includegraphics[width=0.24\linewidth]{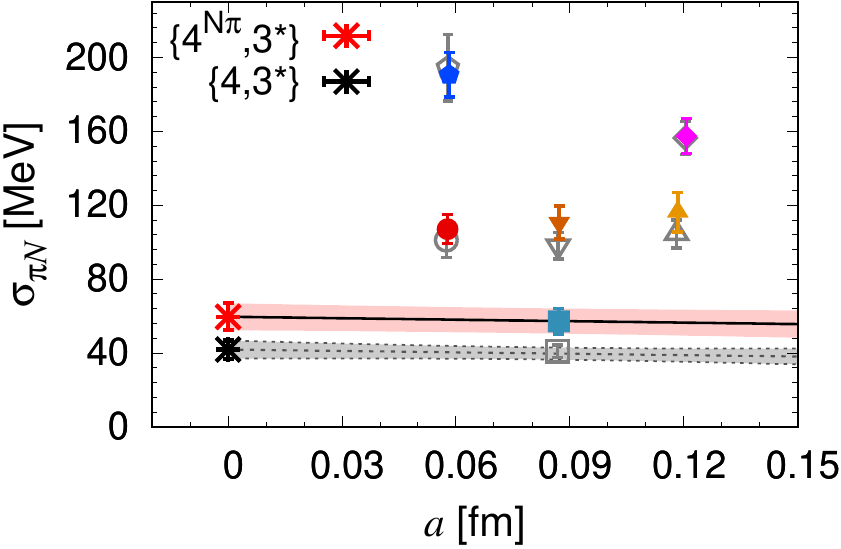}
    \includegraphics[width=0.24\linewidth]{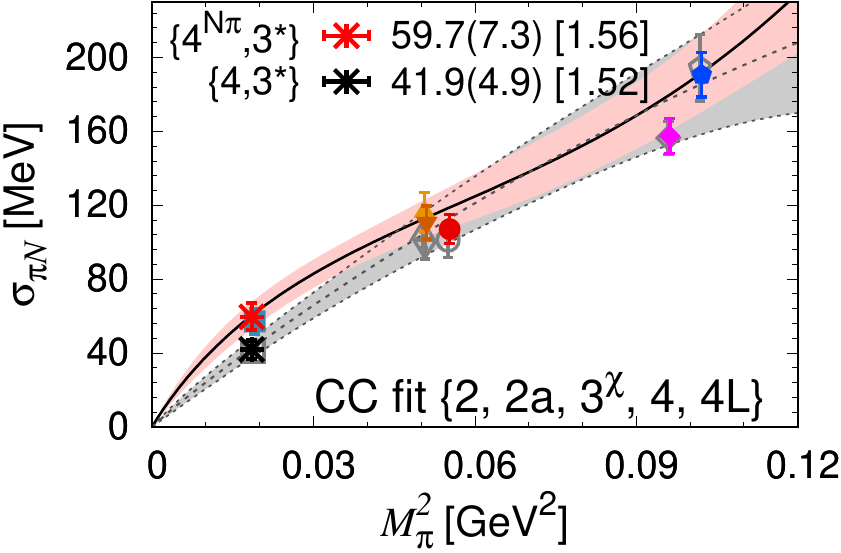}
}
\vspace{-0.1in}
\caption{Data for the $\sigma$-term, $\spiN = {m}_{ud} g_S^{u+d}$,
  from the two ESC strategies $\{4,3^\ast\}$ (gray) and
  $\{4^{N\pi},3^\ast\}$ (color) are shown as a function of $a$ and
  $\mpi^2$. The two left panels show the chiral-continuum (CC) fit  $\{2,2a,3,4\}$ and the two right the CC fit $\{2,2a,3^\chi,4, 4L\}$  described in the text.  The result at $\mpi=135\MeV$
  and [$\chi^2$/dof] of the two fits are given in
  the legend. \looseness-1
\label{fig:CCFV}}
\vspace{-0.1in}
\end{figure*}

It is important to point out that the values of $M_1$ 
and $M_2$ used in both fit
strategies are an effective bundling of the many excited states that
contribute into two.  In fact, as mentioned above, many combinations
of $M_1$ and $M_2$ between $\{4,3^\ast\}$ and
$\{4^{N\pi},3^\ast\}$ (see Table~\ref{tab:results}) give fits with 
equally good $\chi^2$ values. Ref.~\cite{Park:2021ypf} 
showed that for $\tau \gtrsim 1.0$~fm and for both fit strategies,
the dominant ESC in $C^{3\text{pt}}$ comes from the first excited
state. Thus, operationally, our two results for $\spiN$ should
be regarded as: what happens if the first ``effective'' 
excited state has $M_1 \approx 1220\MeV$ 
(motivated by $\chi$PT and corresponding to the
lowest theoretically possible states $N\pi$ or $N\pi\pi$) versus
$1600\MeV$ obtained from the standard fit to the 2-point function.  To further resolve all the excited
states that contribute significantly and their energies in a
finite box requires much higher precision data on additional
$M_\pi \approx 135$~MeV ensembles. In short, while our
$\{4^{N\pi},3^\ast\}$ analysis reconciles 
the lattice and the phenomenological values, it also calls for
validation in future calculations.

\section{Excited states in \texorpdfstring{$\boldsymbol{\chi}$PT}{\textchi PT}}

The contributions of low-momentum $N\pi$ and $N\pi\pi$ states to
 $C^{\rm 2pt}$ and $C^{\rm
  3pt}$ can be studied in $\chi$PT~\cite{Bar:2015zwa,Tiburzi:2015tta,Bar:2016uoj,Bar:2016jof,Bar:2018wco,Bar:2018xyi,Bar:2019gfx,Bar:2021crj}, a
low-energy effective field theory (EFT) of QCD that provides a
systematic expansion of $\mathcal R_S(\tau,t)$ in powers of
$Q/\Lambda_\chi$, where $Q$ denotes a low-energy scale of order of the
pion mass, $Q \in \{\mpi, t^{-1}, (\tau -t)^{-1}\}$, while
$\Lambda_\chi \approx 1\GeV$ is the typical scale of QCD.  In contrast to
the isovector scalar charge considered in Ref.~\cite{Bar:2016uoj}, we
find large contributions from the $N(\bf{k})\pi(-\bf k)$ and
$N(\bf{0}) \pi({\bf{k}}) \pi({-\bf{k}})$ states, which can give up to
$30\%$ corrections to $\mathcal R_S$ and thus affect the extraction of
$g_S^{u+d}$ and $\spiN$ in a significant way. 

The diagrams contributing to $\mathcal R_S$ are shown in
\changed{\suppcite{Fig.}{fig:chipt}},
 where we assume $\mathcal N(x)$ to be a local nucleon source with well defined transformation properties under chiral symmetry. The chiral representation of this class of sources has been derived in Refs.~\cite{Dmitrasinovic:2009vp,Bar:2015zwa,Tiburzi:2015tta}. Details of the calculation  at
next-to-next-to-leading order (N$^2$LO) in $\chi$PT and the expansion of $\mathcal N$ in terms of heavy nucleon and pion fields are summarized in \suppcite{App.}{sec:CPT}.
The crucial observation is that the isoscalar
scalar source couples strongly to two pions, so that loop diagrams
with the scalar source emitting two pions, which are consequently
absorbed by the nucleon, are suppressed by only one chiral order,
$Q/\Lambda_\chi$. These diagrams have both $N\pi$ and $N\pi\pi$ cuts,
which give rise to ESC to Euclidean Green's functions.
A second important effect is that the next-to-leading-order (NLO) couplings of the nucleon to
two pions, parameterized in $\chi$PT by the low-energy constants
(LECs) $c_{1,2,3}$, are sizable, 
reflecting the enhancement by degrees of freedom related to the $\Delta(1232)$. 
When the
pions couple to the isoscalar source, these couplings give rise to
large N$^2$LO corrections that are dominated by $N\pi\pi$ excited
states and have the same sign as the NLO correction.  Since, in the
isospin-symmetric limit, the isovector scalar source does not couple
to two pions, the NLO diagrams and the N$^2$LO diagrams proportional
to $c_{1,2,3}$ do not contribute to the isovector
3-point function, whose leading ESC arises at $\mathcal
O(Q^2/\Lambda^2_\chi)$. A detailed analysis showing that the functional form of the ESC predicted by $\chi$PT matches the lattice data and fits for sufficiently large time separations $\tau$ is given in \suppcite{App.}{sec:CPT}. In particular, the NLO and N$^2$LO ESC can each reduce $\spiN$ at a level of $10\MeV$, thus explaining the $\{4^{N\pi},3^\ast\}$ fits, i.e., a larger value when ESC is taken into account.

\section{Analysis of lattice data}
\label{sec:analysis}

Examples of fits with strategies $\{4,3^\ast\}$ and
$\{4^{N\pi},3^\ast\}$ to remove ESC and obtain $g_S^{u+d, \text{bare}}$ are shown in Fig.~\ref{fig:Sfit} and the results
summarized in Table~\ref{tab:results}.  The final results are obtained
from fits to the sum of the connected and disconnected contributions.
These values overlap in all cases with the sum of estimates from
separate fits to $g_S^{u+d,\,\text{conn}}$ and $g_S^{2l}$. From the
separate fits, we infer that most of the difference between the two
ESC strategies comes from the disconnected diagrams, which we
interpret as due to the $N \pi/N\pi \pi$ contributions through
quark-level diagrams such as shown in \changed{Fig.~\ref{fig:Npisketch}}. \looseness-1

Figure~\ref{fig:CCFV} shows data for $\spiN = {m}_{ud}^\text{bare}\ g_S^{u+d, \text{bare}}$ as a function of $a$ and $\mpi^2$. 
The chiral-continuum (CC) extrapolation  is carried out using
the N$^2$LO $\chi$PT expression~\cite{Hoferichter:2015hva}:\looseness-1
\begin{equation}
\sigma_{\pi N} =  (d_2 +d_{2}^{a} a) \mpi^2 + d_3  \mpi^3 + d_4 \mpi^4 + d_{4L} \mpi^4 \log \frac{\mpi^2}{\mN^2} \,.
\label{eq:CPT}
\end{equation}
The $d_i$ in $\chi$PT (henceforth labeled $d_i^\chi$) are given in
\suppeqcite{Eq.}{sigma_chiral} and  evaluated with $\mN=0.939\GeV$, $g_A=1.276$,
$F_\pi=92.3\MeV$.
Neglected finite-volume corrections can also be estimated in $\chi$PT, see \suppcite{App.}{sec:CPT} and Refs.~\cite{Gupta:2018qil,Lin:2017snn}, indicating a correction of less than $1\MeV$ for the $a09m130$ ensemble. \looseness-1

Figure~\ref{fig:CCFV} shows two chiral fits based on the N$^2$LO
$\chi$PT expression for $\spiN$. The $\{2,2a,3,4\}$ fit keeps terms
proportional to $\{d_2, d_{2}^{a}, d_3, d_{4}\}$ with all coefficients
free. In the $\{2,2a,3^\chi,4, 4L\}$ fit we use the $\chi$PT value for
$d_3 = d_3^\chi$, which does not involve any LECs, and include the
$d_{4L}$ term.  Each panel also shows the six data points obtained
with the $\{4^{},3^\ast\}$ and $\{4^{N\pi},3^\ast\}$ strategies and
the fits to them. In each fit $d_2^a$ comes out consistent with zero. 



The results for $\sigma_{\pi N}$ at the physical point $\mpi =  135\MeV$ 
from the various fits are essentially given by the $a09m130$
point.  We have neglected a correction due to flavor mixing inherent in  
Wilson-clover fermions since it is small as shown 
in \suppcite{App.}{sec:renormalization}. Our final result, 
$\sigma_{\pi N} = 59.6(7.4)$~MeV, is the average of results from the $\{2,2a,3,4\}$ and
$\{2,2a,3^\chi,4, 4L\}$ fits to the $\{4^{N\pi},3^\ast\}$ data \changed{given in Fig.~\ref{fig:CCFV}, which}
overlap.  In \suppcite{App.}{sec:CPT}, we consider more constrained fit
variants, which show that the fit coefficients of the $\mpi^2$ and
$\mpi^4\log \mpi^2$ terms are also broadly consistent with their $\chi$PT
prediction.

\section{Conclusions}
\label{sec:comparison}

\begin{figure}[htb]
\subfigure{
     \includegraphics[width=0.95\linewidth]{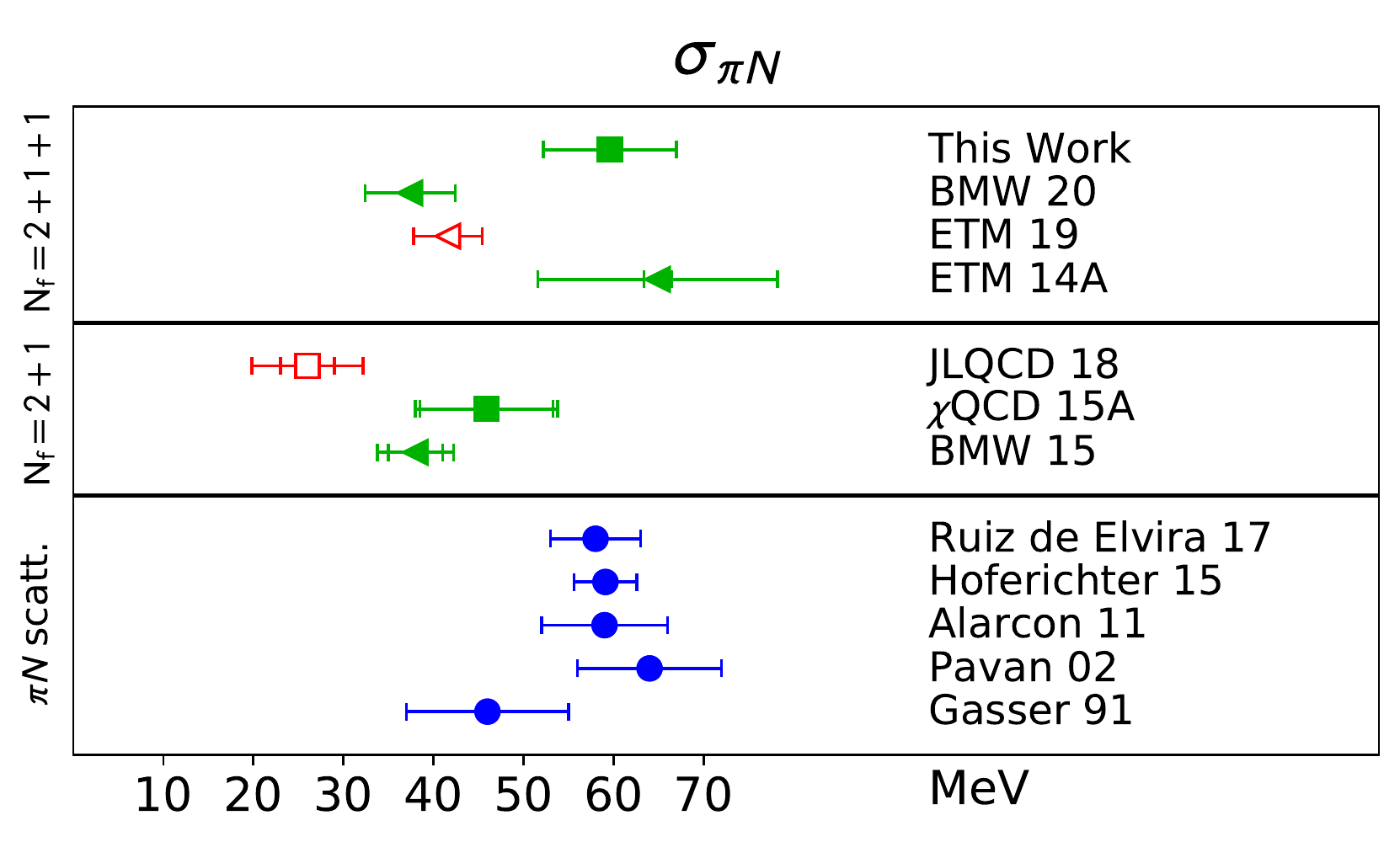}
}
\vspace{-0.1in}
\caption{Results for $\spiN = { m}_{ud}
  g_S^{u+d}$ from 2+1- and 2+1+1-flavor lattice calculations. 
  The BMW 20 result from 1+1+1+1-flavor lattices is listed along with the 
  other 2+1+1-flavor calculations for brevity. 
  Following the FLAG conventions, determinations via the
  direct approach are indicated by squares and the FH
  method by triangles. Also, the symbols used for lattice estimates that satisfy the FLAG criteria for inclusion in averages are filled green, 
  and those not included are  open red. 
The references from which lattice results have been taken are: 
JLQCD~18~\cite{Yamanaka:2018uud}, 
$\chi$QCD~15A~\cite{Yang:2015uis}, 
BMW~15~\cite{Durr:2015dna}, 
ETM~14A~\cite{Alexandrou:2014sha},
ETM 19~\cite{Alexandrou:2019brg}, and
BMW 20~\cite{Borsanyi:2020bpd}. 
  Phenomenological estimates using $\pi N$ scattering data (blue filled circles) are from 
  Gasser~91~\cite{Gasser:1990ce},
  Pavan~02~\cite{Pavan:2001wz},
  Alarcon 11~\cite{Alarcon:2011zs},
  Hoferichter 15~\cite{Hoferichter:2015dsa}, and 
  Ruiz de Elvira 17~\cite{RuizdeElvira:2017stg}. 
\looseness-1
\label{fig:FLAG}}
\vspace{-0.1in}
\end{figure}

Results for $\spiN$ were reviewed by FLAG in 2019~\cite{Aoki:2019cca},
and there have been two new calculations since as summarized in
\suppcite{App.}{sec:FLAG}, \changed{and shown in Fig.~\ref{fig:FLAG}.}  The ETM collaboration~\cite{Alexandrou:2019brg},
using the direct method on one physical mass 2+1+1-flavor twisted mass
clover-improved ensemble, obtained $\spiN = 41.6(3.8)\MeV$; the BMW
collaboration using the FH method and 33 ensembles of 1+1+1+1-flavor
Wilson-clover fermions~\cite{Borsanyi:2020bpd}, but all 
with $\mpi > 199\MeV$, find $\spiN = 37.4(5.1)\MeV$. The $\chi$PT analysis of the
impact of low-lying excited $N\pi$ states in the FH and direct methods
is the same, and as shown in Fig.~\ref{fig:CCFV}, it mainly affects
the behavior for $\mpi \lesssim 135\MeV$.  Our work indicates that
previous lattice calculations give the lower 
value $\spiN \approx 40\MeV$ because in the FH analysis~\cite{Borsanyi:2020bpd} the fit ansatz (Taylor or Pad\'e) parameters are determined using $\mpi \geq  199\MeV$ data, and in the direct method, the $N\pi / N\pi \pi$ states
have not been included when extracting the ground-state matrix
element~\cite{Alexandrou:2019brg}.

\begin{table*}[tbp]    
\begin{center}
\renewcommand{\arraystretch}{1.2} 
\begin{ruledtabular}
\begin{tabular}{l|cc|cc|cccc|ccc}
Ensemble ID     & $a$ (fm)   & $\mpi$ (MeV) & $(L/a)^3\times T/a$   & $\mpi L$ & $N_\text{conf}^{\rm 2pt}$ & $N_\text{conf}^{\rm conn}$ & $N_\text{LP}$ 
                & $N_\text{HP}$   & $N_\text{conf}^{l}$ & $N_\text{src}^l$   &  $N_{\rm LP}^l/N_{\rm HP}^l$  \\
\colrule
$a12m310 $      & 0.1207(11) & 310(3) & $24^3\times 64$ & 4.55      & 1013 & 1013        & 64      &  8  & 1013         & 5000               &  30   \\
$a12m220 $      & 0.1184(10) & 228(2) & $32^3\times 64$ & 4.38      & 959  &  744        & 64      &  4  &  958         & 11000              &  30   \\
\colrule
$a09m310 $      & 0.0888(08) & 313(3) & $32^3\times 96$ & 4.51      & 2263 & 2263        & 64      &  4  & --            & --                  &  --    \\
$a09m220 $      & 0.0872(07) & 226(2) & $48^3\times 96$ & 4.79      & 964  & 964         & 128     &  8  & 712          & 8000               &  30   \\
$a09m130 $      & 0.0871(06) & 138(1) & $64^3\times 96$ & 3.90      & 1274 & 1290        & 128     &  4  & 1270         & 10000              &  50   \\
\colrule
$a06m310 $      & 0.0582(04) & 320(2) & $48^3\times 144$& 4.52      & 977  & 500         & 128     &  4  & 808          & 12000              &  50   \\
$a06m220 $      & 0.0578(04) & 235(2) & $64^3\times 144$& 4.41      & 1010 & 649         & 64      &  4  & 1001         & 10000              &  50   \\
\end{tabular}
\end{ruledtabular}
\caption{Lattice parameters of the six ensembles analyzed for the
  isoscalar scalar charge $g_S^{u+d}$. Columns 6--9 give the
  number of configurations analyzed, and the number of low- (high-)
  precision measurements made per configuration for the connected
  contributions. Columns 10--12 give the number of configurations, low-precision random sources used, and the ratio $N_{\rm LP}/N_{\rm
    HP}$ for the disconnected contributions. The analysis of the
  isovector contributions has been presented in 
  Ref.~\cite{Gupta:2018qil}.  The ensemble $a09m310$ has been used only in the analysis of the nucleon mass in App.~\ref{sec:Mn}. }
\label{tab:ens}
\end{center}
\end{table*}

To conclude, a $\chi$PT analysis shows that the low-lying $N \pi$ and
$N \pi\pi$ states can make a significant contribution to $g_S^{u+d}$.
Including these states in our analysis (the $\{4^{N\pi},3^\ast\}$
strategy) gives $\spiN = 59.6(7.4)\MeV$, whereas the standard analysis
($\{4,3^\ast\}$ strategy) gives $\spiN = 41.9(4.9)\MeV$ consistent
with previous analyses~\cite{Aoki:2019cca}. These chiral fits are
shown in Fig.~\ref{fig:CCFV}.  Since the $\{4,3^\ast\}$ and
$\{4^{N\pi},3^\ast\}$ strategies to remove ESC are not distinguished
by the $\chi^2$ of the fits, we provide a detailed N$^2$LO $\chi$PT
analysis of ESC, which reveals sizable corrections consistent with the
$\{4^{N\pi},3^\ast\}$ analysis, restoring agreement with
phenomenology.  Since the effect of the $N\pi$ and $N\pi\pi$ states
becomes significant near $\mpi = 135\MeV$, further work on physical
mass ensembles is needed to validate our result and to increase the
precision in the extraction of the nucleon isoscalar scalar
charge.\looseness-1

\begin{acknowledgments}
We thank 
O.~B\"ar, L.~Lellouch, and A.~Walker-Loud for comments on the manuscript and
the MILC collaboration for providing
the 2+1+1-flavor HISQ lattices. The calculations used the Chroma
software suite~\cite{Edwards:2004sx}.  This research used resources at
(i) the National Energy Research Scientific Computing Center, a DOE
Office of Science User Facility supported by the Office of Science
of the U.S.\ Department of Energy under Contract No.\ DE-AC02-05CH11231; (ii) the Oak Ridge Leadership Computing Facility,
which is a DOE Office of Science User Facility supported under
Contract DE-AC05-00OR22725, and was awarded through the ALCC program
project LGT107; (iii) the USQCD collaboration, which is funded by the
Office of Science of the U.S.\ Department of Energy; and (iv)
Institutional Computing at Los Alamos National
Laboratory. T.~Bhattacharya and R.~Gupta were partly supported by the
U.S.\ Department of Energy, Office of Science, Office of High Energy
Physics under Contract No.\ DE-AC52-06NA25396. T.~Bhattacharya,
R.~Gupta, E.~Mereghetti, S.~Park, and B.~Yoon were partly supported by
the LANL LDRD program, and S.~Park by the Center for Nonlinear
Studies. M.~Hoferichter was supported by the Swiss National Science Foundation (Project No.\ PCEFP2\_181117). 
\end{acknowledgments}

\appendix

\section{Details of the Lattice Analysis}
\label{sec:Lparams}

The parameters of the seven 2+1+1-flavor ensembles generated using the
HISQ action~\cite{Follana:2006rc} by
the MILC collaboration~\cite{Bazavov:2012xda} are given in
Table~\ref{tab:ens}.  Note that the seventh ensemble, $a09m310$, listed has been used only 
for the analysis of the nucleon mass in App.~\ref{sec:Mn}. On each of these ensembles, the calculations of
the 2- and 3-point functions was carried out using tadpole improved
Wilson-clover fermions as described in Ref.~\cite{Gupta:2018qil}. 

To reduce ESC, smeared sources using the Wuppertal method~\cite{Gusken:1989ad} 
were used to generate quark propagators with parameters given in Ref.~\cite{Gupta:2018qil}. The same
smearing was used at the source and sink points. 

All correlation functions are constructed using the truncated solver method with
bias correction~\cite{Bali:2009hu,Blum:2012uh,Gupta:2018qil}. In this
method, high statistics are obtained by using a low-precision (LP)
stopping criterion in the inversion of the quark propagators, 
which was taken to be between 
$r_{\rm LP} \equiv |{\rm residue}|_{\rm LP}/|{\rm source}| = 10^{-3}$ and $5 \times 10^{-4}$. These estimates are
corrected for possible bias using high-precision (HP) measurements
with $r_{\rm HP}$ taken to be between $10^{-7}$ and
$10^{-8}$~\cite{Bhattacharya:2015wna,Gupta:2018qil}. The number of
configurations analyzed, and the number of LP and HP measurements made
for the connected and disconnected contributions, are given in
Table~\ref{tab:ens}.  In our data, the bias correction term was found
to be a fraction of the $1\sigma$ error in all quantities and 
for all six ensembles.

For the statistical analysis of the data, we first constructed
bias-corrected values for the 2- and 3-point correlation functions,
then averaged these over the multiple measurements made on each
configuration, and finally binned these. These binned data, 250--320
depending on the ensemble, were analyzed using the single
elimination jackknife process. The analysis was repeated to quantify
model variation of results by choosing data with different set of
source-sink separations $\tau$ and different number of points
$t_\text{skip}$, next to the source and the sink for each $\tau$, skipped
in the excited-state fits. The final result was taken to be the
average over the model values, weighting each by its Akaike
information criteria weight.

The bare quark mass is defined to be 
$m_{ud}^\text{bare} = 1/2\kappa_l - 1/2\kappa_c $, 
with the critical value of the hopping parameter,
$\kappa_c$, determined using a linear fit to $(a\mpi)^2$ versus $1/2\kappa$ at each of the three values of $a$. Results for $\mpi$ are given in 
Table~\ref{tab:ens} and fits to the nucleon mass are discussed in App.~\ref{sec:Mn}. A subtle point in the renormalization of $\spiN$ for Wilson-clover fermions is presented in App.~\ref{sec:renormalization}.

\changed{The final quoted errors are from the chiral fits shown in \mainref{Fig.}{fig:CCFV} and given in the labels. The error in each data point, $\sigma_{\pi N} = m^{\rm bare}_{ud} \times g_S^{u+d,{\rm bare}}$, combines in quadrature those in $m^{\rm bare}_{ud}$ and $g_S^{u+d,{\rm bare}}$ (see \mainref{Table}{tab:results}), with the latter given by the appropriate fit used to remove the ESC as illustrated in \mainref{Fig.}{fig:Sfit}.  }

In addition to the simultaneous fits to $C^{2\text{pt}}$ and  $C^{3\text{pt}}$ to remove ESC, we have also carried out the full analysis by first calculating the $M_i$ from 4-state fits to $C^{2\text{pt}}$ and using these as input in 3-state fits to  $C^{3\text{pt}}$ as described  in Ref.~\cite{Park:2021ypf}. \changed{The priors used for the excited state masses $M_i$ and the amplitudes $\mathcal{A}_i$ are given in Table~\ref{tab:priors} in App.~\ref{sec:Mn}. The two sets of results for $M_i$ from the two approaches (simultaneous versus individual fits) are consistent and the ground state matrix elements agree 
 within $1 \sigma$. This agreement occurs for both strategies, $\{4,3^\ast\}$ and $\{4^{N\pi},3^\ast\}$. The error estimates from the simultaneous fits used to get the final results are slightly larger.}

\begin{figure*}[th]
  \subfigure{ \includegraphics[width=0.98\textwidth]{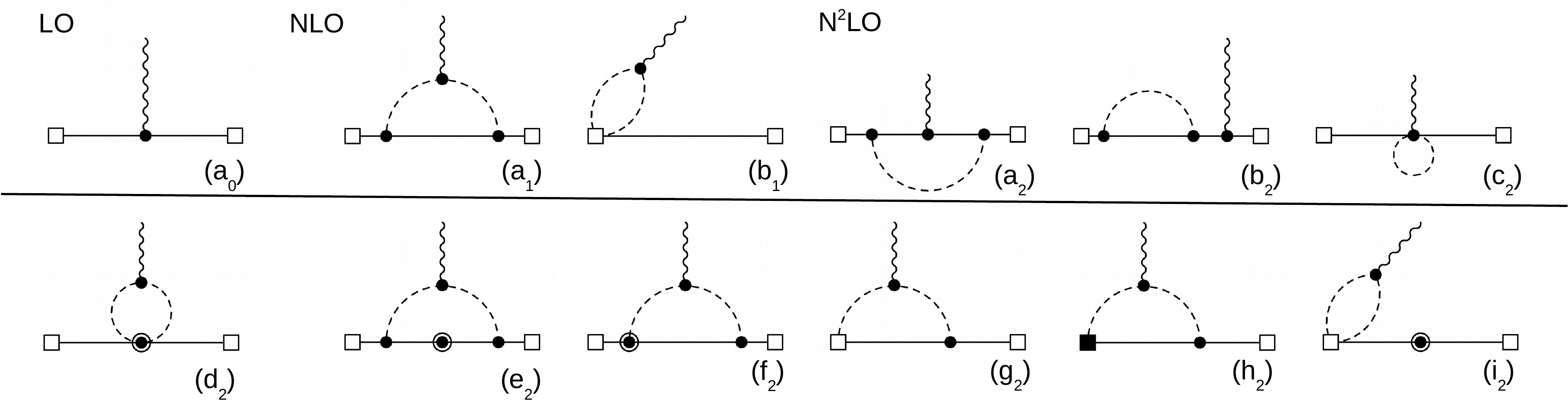}    }
\vspace{-0.1in}
\caption{Corrections to the scalar charge in $\chi$PT. An empty and full square denote, respectively, an insertion of the \changed{LO} and NLO expansion of the source fields $\mathcal N$ and $\bar{\mathcal N}$. 
Plain, dashed, and wavy lines denote, respectively, nucleons, pions, and an insertion of the scalar source. Dots and circled dots denote LO and NLO vertices in the chiral Lagrangian.
We show one diagram for each topology, and the inclusion of all the possible orderings is understood.
Diagrams $(g_2)$, $(h_2)$, and $(i_2)$
are representative of  
N$^2$LO corrections arising from the chiral expansion of $\mathcal N$. We do not consider these diagrams in our calculation,  
as they only produce N$^2$LO recoil corrections (including  $(g_2)$ despite containing only LO interactions~\cite{Bar:2015zwa,Tiburzi:2015tta}).
\looseness-1
  \label{fig:chipt}}
\vspace{-0.1in}
\end{figure*}

\section{Chiral Perturbation Theory}
\label{sec:CPT}

The corrections to the nucleon mass and the $\sigma$-term in $\chi$PT have
a long history in the literature~\cite{Gasser:1987rb,Bernard:1992qa,Bernard:1995dp,Bernard:1996nu,Borasoy:1996bx,Meissner:1997ii,Steininger:1998ya,Kambor:1998pi,McGovern:1998tm,Becher:1999he,Becher:2001hv,McGovern:2006fm,Schindler:2006ha,Schindler:2007dr}. The
LO, NLO, and 
N$^2$LO diagrams contributing to the isoscalar scalar
charge $g_S^{u+d}$ \changed{are shown in Fig.~\ref{fig:chipt}.}
In these diagrams, plain and dashed lines denote pions and nucleons in the  interaction picture of $\chi$PT, and not the nucleon and pion eigenstates of the full theory.

\begin{table*}[htbp]   
\centering
\begin{ruledtabular}
\begin{tabular}{cc|cc|cc|c|| c c|cc|cc}
     &  &    \multicolumn{2}{c|}{NLO}       & \multicolumn{2}{c|}{N$^2$LO} & $\delta \spiN(16,8)$
     &  \multicolumn{2}{c|}{NLO}  &
     \multicolumn{2}{c|}{N$^2$LO} & $\delta \spiN(12,6)$
     \\ 
$|\nn_\text{max}|$  & $m_{\nn}$        & loop  
        & source & $ c_i$ & recoil &  
        total & loop  
        & source & $ c_i$ & recoil &  
        total \\
\colrule                                                   
        $0$ & $1$ & $0$ & $-0.4$ & $-1.0$ & $-0.0$ & $-1.4$ & $0$ & $-0.5$ & $-1.2$ & $-0.1$ & $-1.7$\\
        $1$ & $6$ & $-2.2$ & $-0.6$ & $-2.9$ & $-0.2$ & $-5.9$ & $-2.7$ & $-0.9$ & $-4.3$ & $-0.2$ & $-8.0$\\
        $\sqrt{2}$ & $12$ & $-4.4$ & $-0.8$ & $-4.6$ & $-0.3$ & $-10.0$ & $-5.6$ & $-1.2$ & $-7.4$ & $-0.3$ & $-14.5$\\
        $\sqrt{3}$ & $8$ & $-5.2$ & $-0.8$ & $-5.2$ & $-0.3$ & $-11.6$ & $-6.7$ & $-1.3$ & $-8.7$ & $-0.4$ & $-17.1$\\
        $2$ & $6$ & $-5.6$ & $-0.9$ & $-5.5$ & $-0.3$ & $-12.2$ & $-7.3$ & $-1.3$ & $-9.4$ & $-0.4$ & $-18.4$\\
        $\sqrt{5}$ & $24$ & $-6.5$ & $-0.9$ & $-6.2$ & $-0.3$ & $-14.0$ & $-8.9$ & $-1.4$ & $-11.1$ & $-0.5$ & $-21.9$\\
        $\sqrt{6}$ & $24$ & $-7.2$ & $-0.9$ & $-6.7$ & $-0.4$ & $-15.2$ & $-10.1$ & $-1.5$ & $-12.4$ & $-0.5$ & $-24.5$\\
        $\sqrt{8}$ & $12$ & $-7.4$ & $-0.9$ & $-6.8$ & $-0.4$ & $-15.5$ & $-10.4$ & $-1.5$ & $-12.8$ & $-0.5$ & $-25.2$\\
        $3$ & $30$ & $-7.7$ & $-0.9$ & $-7.0$ & $-0.4$ & $-16.0$ & $-11.1$ & $-1.6$ & $-13.5$ & $-0.5$ & $-26.6$\\\colrule
        $\infty$ & & $-9.2$ & $-0.9$ & $-7.6$ & $-0.4$ & $-18.0$ & $-14.2$ & $-1.5$ & $-16.3$ & $-0.6$ & $-32.6$\\
        \end{tabular}
\end{ruledtabular}
\caption{Excited-state corrections to the $\sigma$-term (in MeV). $\delta \spiN(\tau,t)$, defined in Eq.~\eqref{deltagS}, is evaluated for $t = \tau/2 = 8a$ (left) and $t = \tau/2 = 6a$ (right),   
using the parameters of the $a09m130$ lattice ensemble listed in Table~\ref{tab:ens}.  
$|\nn_\text{max}|$ denotes the maximum momentum included in the sum in Eqs.~\eqref{R1},
\eqref{Rrecoil}, and \eqref{Rci}, $m_{\nn}$ the multiplicity of the momentum state~\cite{Colangelo:2003hf}.
We split the NLO corrections in Eq.~\eqref{R1} into the loop diagram $(a_1)$ in \changed{Fig.~\ref{fig:chipt}}, which also contributes to the ground state, and the diagram in which the scalar source couples to pions emitted by the nucleon source $\mathcal N$. At N$^2$LO, we label by $c_i$ 
the contributions in Eq.~\eqref{Rci}
and by ``recoil'' those in Eq.~\eqref{Rrecoil}. The last line is evaluated in the continuum.
\looseness-1}
\label{tab:chiralexcited}
\end{table*}

These diagrams lead to the expansion 
\begin{equation}\label{sigma1}
g_S^{u+d}  = g_S^{(0)} + g_S^{(1)}\frac{\mpi}{\Lambda_\chi} + g_S^{(2)} \frac{\mpi^2}{\Lambda_\chi^2} + \ldots,  
\end{equation}
where $\Lambda_\chi = 4 \pi F_\pi \approx 1$ GeV is the breakdown scale
of the chiral expansion. The NLO diagrams $(a_1)$ and $(b_1)$ in
\changed{Fig.~\ref{fig:chipt}}  only contribute to the isoscalar channel, implying that the isovector channel has the different expansion
\begin{equation}
g_S^{u-d}  =  \tilde g_S^{(0)}+  \tilde g_S^{(2)}\frac{\mpi^2}{\Lambda_\chi^2} + \ldots 
\end{equation}
We will show that the same loop diagrams responsible for the NLO and
N$^2$LO corrections to $g_S$ also induce a sizable contribution from
$N \pi$ and $N\pi \pi $ intermediate states\changed{.}

To this end, we calculate the ratio $\mathcal R_S(\tau, t)$ using heavy-baryon
$\chi$PT and expand it as
\begin{equation}
\mathcal R(\tau, t)  =  \mathcal R^{(0)}(\tau, t) + \mathcal R^{(1)}(\tau, t) + \mathcal R^{(2)}(\tau, t),
\end{equation}
including in its definition a factor $m_{ud}$ to make the result $\mathcal R = m_{ud} \mathcal R_S$ scale independent and ensure a normalization that facilitates the comparison to $\spiN$.
We assume $\mathcal N$ to be a local nucleon source,
transforming as $\left(\frac{1}{2},0\right)\oplus \left(0,\frac{1}{2}\right)$  under the chiral group $SU(2)_L \otimes SU(2)_R$. The heavy-baryon $\chi$PT realization of $\mathcal N$ was constructed in Ref.~\cite{Tiburzi:2015tta}
\begin{equation}\label{chisource}
    \mathcal N(x) = \left[ \left(1 - \frac{\boldsymbol{\pi}^2}{8 F_\pi^2}\right) - i \frac{\boldsymbol{\pi} \cdot \boldsymbol{\tau}}{2 F_\pi} \gamma_5 \right] N_v 
 + \mathcal O\left(\frac{1}{\mN}\right),
\end{equation}
where $N_v = (1+\gamma_4) N_v/2$ represents a heavy-nucleon field.
At $\mathcal O(1/M_N)$, Eq.~\eqref{chisource} contains additional LECs, which reduce the predictive power of the calculation.

At LO one simply has  $\mathcal R^{(0)} = m_{ud} g_S^{(0)}$.
The NLO diagrams receive contributions from nucleon, $N \pi $, and $N \pi\pi $ excited states, leading to  
\begin{align}\label{R1}
\mathcal R^{(1)}(\tau, t) &= \frac{3 g^2_A \mpi^2 }{8 F_\pi^2  L^3} \sum_{\kk}  \frac{\kk^2}{E_\pi^4}
 \bigg[ 1 -  
  e^{-E_{N \pi} t}  - e^{- E_{N \pi} t_B }  \notag\\
  &+ \frac{1}{2}  e^{ - E_{N \pi} \tau  } 
 + \frac{1}{4}e^{- 2 E_\pi t }  + \frac{1}{4} e^{- 2 E_\pi t_B } 
   \bigg] \nonumber \\
&      - \frac{3 \mpi^2 }{32 F_\pi^2} \frac{1}{L^3} \sum_{\kk}  \frac{1}{E_\pi^2}
    \left(e^{-2 E_\pi t } + e^{-2 E_\pi t_B}\right),
\end{align}
where $E_\pi = \sqrt{\kk^2 + \mpi^2}$, $\widetilde E_N = \sqrt{\mN^2 + \kk^2} - \mN$, $E_{N \pi}=E_\pi+\tilde E_N$,
$\kk = 2 \pi \nn/L$,
and $t_B = \tau - t$.  $F_\pi = 92.3\MeV$ is the pion decay constant, $g_A=1.276$ the axial charge of the nucleon~\cite{Zyla:2020zbs}.
The first term in Eq.~\eqref{R1} is a correction to the ground-state
contribution
\begin{equation}\label{sigma_cont}
    \frac{3 g^2_A \mpi^2 }{8 F_\pi^2} \frac{1}{L^3} \sum_{\kk}  \frac{\kk^2}{E_\pi^4} =  - \frac{9 g^2_A \mpi^3}{64 \pi F_\pi^2} + \Delta_L \spiN,  
\end{equation}
where $\Delta_L \spiN$ is the finite-volume correction to the $\sigma$-term~\cite{Beane:2004tw}.  The remaining terms reflect the excited states, with diagram $(a_1)$ receiving a contribution from 
 $N \pi $ intermediate states (nucleon and pion having opposite momenta) and from an $N \pi\pi $ state (the nucleon at rest and the two pions carrying momenta $\pm \kk$). 
The $N \pi $ and $N \pi\pi $ states with zero pion momentum vanish
due to the prefactor $\kk^2$. The amplitude of the $N \pi $ contribution is suppressed by $(\mpi L)^3$ compared to $g_S^{(0)}$, but enhanced by the large
coupling of a scalar source to the pion, making it suppressed by only a single chiral order. The last line of Eq.~\eqref{R1} originates from diagram $(b_1)$.
In this case the dominant excited state is $N \pi\pi $, with the two pions at zero momentum.
Diagram $(g_2)$ arises from the last term in the square bracket in Eq.~\eqref{chisource}.
Though formally NLO, this diagram vanishes up to $\mathcal O(1/\mN)$ corrections, and thus the topology $(g_2)$ only contributes to N$^2$LO. Similarly, the diagram with the pion emitted by $\mathcal N$ and absorbed by $\bar{\mathcal N}$ vanishes at NLO.

\begin{figure*}[tb]
  \subfigure{ 
\includegraphics[width=0.44\textwidth]{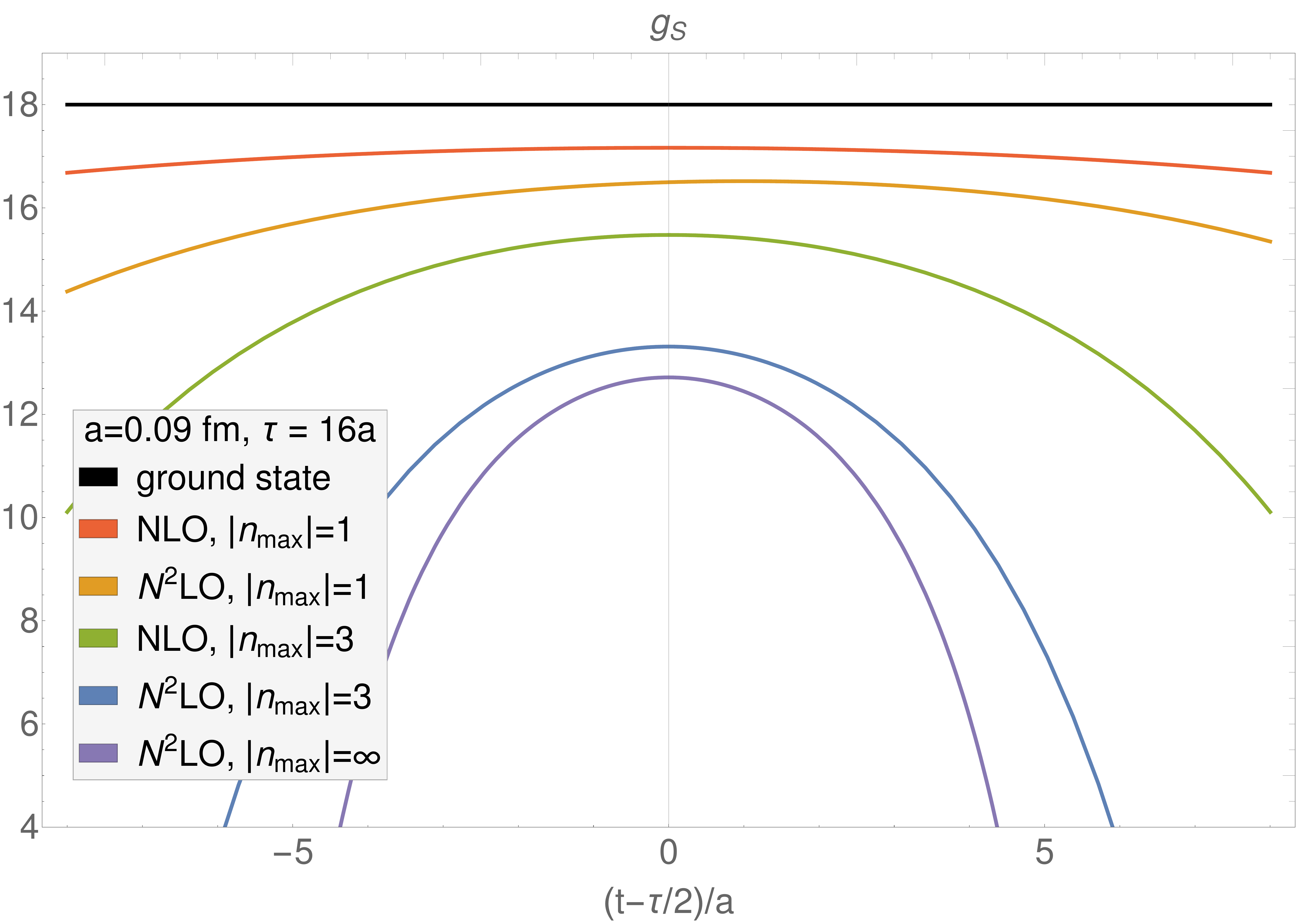}    
\includegraphics[width=0.44\textwidth]{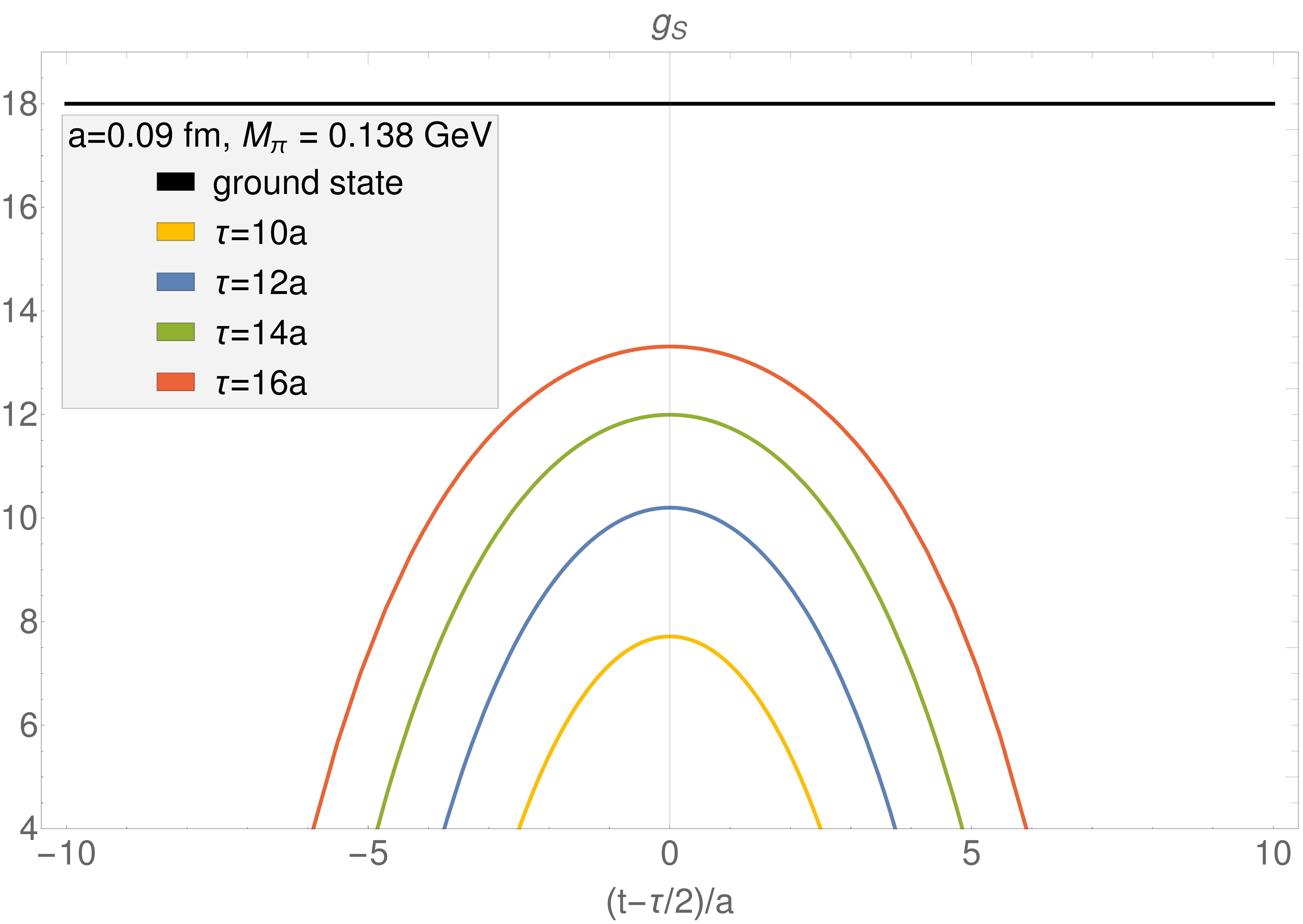}    
}
\vspace{-0.1in}
\caption{(Left) Excited-state corrections from different truncations
  to the isoscalar scalar charge $g_S$ in $\chi$PT.  (Right) Estimates for
  ${\cal R}_S(\tau,t)$ from the N${}^2$LO analysis for the $a09m130$
  ensemble, which should be compared to the data in
  \changed{\mainref{Fig.}{fig:Sfit}}  (and the shape with that of the separate contributions shown in Fig.~\ref{fig:Sdata}). We assume $g_S = 18$ is the asymptotic value in
  both cases. \looseness-1
  \label{fig:excited}}
\end{figure*}

\begin{figure*}[tb]
  \subfigure{
    \includegraphics[width=0.235\linewidth]{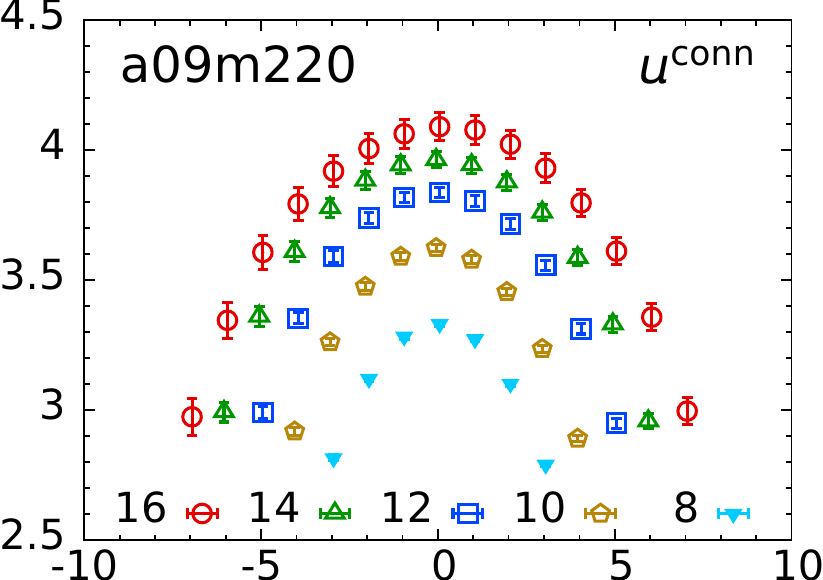}  \hspace{0.001\linewidth}
    \includegraphics[width=0.235\linewidth]{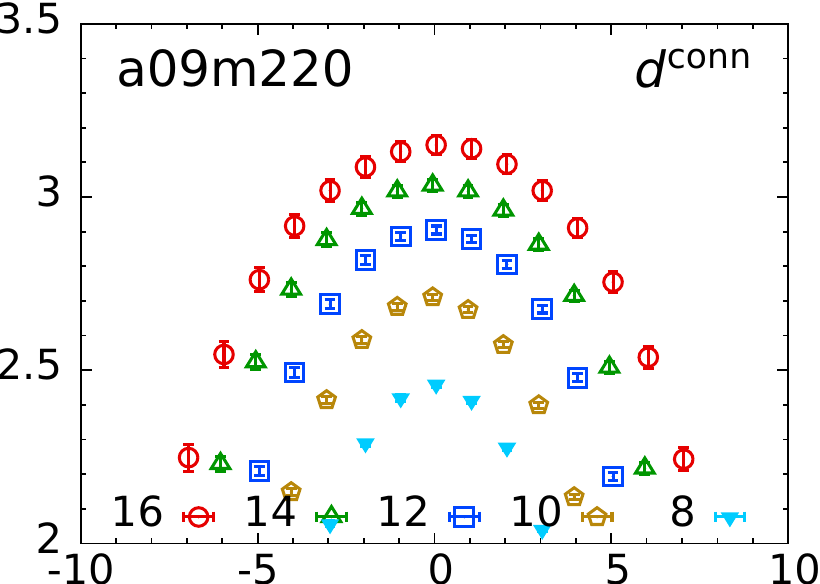}  \hspace{0.001\linewidth}
    \includegraphics[width=0.235\linewidth]{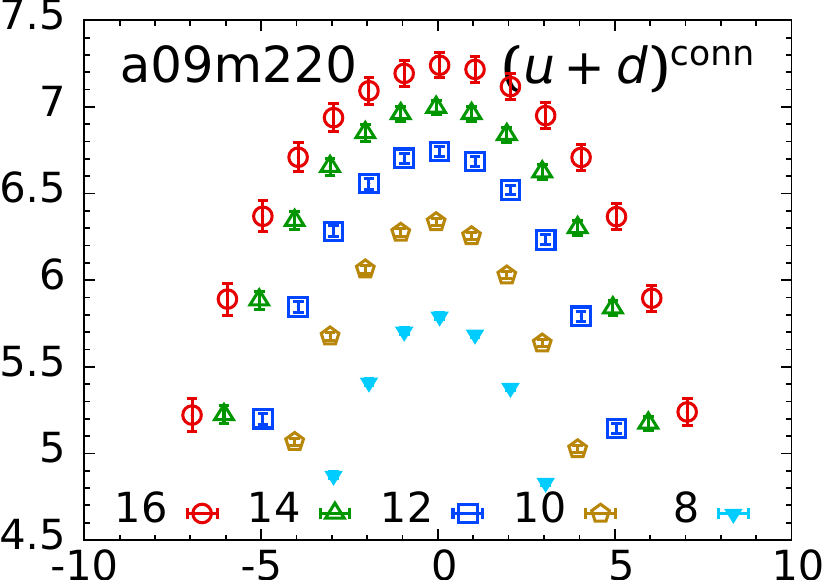}  \hspace{0.001\linewidth}
    \includegraphics[width=0.235\linewidth]{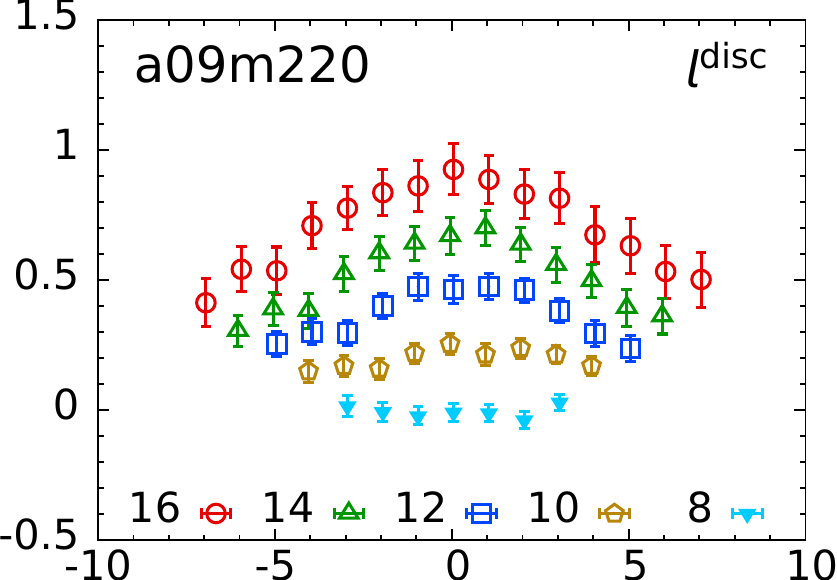}  
  }
  \subfigure{
    \includegraphics[width=0.235\linewidth]{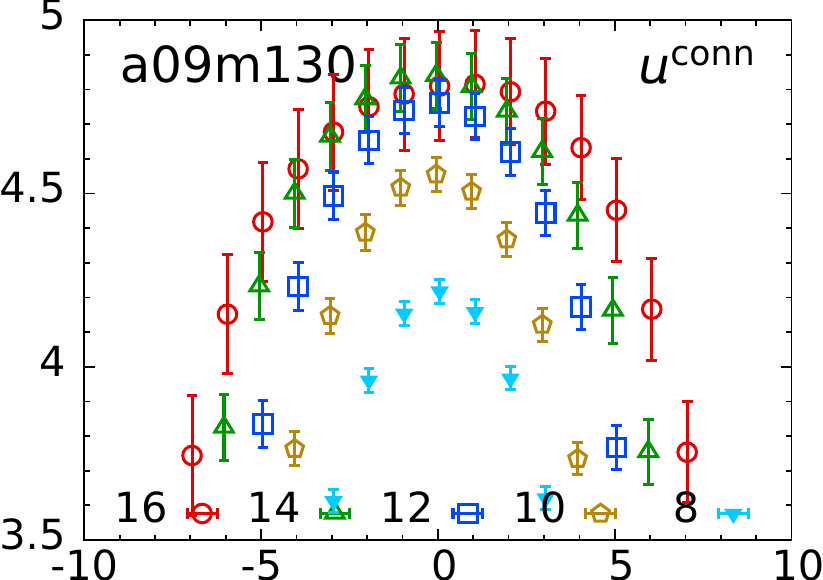}  \hspace{0.001\linewidth}
    \includegraphics[width=0.235\linewidth]{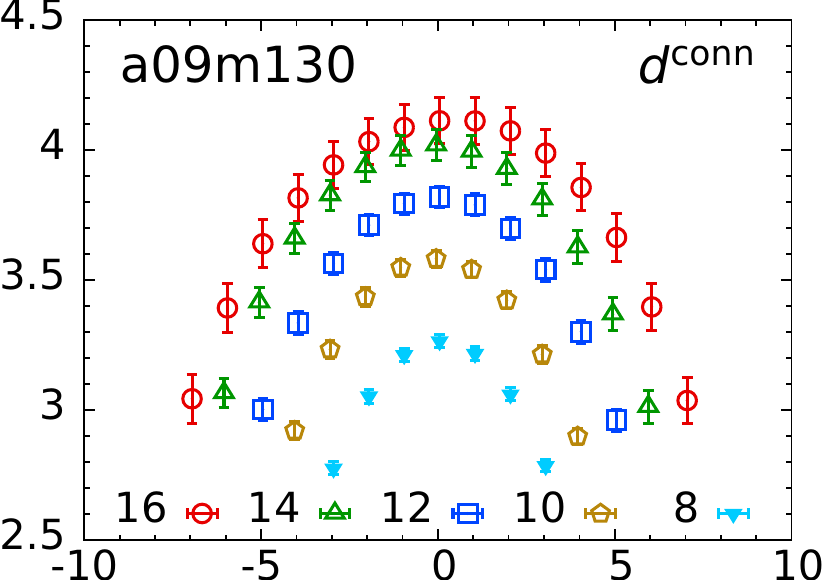}  \hspace{0.001\linewidth}
    \includegraphics[width=0.235\linewidth]{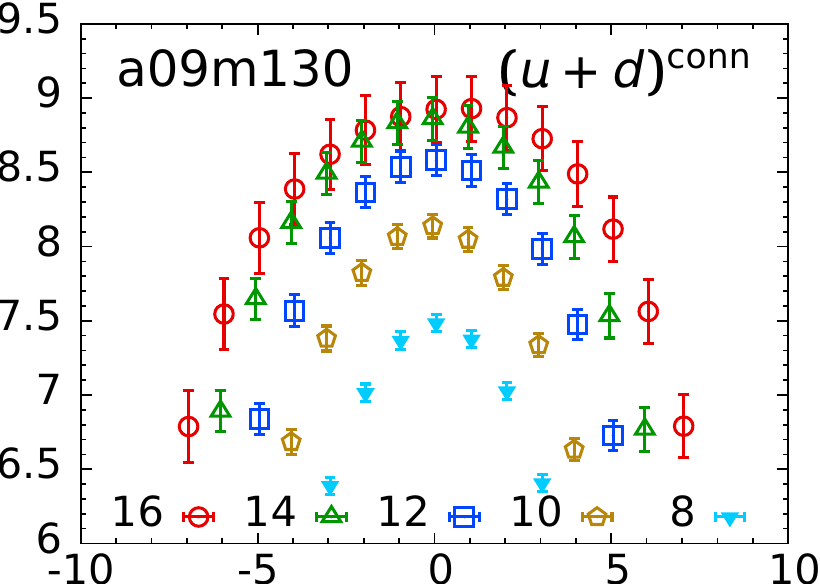}  \hspace{0.001\linewidth}
    \includegraphics[width=0.235\linewidth]{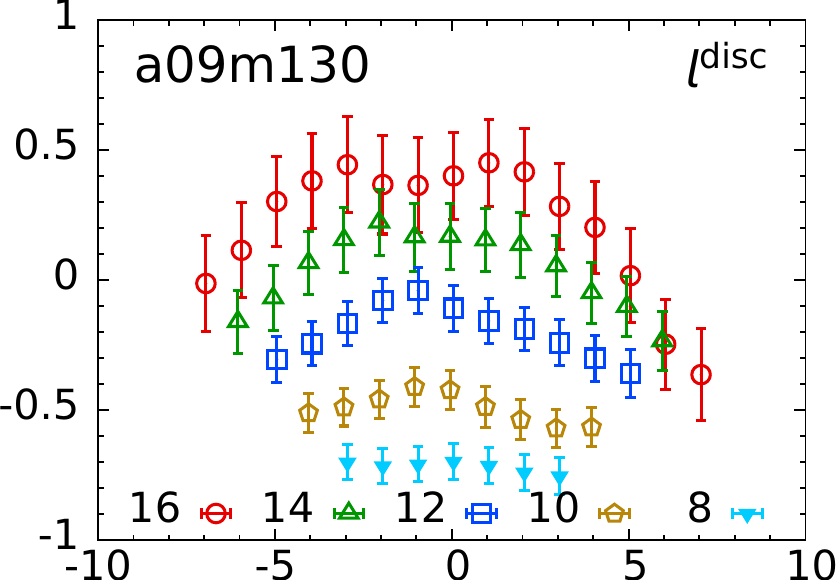}  
  }
\vspace{-0.1in}
\caption{Data for the ratio ${\cal R}_\mathcal{S} = C({\rm 3pt})/C({\rm 2pt})$ for $\tau/a$ listed in the label are shown for the  $u$, $d$, and $(u+d)$ connected, and light quark~$l$ disconnected contributions, and plotted versus $(t - \tau/2)/a$  
  for the $a09m220$ (top) and   $a09m130$ (bottom) ensembles.    \looseness-1
  \label{fig:Sdata}}
\vspace{-0.1in}
\end{figure*}

At N$^2$LO there are several contributions.
They come from loop corrections to the LO (diagrams $(a_2)$, $(b_2)$, and $(c_2)$)
and from diagrams with subleading interactions in the chiral Lagrangian and the scalar source coupling to two pions (diagrams $(d_2)$, $(e_2)$, and $(f_2)$).
Here, 
diagrams $(a_2)$ and $(b_2)$  are exactly canceled by the N$^2$LO corrections to
the 2-point function. This is in contrast to the isovector case, in which they are responsible   
for the leading ESC:
\begin{align}
\mathcal R^{(2)}_\text{isovector}(\tau,t)&= - m_{ud}\tilde g^{(0)}_S 
\frac{ g_A^2  }{2 F_\pi^2} 
  \frac{1}{ L^3} \sum_{\kk}  \frac{  \kk^2}{ E^3_\pi} \notag\\
  &\times \bigg[ 1 
  - e^{-E_\pi t}  - e^{-E_\pi t_B} 
  + e^{-E_\pi \tau} \bigg].
\end{align}
This result agrees with the corrections to the isovector scalar charge
computed in Ref.~\cite{Bar:2016uoj}, once we expand in the limit $\mN
\gg |\kk|$.

\begin{figure*}[htb]
\subfigure{
     \includegraphics[width=0.24\linewidth]{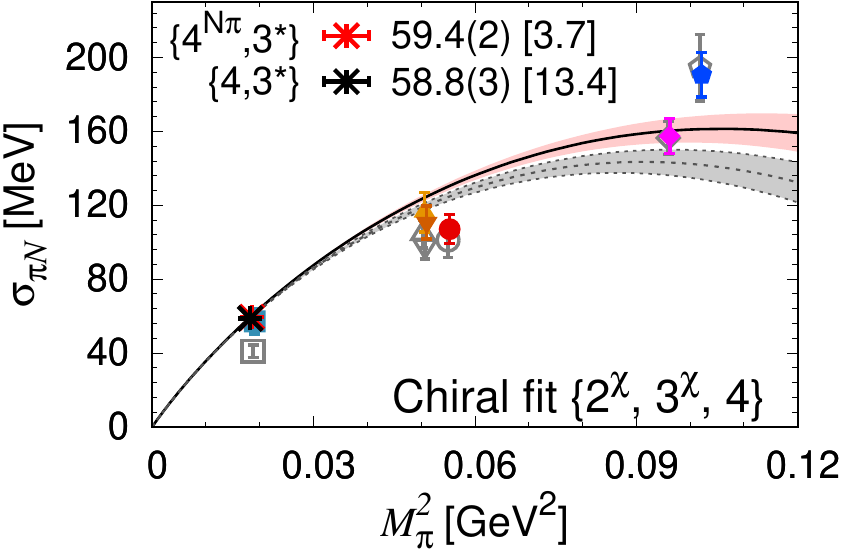}
     \includegraphics[width=0.24\linewidth]{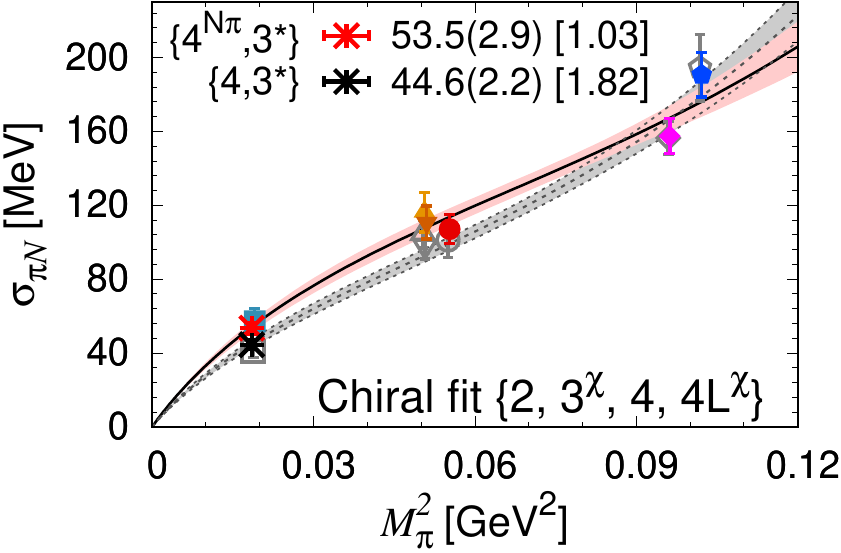}
     \includegraphics[width=0.24\linewidth]{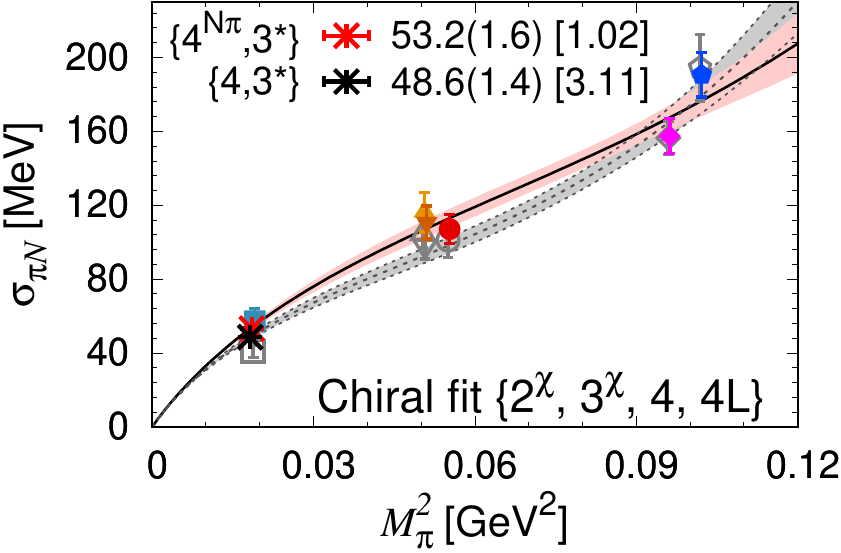}
     \includegraphics[width=0.24\linewidth]{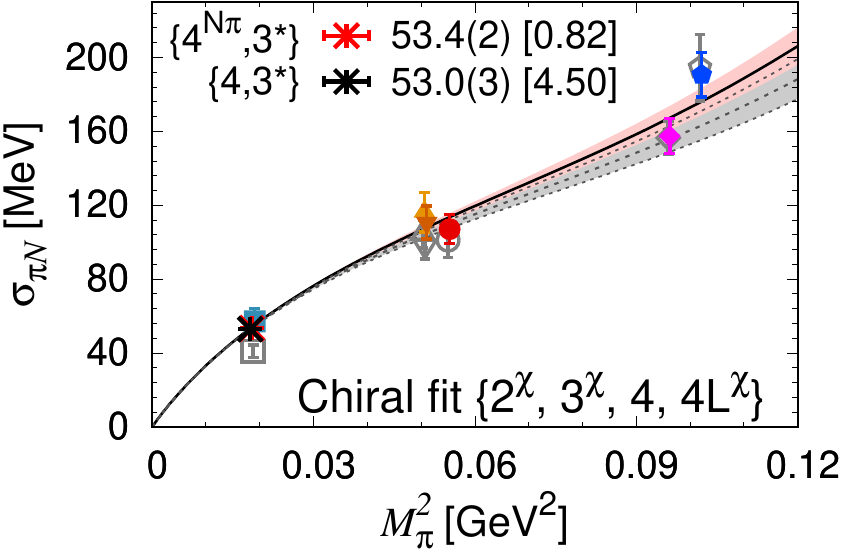}
}
\vspace{-0.1in}
\caption{Chiral fits to the data for the $\sigma$-term, $\spiN = {m}_{ud} g_S^{u+d}$,
  from the two ESC strategies $\{4,3^\ast\}$ (gray) and
  $\{4^{N\pi},3^\ast\}$ (color) shown as a function of $\mpi^2$. The four 
  chiral fits are $\{2^\chi,3^\chi,4\}$,
$\{2,3^\chi,4, 4L^\chi\}$, $\{2^\chi,3^\chi,4,4L\}$, and
$\{2^\chi,3^\chi,4,4L^\chi\}$. 
  The result
  at $\mpi=135\MeV$  and the [$\chi^2$/dof] of the chiral fit for the two ESC strategies are given in   the legend. \looseness-1
\label{fig:CCFV2}}
\vspace{-0.1in}
\end{figure*}

\begin{table*}
{\footnotesize
\begin{ruledtabular}
  \begin{tabular}{c | cccccc | cccccc }
chiral fit & \multicolumn{6}{c|}{$\{4,3^\ast\}$} & \multicolumn{6}{c}{$\{4^{N\pi},3^\ast\}$} \\
   & $d_2$  & $d_3$ & $d_4$ & $d_{4L}$ & $\frac{\chi^2}{\text{dof}}$ & $\spiN $ 
   & $d_2$  & $d_3$ & $d_4$ & $d_{4L}$ & $\frac{\chi^2}{\text{dof}}$ & $\spiN $ 
\\ 
   & $(\text{GeV}^{-1})$ & $(\text{GeV}^{-2})$ & $(\text{GeV}^{-3})$ & $(\text{GeV}^{-3})$ & & (MeV) 
   & $(\text{GeV}^{-1})$ & $(\text{GeV}^{-2})$ & $(\text{GeV}^{-3})$ & $(\text{GeV}^{-3})$ & & (MeV) 
\\ \colrule
$\chi$PT & $4.44$ & $-8.55$ & -- & $11.35$ & -- & -- & $4.44$ & $-8.55$ & -- & $11.35$ & -- & --  \\ \colrule
                    $\{2,3\}$ &   $2.56(26)$ &  $-2.77(98)$ &           -- &           -- & $0.83$ & $39.8(2.4)$ &   $3.30(36)$ &  $-5.1(1.3)$ &           -- &           -- & $1.63$ & $47.7(3.5)$ \\
                  $\{2,3,4\}$ &   $2.49(94)$ &  $-2.1(8.1)$ &     $-1(17)$ &           -- & $1.11$ & $39.7(3.3)$ &   $6.0(1.7)$ &    $-27(13)$ &     $42(26)$ &           -- & $1.28$ & $57.4(6.9)$ \\
        $\{2^\chi,3^\chi,4\}$ &       $4.44$ &      $-8.55$ &  $-3.11(75)$ &           -- & $13.4$ & $58.82(25)$ &       $4.44$ &      $-8.55$ &  $-1.25(70)$ &           -- & $3.71$ & $59.44(23)$ \\
          $\{2,3^\chi,4,4L\}$ &   $2.84(53)$ &      $-8.55$ &  $0.2(15.6)$ &  $-7.0(9.3)$ & $1.13$ & $39.8(3.3)$ &   $5.1(1.0)$ &      $-8.55$ &     $41(23)$ &     $22(15)$ & $1.21$ & $57.7(6.8)$ \\
     $\{2,3^\chi,4,4L^\chi\}$ &   $3.84(15)$ &      $-8.55$ &  $30.8(2.0)$ &      $11.35$ & $1.82$ & $44.6(2.2)$ &   $4.45(20)$ &      $-8.55$ &  $24.6(2.4)$ &      $11.35$ & $1.03$ & $53.5(2.9)$ \\
     $\{2^\chi,3^\chi,4,4L\}$ &       $4.44$ &      $-8.55$ &  $43.3(6.3)$ &  $19.9(2.7)$ & $3.11$ & $48.6(1.4)$ &       $4.44$ &      $-8.55$ &  $25.4(7.0)$ &  $11.7(3.1)$ & $1.02$ & $53.2(1.6)$ \\
$\{2^\chi,3^\chi,4,4L^\chi\}$ &       $4.44$ &      $-8.55$ &  $23.40(75)$ &      $11.35$ & $4.50$ & $53.01(25)$ &       $4.44$ &      $-8.55$ &  $24.64(70)$ &      $11.35$ & $0.82$ & $53.42(23)$ \\
\end{tabular}  
\end{ruledtabular}
}
\vspace{0.1cm}
{\tiny
\begin{ruledtabular}
  \begin{tabular}{c | ccccccc | ccccccc }
   & $d_2$ & $d_2^a$  & $d_3$ & $d_4$ & $d_{4L}$ & $\frac{\chi^2}{\text{dof}}$ & $\spiN $ 
   & $d_2$ & $d_2^a$  & $d_3$ & $d_4$ & $d_{4L}$ & $\frac{\chi^2}{\text{dof}}$ & $\spiN $ 
\\ 
   & $(\text{GeV}^{-1})$ & & $(\text{GeV}^{-2})$ & $(\text{GeV}^{-3})$ & $(\text{GeV}^{-3})$ & & (MeV) 
   & $(\text{GeV}^{-1})$ & & $(\text{GeV}^{-2})$ & $(\text{GeV}^{-3})$ & $(\text{GeV}^{-3})$ & & (MeV) 
\\ \colrule
                  $\{2,2a,3,4\}$ &   $2.61(96)$ & $-0.3(5)$ &  $-2.2(8.1)$ &     $-1(17)$ &           -- & $1.49$ & $41.8(4.9)$ &   $6.1(1.7)$ & $-0.3(4)$ &    $-27(13)$ &     $42(26)$ &           -- & $1.65$ & $59.5(7.4)$ \\
          $\{2,2a,3^\chi,4,4L\}$ &   $2.95(57)$ & $-0.3(5)$ &      $-8.55$ &  $0.6(15.6)$ &  $-7.0(9.3)$ & $1.52$ & $41.9(4.9)$ &   $5.2(1.0)$ & $-0.3(4)$ &      $-8.55$ &     $40(23)$ &     $21(15)$ & $1.56$ & $59.7(7.3)$ \\
\end{tabular}  
\end{ruledtabular}
}
\caption{Chiral fit coefficients with the ansatz $d_2 \mpi^2 + d_3
  \mpi^3 + d_4 \mpi^4 + d_{4L} \mpi^4 \log (\mpi/\mN)^2 $. Here
  the $\chi$PT values for $d_i$ are calculated using Eq.~\eqref{ci} and 
  with $\mN=0.939\GeV$, $g_A=1.276$, $F_\pi=92.3\MeV$. $d_{4L}$ includes the chiral logarithm in $\bar l_3=-\log \mpi^2+$ finite. The lower table gives the coefficients for the fits including an $a\mpi^2$ term, as discussed in Sec.~\ref{sec:analysis}.}
\label{tab:CPTcoeffs}
\end{table*}

Diagram $(c_2)$ only contributes to the ground state, and diagrams $(d_2)$
and $(e_2)$ are recoil corrections. A first effect of these diagrams is to shift the energy excitation of the $N \pi$ state from $E_\pi$ to $E_\pi + \widetilde E_N$. The remaining recoil corrections are 
\begin{align}\label{Rrecoil}
 \mathcal R^{(2)}_\text{recoil}&=-   \frac{9 g^2_A \mpi^2}{32 \mN F_\pi^2L^3} \sum_{\kk} \frac{(\kk^2)^2}{E_\pi^{5}}
    \bigg[ 1 - \frac{2}{3} e^{-E_\pi t} \notag\\
    &- \frac{2}{3}
    e^{- E_\pi t_B} + \frac{2}{3} e^{- E_\pi \tau} - \frac{1}{6} 
    e^{- 2 E_\pi t} - \frac{1}{6} e^{-2 E_\pi t_B}
    \bigg] \notag \\
  &+ \frac{3 g_A^2 \mpi^2}{32 \mN F_\pi^2 L^3} \sum_{\kk} \frac{1}{E_\pi} \bigg[1 -
  \frac{1}{2}e^{-2 E_\pi t} - \frac{1}{2}e^{-2 E_\pi t_B}\notag\\
  &+\frac{2\kk^2}{E_\pi^2} \Big(1 -
  e^{-E_\pi t} - e^{-E_\pi t_B}+ e^{-E_\pi \tau}
  \Big)
  \bigg].
\end{align}
Finally, diagram $(d_2)$ receives 
contributions from the LECs $c_{1,2,3}$, which contribute to $\pi N$ scattering at NLO in $\chi$PT. Including diagram $(c_2)$, they give 
\begin{align}\label{Rci}
\mathcal R^{(2)}_{c_i} &= -   \frac{3 \mpi^2}{4 F_\pi^2}
 \frac{1}{  L^3} \sum_{\kk}  \frac{ 1  }{E_\pi^3} 
\Big(  (c_2 + 2 c_3) E_\pi^2 \notag\\
&+ (2 c_1 -c_3) \mpi^2\Big)
 \bigg[ 1 -  
 \frac{1}{2} e^{-2 E_\pi  t} - \frac{1}{2}e^{-2 E_\pi t_B }  
  \bigg] \notag\\
  &+ \frac{3 \mpi^2}{ F_\pi^2}
 \frac{1}{  L^3} \sum_{\kk}  \frac{ 1  }{E_\pi} c_1.
\end{align}
The energy gap in this case is $\approx 2 \mpi$, since $\kk = \boldsymbol{0}$ is allowed. 
The LECs $c_{1,2,3}$ have been determined most reliably by an analysis of $\pi N$ scattering using Roy--Steiner equations, matched to $\chi$PT in the subthreshold region~\cite{Hoferichter:2015tha,Hoferichter:2015hva}. We will use the N$^3$LO values
\begin{align}
\label{ci}
    c_1 &= -1.11(3)\GeV^{-1}, \qquad
    c_2 = 3.13(3)\GeV^{-1}, \notag\\
   c_3 &= - 5.61(6)\GeV^{-1},
\end{align}
which correspond to N$^2$LO in the scalar form factor.  
We neglect the N$^2$LO diagrams with pions emitted by the nucleon source, represented by $(g_2)$,
$(h_2)$, and $(i_2)$ in \changed{Fig.~\ref{fig:chipt}},  given that for a local source
the NLO contribution is already small.
Of these diagrams, $(g_2)$ 
produces N$^2$LO recoil corrections,
$(h_2)$ contains unknown LECs that appear in the expansion of the source, and $(i_2)$
cancels in the ratio between the 2- and 3-point function.
While we have assumed local nucleon sources, the relative importance of the diagrams in \changed{Fig.~\ref{fig:chipt}}  depends on the details of the lattice realization of $\mathcal N$. However, for the Gaussian smearing applied in this work with $r_s\sim 0.6\,\text{fm}$, corrections in addition to diagram $(b_1)$ scale as $(r_s \mpi)^2\sim 0.2$ and can therefore be neglected, see also Refs.~\cite{Bar:2015zwa,Bali:2019yiy}.

In the infinite-volume limit, the ground-state pieces of Eqs.~\eqref{R1}, \eqref{Rrecoil}, and \eqref{Rci}
reproduce the N$^2$LO expression for the $\sigma$-term (in the form given in Ref.~\cite{Hoferichter:2015hva})
\begin{align}
\label{sigma_chiral}
\spiN&=   m_{ud}\frac{\partial {\mN}}{\partial  m_{ud}} =   m_{ud} g_S^{u+d}\\
&=-4c_1 \mpi^2  -\frac{9 g_A^2 \mpi^3}{64\pi F_\pi^2} 
-\frac{3 \mpi^4}{64\pi^2 F^2_\pi}\Big(2\log \frac{\mpi^2}{\mN^2}+1\Big)\notag\\
&\times \Big(\frac{g_A^2}{\mN}-8c_1+c_2+4c_3\Big)\notag\\
&+2\mpi^4\bigg\{e_1+\frac{3}{128\pi^2 F^2_\pi}\Big(c_2-\frac{2 g_A^2}{\mN}\Big)+\frac{c_1\big(\bar l_3-1\big)}{16\pi^2 F_\pi^2}\bigg\},\notag
\end{align}
except for the LO contribution proportional to $c_1\mpi^2$, its quark-mass renormalization proportional to $\bar l_3-1$, and the N$^2$LO LEC $e_1$, all of which only contribute to the ground state. 
All other terms in Eq.~\eqref{sigma_chiral} can be obtained by replacing the finite-volume sum  $1/L^3 \sum_{\kk}$ with infinite-volume integrals in Eqs.~\eqref{R1}, \eqref{Rrecoil}, and \eqref{Rci}. 

To assess the importance of the $N \pi$
and $N \pi\pi $ contributions we define
\begin{equation}\label{deltagS}
    \delta \spiN(\tau,t) =
    \mathcal R(\tau, t) - \lim_{t,\tau \rightarrow \infty} \mathcal R(\tau, t).
\end{equation}
The contributions to $\delta \spiN$ from the NLO diagrams
(including the formally N$^2$LO correction from $\widetilde E_N$) and from
the N$^2$LO diagrams are evaluated in Table~\ref{tab:chiralexcited},
using the parameters of the $a09m130$ lattice ensemble, for two
choices of source--sink separation, $\tau=16a$ and $\tau=12a$.  We list
separately the corrections from diagram $(a_1)$ and $(b_1)$ in
\changed{Fig.~\ref{fig:chipt}}, as the first is dominated by $N \pi$ excited
states, the second by $N \pi\pi $. Similarly, the N$^2$LO corrections
from the diagrams proportional to the LECs $c_{1,2,3}$ receive
contributions from $N \pi\pi $ states, while the recoil corrections in
Eq.~\eqref{Rrecoil} are dominated by $N \pi $.  From
Table~\ref{tab:chiralexcited} we see that $\mathcal R^{(1)}$ and
$\mathcal R^{(2)}$ are of similar size and the most important
contributions come from diagram $(a_1)$ and from Eq.~\eqref{Rci}. The
nucleon source and recoil contributions are substantially smaller.

We also note that the ESC is amplified by the presence of several
states close in energy.  We can see this in the left panel of Fig.~\ref{fig:excited},
where the red and orange curves denote the NLO and N$^2$LO
corrections, including states up to $|\nn_\text{max}| = 1$, while the
green and blue curves include states up to $|\nn_\text{max}| = 3$. In the continuum, the effect of the entire tower of excited states can be resummed, with the result shown in the last line of Table~\ref{tab:chiralexcited}
and by the purple line in Fig. ~\ref{fig:excited}. The comparison to the different $|\nn_\text{max}|$ values indicates the degree of convergence, which is ultimately determined by $t$ and $\tau$ via the exponential suppression of the continuum integrals. Indeed, we see that for $\tau=16a$ the corrections at $t=8a$ beyond $|\nn_\text{max}|=3$ are around $10\%$, but twice as large for $\tau=12a$, $t=6a$. 

Away from $t \sim \tau/2$, the integrals become increasingly dominated
by large-momentum modes, leading to the eventual breakdown of the
chiral expansion. For this reason, the expansion becomes less reliable
for small and large $t$, which explains why the functional form
predicted by $\chi$PT, see the right panel of Fig.~\ref{fig:excited},
suggests a faster decrease towards the edges than observed in the
lattice data, see Figs.~\ref{fig:Sfit} and~\ref{fig:Sdata}. (Note that
in fits to remove the excited-state effects in the lattice data, we
neglect $t_{\rm skip}$ points next to the source and the sink for each
$\tau$.) In the center, however, the EFT expansion should be reliable,
in particular for the $\tau=16a$ variant, which leads to the
conclusion that NLO and N$^2$LO contributions can each lead to a
reduction of $\sigma_{\pi N}$ at the level of $10\MeV$. Note that
Fig.~\ref{fig:Sdata} shows the behavior versus $t$ and $\tau$ for the
individual contributions, i.e., insertions on the $u$, the $d$, and
the disconnected loop~$l$. \looseness-1

While in Table~\ref{tab:chiralexcited} and Fig.~\ref{fig:excited} we
focused on the ensemble with the lightest pion mass, we find good
agreement between the $\chi$PT expectations and the fits in
Fig.~\ref{fig:Sfit} also for the remaining ensembles.  For example,
with the parameters of the $a09m220$ ensemble, $\chi$PT predicts the
difference between the lattice data and the extrapolated value of
$g_S$ to be $3$ at $t=\tau/2=8a$, in good agreement with the left
panel in the second row of Fig.~\ref{fig:Sfit}. Similarly, for the
$a06m220$ and $a06m310$ ensembles we obtain that $g_S$ is shifted by
$2.8$ and $1.6$, at $t=\tau/2 = 12a$.

\begin{figure*}[htb]
\subfigure{
  \includegraphics[width=0.32\linewidth]{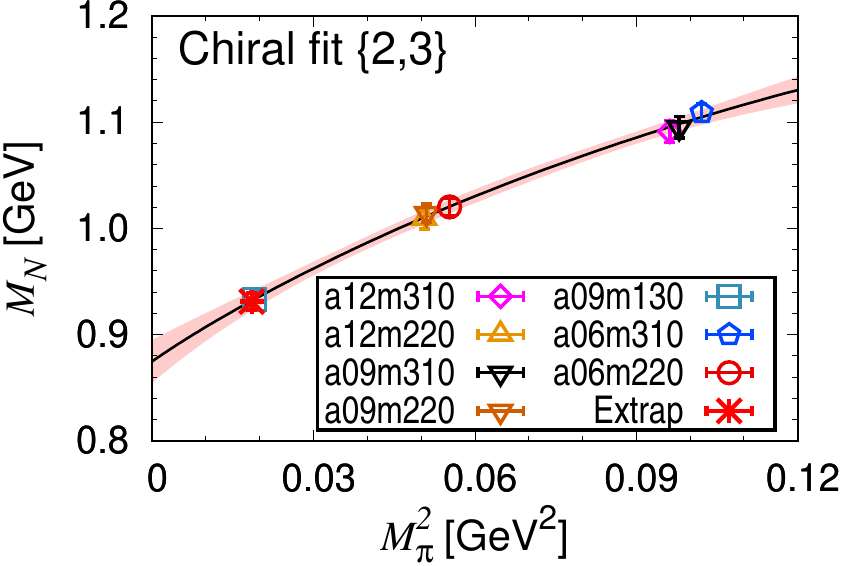}
  \includegraphics[width=0.32\linewidth]{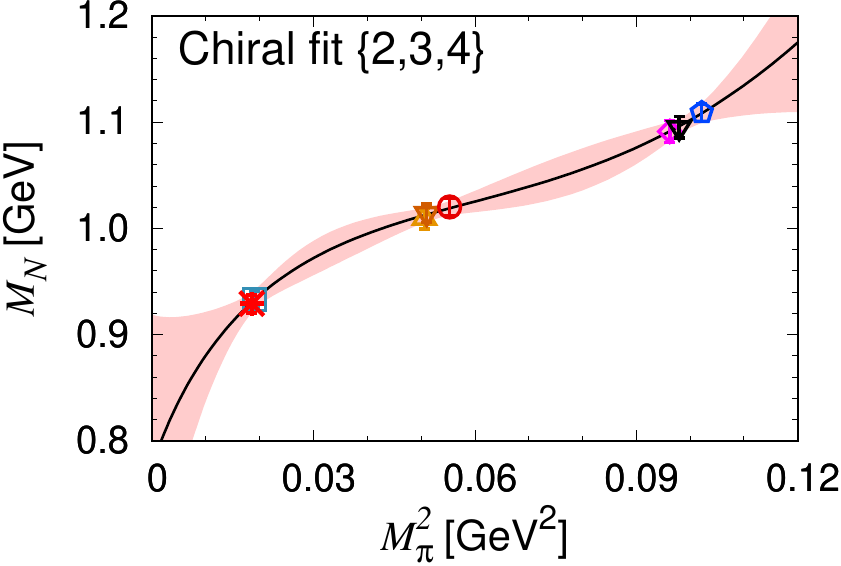}
  \includegraphics[width=0.32\linewidth]{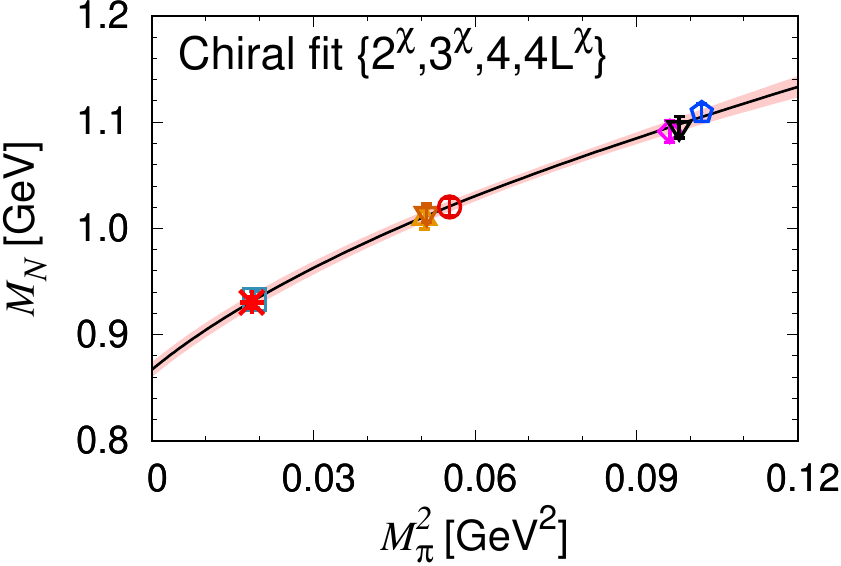}
}
\vspace{-0.1in}
\caption{Chiral fits to the $\{4^{N\pi},3^\ast\}$ data for the nucleon mass $M_0$ given in \mainref{Table}{tab:results}. Data at a seventh point, $a09m310$, is included from  Ref.~\protect\cite{Gupta:2018qil}. The fit in the middle panel is overparameterized. The right panel shows that the $\{2^\chi,3^\chi,4,4L^\chi\}$ fit, with $\chi$PT values for three of the coefficients taken from  Ref.~\protect\cite{Hoferichter:2015hva}, is reasonable and comparable to the fit in the left panel. \looseness-1
\label{fig:CPTMn}}
\vspace{-0.1in}
\end{figure*}

Finally, using Eq.~\eqref{sigma_chiral}, we have carried out the
following chiral fits to the lattice data: $\{2,3\}$,
$\{2^\chi,3^\chi,4\}$, $\{2,3^\chi,4, 4L^\chi\}$,
$\{2^\chi,3^\chi,4,4L\}$, and $\{2^\chi,3^\chi,4,4L^\chi\}$, in
addition to the fits $\{2,3,4\}$, $\{2,3^\chi,4,4L\}$ already shown in
Fig.~\ref{fig:CCFV}. Four of these additional fits are shown in
Fig.~\ref{fig:CCFV2}, with resulting fit parameters given in
Table~\ref{tab:CPTcoeffs}. (We neglect possible discretization errors
in these fits as they are not resolved in our best fits shown in
Fig.~\ref{fig:CCFV}.)  We see that for the $\{4^{N\pi},3^\ast\}$
strategy all fit variants lead to parameters that agree with the
$\chi$PT prediction (including evidence for the nonanalytic $\mpi^3$
term from the $\{2,3\}$ and $\{2,3,4\}$ fits), with changes in
$\sigma_{\pi N}$ consistent with the uncertainty assigned to the least
constrained fits ($\{2,3,4\}$ and $\{2,3^\chi,4,4L\}$), whose average
we quoted as our central result. Even in the most-constrained case
$\{2^\chi,3^\chi,4,4L^\chi\}$ a good fit of the data is obtained, in
marked contrast to the $\{4,3^\ast\}$ strategy, in which case the
$\chi^2/\text{dof}$ becomes unacceptable when imposing the maximum
amount of chiral constraints.

The expressions 
in Eqs.~\eqref{R1}, \eqref{sigma_cont},
and \eqref{Rci} can also be used to assess the importance of finite-volume corrections to $g_S$.
Focusing on the ground-state contributions, we can write     $\Delta_L \spiN$ in Eq.~\eqref{sigma_cont} as \cite{Beane:2004tw} 
\begin{equation}
    \Delta_L \spiN = - \frac{3 g_A^2 \mpi^3}{64 \pi F_\pi^2} \sum_{\nn \neq {\bf 0} }  e^{-  \mpi L |\nn|} \left(1 - \frac{2}{ \mpi L |\nn|}\right). 
\end{equation}
Similarly, for the $c_i$ contributions~\eqref{Rci} we obtain
\begin{align}
    \Delta^{(2)}_L \spiN &= - \frac{3  \mpi^4}{8 \pi^2 F_\pi^2} \sum_{\nn \neq {\bf 0} } \Big[ (2 c_1 - c_3) K_{0}( \mpi L |\nn|) \notag \\ 
    & +(c_2+ 2c_3-4c_1) \frac{K_1(\mpi L |\nn|)}{\mpi L |\nn| } \Big], \end{align}
in terms of the Bessel functions $K_0$ and $K_1$.
Using the parameters of the $a09m130$ lattice ensemble, we get
\begin{equation}
 \Delta_L \spiN = -0.77\MeV, \quad \Delta^{(2)}_L \spiN = -0.43 \MeV,
\end{equation}
implying that the finite-volume corrections are controlled at the level of about $1\MeV$.

We end this appendix by pointing out a subtlety. We have used the
lowest-order (linear) relation between $\mpi^2$ and $1/2\kappa$ to get
$m_{ud}^\text{bare}$ as we have ensembles at only two values of $\mpi$
at each $a$. Higher-order corrections give the LEC $\bar l_3$ term in
Eq.~\eqref{sigma_chiral}. Removing it changes the $\chi$PT prediction
of the $d_{4L}$ term in Table~\ref{tab:CPTcoeffs} from 11.35 to 9.70,
however, the results of the $\{2,3^\chi,4,4L^\chi\}$ and
$\{2^\chi,3^\chi,4,4L^\chi\}$ fits do not change significantly.

\section{Chiral fits to the Nucleon Mass}
\label{sec:Mn}

\begin{table*}[t]
\begin{ruledtabular}
  \begin{tabular}{c | c cccccc}
Ensemble ID & Fit strategy & $r_{1}$ & $a\Delta M_1$  & $r_{2}$ & $a\Delta M_2$  & $r_{3}$ & $a\Delta M_3$
\\
\hline
\multirow{2}{*}{$a12m310$}
 &         $\{4\}$ &   0.20(10) &   0.40(15) &   0.8(4) &   0.6(2) &   0.6(4) &   0.4(2)  \\ 
 &  $\{4^{N\pi}\}$ &   0.20(10) &   0.37(05) &   0.8(4) &   0.6(2) &   0.6(4) &   0.4(2)  \\ 
\hline
\multirow{2}{*}{$a12m220$}
 &         $\{4\}$ &   0.40(20) &   0.40(10) &   1.0(6) &   0.8(4) &   0.8(6) &   0.4(2)  \\ 
 &  $\{4^{N\pi}\}$ &   0.40(30) &   0.27(05) &   1.0(8) &   0.8(4) &   0.8(6) &   0.4(2)  \\ 
\hline
\multirow{2}{*}{$a09m310$}
 &         $\{4\}$ &   0.70(30) &   0.30(05) &   1.0(5) &   0.5(2) &   1.0(6) &   0.5(3)  \\ 
 &  $\{4^{N\pi}\}$ &   0.40(20) &   0.28(05) &   1.0(5) &   0.5(3) &   1.0(6) &   0.5(3)  \\ 
\hline
\multirow{2}{*}{$a09m220$}
 &         $\{4\}$ &   0.60(30) &   0.30(10) &   0.8(5) &   0.4(2) &   0.7(4) &   0.4(2)  \\ 
 &  $\{4^{N\pi}\}$ &   0.35(20) &   0.18(05) &   0.8(5) &   0.4(2) &   0.7(4) &   0.4(2)  \\ 
\hline
\multirow{2}{*}{$a09m130$}
 &         $\{4\}$ &   0.70(40) &   0.30(10) &   0.7(5) &   0.5(3) &   1.0(6) &   0.3(2)  \\ 
 &  $\{4^{N\pi}\}$ &   0.40(20) &   0.12(02) &   0.7(4) &   0.3(1) &   1.0(6) &   0.3(2)  \\ 
\hline
\multirow{2}{*}{$a06m310$}
 &         $\{4\}$ &   0.50(30) &   0.20(05) &   1.0(6) &   0.3(2) &   1.0(6) &   0.3(2)  \\ 
 &  $\{4^{N\pi}\}$ &   0.50(30) &   0.19(03) &   1.0(6) &   0.3(2) &   1.0(6) &   0.3(2)  \\ 
\hline
\multirow{2}{*}{$a06m220$}
 &         $\{4\}$ &   1.00(50) &   0.25(10) &   1.0(5) &   0.3(1) & 1.5(1.0) &   0.3(2)  \\ 
 &  $\{4^{N\pi}\}$ &   1.00(50) &   0.14(03) & 1.5(1.0) &   0.2(1) & 1.5(1.0) &   0.3(2)  \\ 
\end{tabular}
\end{ruledtabular}

\caption{\changed{The Bayesian priors and their width given within parenthesis used for the 
amplitude ratios $r_i\equiv |\mathcal{A}_i / \mathcal{A}_0|^2$ and
mass differences $\Delta M_i \equiv M_i-M_{i-1}$ in the simultaneous fits  to
$C^{2\text{pt}}$ and $C^{3\text{pt}}$ for the two strategies $\{4,3^\ast\}$ and $\{4^{N\pi},3^\ast\}$.}}
\label{tab:priors} 
\end{table*}

\changed{In both types of analyses,  simultaneous fits to $C^{2\text{pt}}$ and  $C^{3\text{pt}}$ 
and individual fits to them, we have used the same empirical Bayesian priors 
for the amplitudes and masses of the three excited states determined using the 
procedure described in Ref.~\cite{Gupta:2018qil}. The central values of these priors and their 
widths for $\{4,3^\ast\}$ and $\{4^{N\pi},3^\ast\}$ fits are given in Table~\ref{tab:priors}.  
The resulting $M_0$, $M_1$, and $M_2$ that enter in $\{3^\ast\}$ fits to $C^{3\text{pt}}$ are given 
in \mainref{Table}{tab:results}, and are essentially the same from the 
two types of analyses. The result for the nucleon mass on the seventh ensemble, $a09m310$, used only 
for the analysis of $M_N$ in this section, is $1.09(1)$~GeV from both analyses. }

\changed{In this section, we study the 
chiral behavior of the  nucleon mass $\mN \equiv M_0$ using the 
 N$^2$LO $\chi$PT result, which has a form similar to \maineqref{Eq.}{eq:CPT}:}
\begin{align}
\mN &=  e_0 + e_2  \mpi^2 + e_3  \mpi^3  \nonumber \\
 &\quad + e_4 \mpi^4 + e_{4L} \mpi^4 \log \frac{\mpi^2}{\mN^2} \,,
\label{eq:CPTMn}  
\end{align}
with the $\chi$PT expressions for the $c_i$ given, e.g., in Ref.~\cite{Hoferichter:2015hva}. Even ignoring discretization and finite-volume corrections and fitting the lattice data using~Eq.~\eqref{eq:CPTMn} we face two challenges. First, as evident from the data for $M_0$ in Table~\ref{tab:results}, there is no significant difference between the $\{4,3^\ast\}$ and $\{4^{N\pi},3^\ast\}$ results. Therefore, we cannot comment on the impact of $N \pi$ states in the analysis of $M_N$. Second, as shown in Fig.~\ref{fig:CPTMn}, at least 
three parameters ($c_0$ and two more) are needed to fit the data. With data at essentially only three values of $\mpi$, even these fits are overparameterized. In short, data at many more values of the lattice parameters, especially in $\mpi^2$, are needed to quantify the lattice artifacts and check the prediction of 
$\chi$PT. The most constrained fit, $\{2^\chi,3^\chi,4,4L^\chi\}$ in Fig.~\ref{fig:CPTMn}, does yield a good description of the data, but the resulting uncertainty in $\spiN$ via the FH method is too large to compete with the direct method.

\newif\ifstandalone\standalonefalse
\ifx\ifstandalone\undefined
  \newif\ifstandalone\noexpand\fi\standalonetrue
\fi
\ifstandalone
\ifx\texorpdfstring\undefined\def\texorpdfstring#1#2{#1}\fi
\ifx\pdfsuppresswarningpagegroup\undefined\else\pdfsuppresswarningpagegroup=1\fi
\documentclass[aps,twocolumn,showpacs,showkeys,preprintnumbers,floatfix,nofootinbib,superscriptaddress,longbibliography]{revtex4-2}
\usepackage[unicode]{hyperref} 
\bibliographystyle{apsrev4-1}
\begin{document}
\fi
\section{Renormalization}
\label{sec:renormalization}

In this appendix we discuss the renormalization of the quark mass and
scalar charge for fermion schemes that break chiral symmetry such as
Wilson-clover fermions.  We will do this using the notation and
results given in Ref.~\cite{Bhattacharya:2005rb} for an $N_f$ flavor
theory with two light degenerate flavors, $m_u = m_d \equiv m_l$,
and $N_f-2$ heavier flavors denoted generically by $m_s$. Here
$m_i$ and $\hat{m}_i$ denote the bare and
renormalized quark masses for flavor $i$ and we will neglect all ${\mathcal O}(a)$ terms as
they do not effect the continuum limit. \changed{These  
discretization errors start at ${\mathcal O}(a)$ in our calculation.}

We start with Eq.~(26) in Ref.~\cite{Bhattacharya:2005rb}:
\begin{equation}
  \hat m_j = Z_m\left[ m_j + (r_m-1)\frac{\sum_{i=1}^{N_f} m_i}{N_f}\right],
  \label{eq:renmass1}
\end{equation}
where $m_j$ are defined as $(1/2\kappa_i) - (1/2\kappa_c^0)$
with $\kappa_i$ the Wilson hopping parameter and $\kappa_c^0$ its
critical value defined to be the point at which all 
pseudoscalar masses vanish, i.e., the $SU(N_f)$
symmetric limit. Using $Z_m$ to denote the
flavor nonsinglet  and $Z_m r_m$ the
isosinglet renormalization constants, one has the relation~\cite{Bhattacharya:2005rb}
\begin{align}
  \sum_s m_s &= \frac{N_f}{Z_m[2+(N_f-2)r_m]} \sum_s \hat m_s\nonumber\\
             &\qquad +\frac{2(N_f-2)(1-r_m)}{2+(N_f-2)r_m} m_l,
\end{align}
and
\begin{equation}
  \hat m_l = \frac{N_f Z_m r_m}{2+(N_f-2)r_m} m_l + \frac{r_m-1}{2+(N_f-2)r_m} \sum \hat m_s,
\end{equation}
from which we obtain,
\begin{equation}
  m_l(\hat m_l=0) = \frac{1-r_m}{N_fZ_mr_m} \sum_s\hat m_s.
\end{equation}
So, we have
\begin{equation}
  \hat m_l = \frac{N_f Z_m r_m}{2+(N_f-2)r_m} \left[m_l - m_l(\hat m_l=0)\right]_{{\rm fixed} \sum_s \hat m_s}.
\end{equation}

Similarly, for the scalar density $S_i$ with flavor $i$, and its expectation
value $\langle S_i \rangle$ in any fixed state
we have from Eqs.~(22--23) in~\cite{Bhattacharya:2005rb}
\begin{equation}
  \langle \hat S_l\rangle = Z_S\left[\frac{(N_f-2)+2r_S}{N_f}\langle S_l\rangle + \frac{r_S-1}{N_f}\langle \sum_s S_s\rangle\right],
\end{equation}
with $Z_S$  and $Z_Sr_S$ 
the nonsinglet and the singlet renormalization constants.
Using $Z_SZ_m=1$
and $r_Sr_m=1$~\cite{Bhattacharya:2005rb}, we get the desired result \looseness-1
\begin{align}
  \hat m_l \langle \hat S_l\rangle &= \left[m_l - m_l(\hat m_l=0)\right]_{{\rm fixed} \sum_s \hat m_s}\nonumber\\
&\times\left(\langle S_l\rangle + \frac {1-r_m}{2+(N_f-2)r_m} \langle \sum_s S_s\rangle\right) ,
\label{eq:sigma_correction}
\end{align}
showing that mixing between flavors in Wilson-like
formulations gives rise to a correction to $\sigma_{\pi N}$, i.e., the
second term on the second line. To explain this, we need to clarify an
important point about the notation. In this work we have defined
$\kappa_c$ as the point at which $\mpi^2$ vanishes with all the $m_s$
kept at their physical values. The difference in the
definition of the chiral point, $\kappa_c^0$ versus $\kappa_c$, leads to a 
difference in the definition of the bare quark masses. 
The connection is that the bare quark mass used 
in this work is the same as $m_l - m_l(\hat m_l=0)|_{{\rm fixed} \sum_s \hat m_s}$ in
Eq.~\eqref{eq:sigma_correction}.  

We have not calculated
$\langle \sum_s S_s\rangle$ that is needed to evaluate the correction to 
the isoscalar scalar charge,
however, it is relatively small since $|1-r_m| \lesssim 2\%$ for the
six ensembles and $\langle \sum_s S_s\rangle < \langle S_l\rangle$. At
the precision at which we are working in this paper,  
and since the focus is on showing that there is a difference in the result 
depending on the excited state analyses, i.e., between $\{4,3^\ast\}$ and
$\{4^{N \pi},3^\ast\}$,
this correction is neglected. Also,
note that $r_m = 1$ for lattice QCD formulations that preserve chiral
symmetry, in which case there is no correction.

\ifstandalone
\bibliography{ref}
\AtEndDocument{\noexpand\iftrue\fi}
\end{document}
\fi

\section{Update of FLAG 2019 Results for \texorpdfstring{$\boldsymbol{\spiN}$}{\textsigma(\textpi N)}}
\label{sec:FLAG}

\changed{\mainref{Figure}{fig:FLAG}} gives an update on the summary of results for
$\spiN$ presented in the FLAG review 2019~\cite{Aoki:2019cca}
using the same notation. We focus on a comparison of results from
lattice QCD and those from $\pi N$ scattering data.  For the lattice
data, we retain only the 2+1 and 2+1+1 (and 1+1+1+1) flavor results
obtained since 2015. (See Refs.~\cite{Abdel-Rehim:2016won,Bali:2016lvx,Bali:2012qs,Durr:2011mp,Ohki:2008ff,Ishikawa:2009vc,Horsley:2011wr,Bali:2012qs,Abdel-Rehim:2016won,MartinCamalich:2010fp,Shanahan:2012wh,Oksuzian:2012rzb,Junnarkar:2013ac} for other determinations included in the FLAG review.) Moreover, \changed{\mainref{Fig.}{fig:FLAG}} does not include results from calculations that analyze more than one lattice data set within the FH approach~\cite{Procura:2006bj,WalkerLoud:2008bp,WalkerLoud:2008pj,Young:2009zb,Ren:2012aj,Walker-Loud:2013yua,Alvarez-Ruso:2013fza,Lutz:2014oxa,Ren:2014vea,Ren:2016aeo,Alexandrou:2017xwd,Ling:2017jyz,Lutz:2018cqo}, or results that use a mixture of
lattice QCD and phenomenological analyses~\cite{Chen:2012nx}. Most of the recent lattice
results are clustered around $\spiN \approx 40\MeV$, while the
phenomenological estimates are at $\spiN \approx 60\MeV$, as
is our result with $N\pi / N\pi \pi$ included when removing ESC.

\bibliography{ref} 

\begin{thebibliography}{131}%
\makeatletter
\providecommand \@ifxundefined [1]{%
 \@ifx{#1\undefined}
}%
\providecommand \@ifnum [1]{%
 \ifnum #1\expandafter \@firstoftwo
 \else \expandafter \@secondoftwo
 \fi
}%
\providecommand \@ifx [1]{%
 \ifx #1\expandafter \@firstoftwo
 \else \expandafter \@secondoftwo
 \fi
}%
\providecommand \natexlab [1]{#1}%
\providecommand \enquote  [1]{``#1''}%
\providecommand \bibnamefont  [1]{#1}%
\providecommand \bibfnamefont [1]{#1}%
\providecommand \citenamefont [1]{#1}%
\providecommand \href@noop [0]{\@secondoftwo}%
\providecommand \href [0]{\begingroup \@sanitize@url \@href}%
\providecommand \@href[1]{\@@startlink{#1}\@@href}%
\providecommand \@@href[1]{\endgroup#1\@@endlink}%
\providecommand \@sanitize@url [0]{\catcode `\\12\catcode `\$12\catcode
  `\&12\catcode `\#12\catcode `\^12\catcode `\_12\catcode `\%12\relax}%
\providecommand \@@startlink[1]{}%
\providecommand \@@endlink[0]{}%
\providecommand \url  [0]{\begingroup\@sanitize@url \@url }%
\providecommand \@url [1]{\endgroup\@href {#1}{\urlprefix }}%
\providecommand \urlprefix  [0]{URL }%
\providecommand \Eprint [0]{\href }%
\providecommand \doibase [0]{http://dx.doi.org/}%
\providecommand \selectlanguage [0]{\@gobble}%
\providecommand \bibinfo  [0]{\@secondoftwo}%
\providecommand \bibfield  [0]{\@secondoftwo}%
\providecommand \translation [1]{[#1]}%
\providecommand \BibitemOpen [0]{}%
\providecommand \bibitemStop [0]{}%
\providecommand \bibitemNoStop [0]{.\EOS\space}%
\providecommand \EOS [0]{\spacefactor3000\relax}%
\providecommand \BibitemShut  [1]{\csname bibitem#1\endcsname}%
\let\auto@bib@innerbib\@empty
\bibitem [{\citenamefont {Hellman}(1937)}]{Hellmann:1937}%
  \BibitemOpen
  \bibfield  {author} {\bibinfo {author} {\bibfnamefont {H.}~\bibnamefont
  {Hellman}},\ }\href@noop {} {\emph {\bibinfo {title} {Einf{\"u}hrung in die
  Quantenchemie}}}\ (\bibinfo  {publisher} {Franz Deuticke},\ \bibinfo
  {address} {Leipzig und Wein},\ \bibinfo {year} {1937})\BibitemShut {NoStop}%
\bibitem [{\citenamefont {Feynman}(1939)}]{Feynman:1939zza}%
  \BibitemOpen
  \bibfield  {author} {\bibinfo {author} {\bibfnamefont {R.~P.}\ \bibnamefont
  {Feynman}},\ }\href {\doibase 10.1103/PhysRev.56.340} {\bibfield  {journal}
  {\bibinfo  {journal} {Phys. Rev.}\ }\textbf {\bibinfo {volume} {56}},\
  \bibinfo {pages} {340} (\bibinfo {year} {1939})}\BibitemShut {NoStop}%
\bibitem [{\citenamefont {Gasser}\ and\ \citenamefont
  {Zepeda}(1980)}]{Gasser:1979hf}%
  \BibitemOpen
  \bibfield  {author} {\bibinfo {author} {\bibfnamefont {J.}~\bibnamefont
  {Gasser}}\ and\ \bibinfo {author} {\bibfnamefont {A.}~\bibnamefont
  {Zepeda}},\ }\href {\doibase 10.1016/0550-3213(80)90294-1} {\bibfield
  {journal} {\bibinfo  {journal} {Nucl. Phys. B}\ }\textbf {\bibinfo {volume}
  {174}},\ \bibinfo {pages} {445} (\bibinfo {year} {1980})}\BibitemShut
  {NoStop}%
\bibitem [{\citenamefont {Bottino}\ \emph {et~al.}(2000)\citenamefont
  {Bottino}, \citenamefont {Donato}, \citenamefont {Fornengo},\ and\
  \citenamefont {Scopel}}]{Bottino:1999ei}%
  \BibitemOpen
  \bibfield  {author} {\bibinfo {author} {\bibfnamefont {A.}~\bibnamefont
  {Bottino}}, \bibinfo {author} {\bibfnamefont {F.}~\bibnamefont {Donato}},
  \bibinfo {author} {\bibfnamefont {N.}~\bibnamefont {Fornengo}}, \ and\
  \bibinfo {author} {\bibfnamefont {S.}~\bibnamefont {Scopel}},\ }\href
  {\doibase 10.1016/S0927-6505(99)00122-X} {\bibfield  {journal} {\bibinfo
  {journal} {Astropart. Phys.}\ }\textbf {\bibinfo {volume} {13}},\ \bibinfo
  {pages} {215} (\bibinfo {year} {2000})},\ \Eprint
  {http://arxiv.org/abs/hep-ph/9909228} {arXiv:hep-ph/9909228} \BibitemShut
  {NoStop}%
\bibitem [{\citenamefont {Bottino}\ \emph {et~al.}(2002)\citenamefont
  {Bottino}, \citenamefont {Donato}, \citenamefont {Fornengo},\ and\
  \citenamefont {Scopel}}]{Bottino:2001dj}%
  \BibitemOpen
  \bibfield  {author} {\bibinfo {author} {\bibfnamefont {A.}~\bibnamefont
  {Bottino}}, \bibinfo {author} {\bibfnamefont {F.}~\bibnamefont {Donato}},
  \bibinfo {author} {\bibfnamefont {N.}~\bibnamefont {Fornengo}}, \ and\
  \bibinfo {author} {\bibfnamefont {S.}~\bibnamefont {Scopel}},\ }\href
  {\doibase 10.1016/S0927-6505(02)00107-X} {\bibfield  {journal} {\bibinfo
  {journal} {Astropart. Phys.}\ }\textbf {\bibinfo {volume} {18}},\ \bibinfo
  {pages} {205} (\bibinfo {year} {2002})},\ \Eprint
  {http://arxiv.org/abs/hep-ph/0111229} {arXiv:hep-ph/0111229} \BibitemShut
  {NoStop}%
\bibitem [{\citenamefont {Ellis}\ \emph {et~al.}(2008)\citenamefont {Ellis},
  \citenamefont {Olive},\ and\ \citenamefont {Savage}}]{Ellis:2008hf}%
  \BibitemOpen
  \bibfield  {author} {\bibinfo {author} {\bibfnamefont {J.~R.}\ \bibnamefont
  {Ellis}}, \bibinfo {author} {\bibfnamefont {K.~A.}\ \bibnamefont {Olive}}, \
  and\ \bibinfo {author} {\bibfnamefont {C.}~\bibnamefont {Savage}},\ }\href
  {\doibase 10.1103/PhysRevD.77.065026} {\bibfield  {journal} {\bibinfo
  {journal} {Phys. Rev. D}\ }\textbf {\bibinfo {volume} {77}},\ \bibinfo
  {pages} {065026} (\bibinfo {year} {2008})},\ \Eprint
  {http://arxiv.org/abs/0801.3656} {arXiv:0801.3656 [hep-ph]} \BibitemShut
  {NoStop}%
\bibitem [{\citenamefont {Crivellin}\ \emph
  {et~al.}(2014{\natexlab{a}})\citenamefont {Crivellin}, \citenamefont
  {Hoferichter},\ and\ \citenamefont {Procura}}]{Crivellin:2013ipa}%
  \BibitemOpen
  \bibfield  {author} {\bibinfo {author} {\bibfnamefont {A.}~\bibnamefont
  {Crivellin}}, \bibinfo {author} {\bibfnamefont {M.}~\bibnamefont
  {Hoferichter}}, \ and\ \bibinfo {author} {\bibfnamefont {M.}~\bibnamefont
  {Procura}},\ }\href {\doibase 10.1103/PhysRevD.89.054021} {\bibfield
  {journal} {\bibinfo  {journal} {Phys. Rev. D}\ }\textbf {\bibinfo {volume}
  {89}},\ \bibinfo {pages} {054021} (\bibinfo {year} {2014}{\natexlab{a}})},\
  \Eprint {http://arxiv.org/abs/1312.4951} {arXiv:1312.4951 [hep-ph]}
  \BibitemShut {NoStop}%
\bibitem [{\citenamefont {Hoferichter}\ \emph {et~al.}(2017)\citenamefont
  {Hoferichter}, \citenamefont {Klos}, \citenamefont {Men\'endez},\ and\
  \citenamefont {Schwenk}}]{Hoferichter:2017olk}%
  \BibitemOpen
  \bibfield  {author} {\bibinfo {author} {\bibfnamefont {M.}~\bibnamefont
  {Hoferichter}}, \bibinfo {author} {\bibfnamefont {P.}~\bibnamefont {Klos}},
  \bibinfo {author} {\bibfnamefont {J.}~\bibnamefont {Men\'endez}}, \ and\
  \bibinfo {author} {\bibfnamefont {A.}~\bibnamefont {Schwenk}},\ }\href
  {\doibase 10.1103/PhysRevLett.119.181803} {\bibfield  {journal} {\bibinfo
  {journal} {Phys. Rev. Lett.}\ }\textbf {\bibinfo {volume} {119}},\ \bibinfo
  {pages} {181803} (\bibinfo {year} {2017})},\ \Eprint
  {http://arxiv.org/abs/1708.02245} {arXiv:1708.02245 [hep-ph]} \BibitemShut
  {NoStop}%
\bibitem [{\citenamefont {Cirigliano}\ \emph {et~al.}(2009)\citenamefont
  {Cirigliano}, \citenamefont {Kitano}, \citenamefont {Okada},\ and\
  \citenamefont {Tuzon}}]{Cirigliano:2009bz}%
  \BibitemOpen
  \bibfield  {author} {\bibinfo {author} {\bibfnamefont {V.}~\bibnamefont
  {Cirigliano}}, \bibinfo {author} {\bibfnamefont {R.}~\bibnamefont {Kitano}},
  \bibinfo {author} {\bibfnamefont {Y.}~\bibnamefont {Okada}}, \ and\ \bibinfo
  {author} {\bibfnamefont {P.}~\bibnamefont {Tuzon}},\ }\href {\doibase
  10.1103/PhysRevD.80.013002} {\bibfield  {journal} {\bibinfo  {journal} {Phys.
  Rev. D}\ }\textbf {\bibinfo {volume} {80}},\ \bibinfo {pages} {013002}
  (\bibinfo {year} {2009})},\ \Eprint {http://arxiv.org/abs/0904.0957}
  {arXiv:0904.0957 [hep-ph]} \BibitemShut {NoStop}%
\bibitem [{\citenamefont {Crivellin}\ \emph
  {et~al.}(2014{\natexlab{b}})\citenamefont {Crivellin}, \citenamefont
  {Hoferichter},\ and\ \citenamefont {Procura}}]{Crivellin:2014cta}%
  \BibitemOpen
  \bibfield  {author} {\bibinfo {author} {\bibfnamefont {A.}~\bibnamefont
  {Crivellin}}, \bibinfo {author} {\bibfnamefont {M.}~\bibnamefont
  {Hoferichter}}, \ and\ \bibinfo {author} {\bibfnamefont {M.}~\bibnamefont
  {Procura}},\ }\href {\doibase 10.1103/PhysRevD.89.093024} {\bibfield
  {journal} {\bibinfo  {journal} {Phys. Rev. D}\ }\textbf {\bibinfo {volume}
  {89}},\ \bibinfo {pages} {093024} (\bibinfo {year} {2014}{\natexlab{b}})},\
  \Eprint {http://arxiv.org/abs/1404.7134} {arXiv:1404.7134 [hep-ph]}
  \BibitemShut {NoStop}%
\bibitem [{\citenamefont {Engel}\ \emph {et~al.}(2013)\citenamefont {Engel},
  \citenamefont {Ramsey-Musolf},\ and\ \citenamefont {van
  Kolck}}]{Engel:2013lsa}%
  \BibitemOpen
  \bibfield  {author} {\bibinfo {author} {\bibfnamefont {J.}~\bibnamefont
  {Engel}}, \bibinfo {author} {\bibfnamefont {M.~J.}\ \bibnamefont
  {Ramsey-Musolf}}, \ and\ \bibinfo {author} {\bibfnamefont {U.}~\bibnamefont
  {van Kolck}},\ }\href {\doibase 10.1016/j.ppnp.2013.03.003} {\bibfield
  {journal} {\bibinfo  {journal} {Prog. Part. Nucl. Phys.}\ }\textbf {\bibinfo
  {volume} {71}},\ \bibinfo {pages} {21} (\bibinfo {year} {2013})},\ \Eprint
  {http://arxiv.org/abs/1303.2371} {arXiv:1303.2371 [nucl-th]} \BibitemShut
  {NoStop}%
\bibitem [{\citenamefont {de~Vries}\ and\ \citenamefont
  {Mei\ss{}ner}(2016)}]{deVries:2015gea}%
  \BibitemOpen
  \bibfield  {author} {\bibinfo {author} {\bibfnamefont {J.}~\bibnamefont
  {de~Vries}}\ and\ \bibinfo {author} {\bibfnamefont {{\relax
  Ulf-G}.}~\bibnamefont {Mei\ss{}ner}},\ }\href {\doibase
  10.1142/S0218301316410081} {\bibfield  {journal} {\bibinfo  {journal} {Int.
  J. Mod. Phys. E}\ }\textbf {\bibinfo {volume} {25}},\ \bibinfo {pages}
  {1641008} (\bibinfo {year} {2016})},\ \Eprint
  {http://arxiv.org/abs/1509.07331} {arXiv:1509.07331 [hep-ph]} \BibitemShut
  {NoStop}%
\bibitem [{\citenamefont {de~Vries}\ \emph {et~al.}(2017)\citenamefont
  {de~Vries}, \citenamefont {Mereghetti}, \citenamefont {Seng},\ and\
  \citenamefont {Walker-Loud}}]{deVries:2016jox}%
  \BibitemOpen
  \bibfield  {author} {\bibinfo {author} {\bibfnamefont {J.}~\bibnamefont
  {de~Vries}}, \bibinfo {author} {\bibfnamefont {E.}~\bibnamefont
  {Mereghetti}}, \bibinfo {author} {\bibfnamefont {C.-Y.}\ \bibnamefont
  {Seng}}, \ and\ \bibinfo {author} {\bibfnamefont {A.}~\bibnamefont
  {Walker-Loud}},\ }\href {\doibase 10.1016/j.physletb.2017.01.017} {\bibfield
  {journal} {\bibinfo  {journal} {Phys. Lett. B}\ }\textbf {\bibinfo {volume}
  {766}},\ \bibinfo {pages} {254} (\bibinfo {year} {2017})},\ \Eprint
  {http://arxiv.org/abs/1612.01567} {arXiv:1612.01567 [hep-lat]} \BibitemShut
  {NoStop}%
\bibitem [{\citenamefont {Yamanaka}\ \emph {et~al.}(2017)\citenamefont
  {Yamanaka}, \citenamefont {Sahoo}, \citenamefont {Yoshinaga}, \citenamefont
  {Sato}, \citenamefont {Asahi},\ and\ \citenamefont {Das}}]{Yamanaka:2017mef}%
  \BibitemOpen
  \bibfield  {author} {\bibinfo {author} {\bibfnamefont {N.}~\bibnamefont
  {Yamanaka}}, \bibinfo {author} {\bibfnamefont {B.~K.}\ \bibnamefont {Sahoo}},
  \bibinfo {author} {\bibfnamefont {N.}~\bibnamefont {Yoshinaga}}, \bibinfo
  {author} {\bibfnamefont {T.}~\bibnamefont {Sato}}, \bibinfo {author}
  {\bibfnamefont {K.}~\bibnamefont {Asahi}}, \ and\ \bibinfo {author}
  {\bibfnamefont {B.~P.}\ \bibnamefont {Das}},\ }\href {\doibase
  10.1140/epja/i2017-12237-2} {\bibfield  {journal} {\bibinfo  {journal} {Eur.
  Phys. J. A}\ }\textbf {\bibinfo {volume} {53}},\ \bibinfo {pages} {54}
  (\bibinfo {year} {2017})},\ \Eprint {http://arxiv.org/abs/1703.01570}
  {arXiv:1703.01570 [hep-ph]} \BibitemShut {NoStop}%
\bibitem [{\citenamefont {Cheng}\ and\ \citenamefont
  {Dashen}(1971)}]{Cheng:1970mx}%
  \BibitemOpen
  \bibfield  {author} {\bibinfo {author} {\bibfnamefont {T.~P.}\ \bibnamefont
  {Cheng}}\ and\ \bibinfo {author} {\bibfnamefont {R.~F.}\ \bibnamefont
  {Dashen}},\ }\href {\doibase 10.1103/PhysRevLett.26.594} {\bibfield
  {journal} {\bibinfo  {journal} {Phys. Rev. Lett.}\ }\textbf {\bibinfo
  {volume} {26}},\ \bibinfo {pages} {594} (\bibinfo {year} {1971})}\BibitemShut
  {NoStop}%
\bibitem [{\citenamefont {Brown}\ \emph {et~al.}(1971)\citenamefont {Brown},
  \citenamefont {Pardee},\ and\ \citenamefont {Peccei}}]{Brown:1971pn}%
  \BibitemOpen
  \bibfield  {author} {\bibinfo {author} {\bibfnamefont {L.~S.}\ \bibnamefont
  {Brown}}, \bibinfo {author} {\bibfnamefont {W.~J.}\ \bibnamefont {Pardee}}, \
  and\ \bibinfo {author} {\bibfnamefont {R.~D.}\ \bibnamefont {Peccei}},\
  }\href {\doibase 10.1103/PhysRevD.4.2801} {\bibfield  {journal} {\bibinfo
  {journal} {Phys. Rev. D}\ }\textbf {\bibinfo {volume} {4}},\ \bibinfo {pages}
  {2801} (\bibinfo {year} {1971})}\BibitemShut {NoStop}%
\bibitem [{\citenamefont {Bernard}\ \emph {et~al.}(1996)\citenamefont
  {Bernard}, \citenamefont {Kaiser},\ and\ \citenamefont
  {Mei\ss{}ner}}]{Bernard:1996nu}%
  \BibitemOpen
  \bibfield  {author} {\bibinfo {author} {\bibfnamefont {V.}~\bibnamefont
  {Bernard}}, \bibinfo {author} {\bibfnamefont {N.}~\bibnamefont {Kaiser}}, \
  and\ \bibinfo {author} {\bibfnamefont {{\relax Ulf-G}.}~\bibnamefont
  {Mei\ss{}ner}},\ }\href {\doibase 10.1016/S0370-2693(96)01243-9} {\bibfield
  {journal} {\bibinfo  {journal} {Phys. Lett. B}\ }\textbf {\bibinfo {volume}
  {389}},\ \bibinfo {pages} {144} (\bibinfo {year} {1996})},\ \Eprint
  {http://arxiv.org/abs/hep-ph/9607245} {arXiv:hep-ph/9607245} \BibitemShut
  {NoStop}%
\bibitem [{\citenamefont {Becher}\ and\ \citenamefont
  {Leutwyler}(2001)}]{Becher:2001hv}%
  \BibitemOpen
  \bibfield  {author} {\bibinfo {author} {\bibfnamefont {T.}~\bibnamefont
  {Becher}}\ and\ \bibinfo {author} {\bibfnamefont {H.}~\bibnamefont
  {Leutwyler}},\ }\href {\doibase 10.1088/1126-6708/2001/06/017} {\bibfield
  {journal} {\bibinfo  {journal} {JHEP}\ }\textbf {\bibinfo {volume} {06}},\
  \bibinfo {pages} {017} (\bibinfo {year} {2001})},\ \Eprint
  {http://arxiv.org/abs/hep-ph/0103263} {arXiv:hep-ph/0103263} \BibitemShut
  {NoStop}%
\bibitem [{\citenamefont {Gasser}\ \emph
  {et~al.}(1988{\natexlab{a}})\citenamefont {Gasser}, \citenamefont
  {Leutwyler}, \citenamefont {Locher},\ and\ \citenamefont
  {Sainio}}]{Gasser:1988jt}%
  \BibitemOpen
  \bibfield  {author} {\bibinfo {author} {\bibfnamefont {J.}~\bibnamefont
  {Gasser}}, \bibinfo {author} {\bibfnamefont {H.}~\bibnamefont {Leutwyler}},
  \bibinfo {author} {\bibfnamefont {M.~P.}\ \bibnamefont {Locher}}, \ and\
  \bibinfo {author} {\bibfnamefont {M.~E.}\ \bibnamefont {Sainio}},\ }\href
  {\doibase 10.1016/0370-2693(88)91052-0} {\bibfield  {journal} {\bibinfo
  {journal} {Phys. Lett. B}\ }\textbf {\bibinfo {volume} {213}},\ \bibinfo
  {pages} {85} (\bibinfo {year} {1988}{\natexlab{a}})}\BibitemShut {NoStop}%
\bibitem [{\citenamefont {Gasser}\ \emph
  {et~al.}(1991{\natexlab{a}})\citenamefont {Gasser}, \citenamefont
  {Leutwyler},\ and\ \citenamefont {Sainio}}]{Gasser:1990ce}%
  \BibitemOpen
  \bibfield  {author} {\bibinfo {author} {\bibfnamefont {J.}~\bibnamefont
  {Gasser}}, \bibinfo {author} {\bibfnamefont {H.}~\bibnamefont {Leutwyler}}, \
  and\ \bibinfo {author} {\bibfnamefont {M.~E.}\ \bibnamefont {Sainio}},\
  }\href {\doibase 10.1016/0370-2693(91)91393-A} {\bibfield  {journal}
  {\bibinfo  {journal} {Phys. Lett. B}\ }\textbf {\bibinfo {volume} {253}},\
  \bibinfo {pages} {252} (\bibinfo {year} {1991}{\natexlab{a}})}\BibitemShut
  {NoStop}%
\bibitem [{\citenamefont {Gasser}\ \emph
  {et~al.}(1991{\natexlab{b}})\citenamefont {Gasser}, \citenamefont
  {Leutwyler},\ and\ \citenamefont {Sainio}}]{Gasser:1990ap}%
  \BibitemOpen
  \bibfield  {author} {\bibinfo {author} {\bibfnamefont {J.}~\bibnamefont
  {Gasser}}, \bibinfo {author} {\bibfnamefont {H.}~\bibnamefont {Leutwyler}}, \
  and\ \bibinfo {author} {\bibfnamefont {M.~E.}\ \bibnamefont {Sainio}},\
  }\href {\doibase 10.1016/0370-2693(91)91394-B} {\bibfield  {journal}
  {\bibinfo  {journal} {Phys. Lett. B}\ }\textbf {\bibinfo {volume} {253}},\
  \bibinfo {pages} {260} (\bibinfo {year} {1991}{\natexlab{b}})}\BibitemShut
  {NoStop}%
\bibitem [{\citenamefont {Koch}\ and\ \citenamefont
  {Pietarinen}(1980)}]{Koch:1980ay}%
  \BibitemOpen
  \bibfield  {author} {\bibinfo {author} {\bibfnamefont {R.}~\bibnamefont
  {Koch}}\ and\ \bibinfo {author} {\bibfnamefont {E.}~\bibnamefont
  {Pietarinen}},\ }\href {\doibase 10.1016/0375-9474(80)90214-6} {\bibfield
  {journal} {\bibinfo  {journal} {Nucl. Phys. A}\ }\textbf {\bibinfo {volume}
  {336}},\ \bibinfo {pages} {331} (\bibinfo {year} {1980})}\BibitemShut
  {NoStop}%
\bibitem [{\citenamefont {H\"ohler}(1983)}]{Hohler:1984ux}%
  \BibitemOpen
  \bibfield  {author} {\bibinfo {author} {\bibfnamefont {G.}~\bibnamefont
  {H\"ohler}},\ }\href {\doibase 10.1007/b19946} {\emph {\bibinfo {title}
  {Methods and Results of Phenomenological Analyses}}},\ edited by\ \bibinfo
  {editor} {\bibfnamefont {H.}~\bibnamefont {Schopper}},\ \bibinfo {series}
  {Landolt-Boernstein - Group I Elementary Particles, Nuclei and Atoms}, Vol.\
  \bibinfo {volume} {9b2}\ (\bibinfo  {publisher} {Springer-Verlag Berlin},\
  \bibinfo {address} {Heidelberg},\ \bibinfo {year} {1983})\BibitemShut
  {NoStop}%
\bibitem [{\citenamefont {Arndt}\ \emph {et~al.}(2006)\citenamefont {Arndt},
  \citenamefont {Briscoe}, \citenamefont {Strakovsky},\ and\ \citenamefont
  {Workman}}]{Arndt:2006bf}%
  \BibitemOpen
  \bibfield  {author} {\bibinfo {author} {\bibfnamefont {R.~A.}\ \bibnamefont
  {Arndt}}, \bibinfo {author} {\bibfnamefont {W.~J.}\ \bibnamefont {Briscoe}},
  \bibinfo {author} {\bibfnamefont {I.~I.}\ \bibnamefont {Strakovsky}}, \ and\
  \bibinfo {author} {\bibfnamefont {R.~L.}\ \bibnamefont {Workman}},\ }\href
  {\doibase 10.1103/PhysRevC.74.045205} {\bibfield  {journal} {\bibinfo
  {journal} {Phys. Rev. C}\ }\textbf {\bibinfo {volume} {74}},\ \bibinfo
  {pages} {045205} (\bibinfo {year} {2006})},\ \Eprint
  {http://arxiv.org/abs/nucl-th/0605082} {arXiv:nucl-th/0605082} \BibitemShut
  {NoStop}%
\bibitem [{\citenamefont {Workman}\ \emph {et~al.}(2012)\citenamefont
  {Workman}, \citenamefont {Arndt}, \citenamefont {Briscoe}, \citenamefont
  {Paris},\ and\ \citenamefont {Strakovsky}}]{Workman:2012hx}%
  \BibitemOpen
  \bibfield  {author} {\bibinfo {author} {\bibfnamefont {R.~L.}\ \bibnamefont
  {Workman}}, \bibinfo {author} {\bibfnamefont {R.~A.}\ \bibnamefont {Arndt}},
  \bibinfo {author} {\bibfnamefont {W.~J.}\ \bibnamefont {Briscoe}}, \bibinfo
  {author} {\bibfnamefont {M.~W.}\ \bibnamefont {Paris}}, \ and\ \bibinfo
  {author} {\bibfnamefont {I.~I.}\ \bibnamefont {Strakovsky}},\ }\href
  {\doibase 10.1103/PhysRevC.86.035202} {\bibfield  {journal} {\bibinfo
  {journal} {Phys. Rev. C}\ }\textbf {\bibinfo {volume} {86}},\ \bibinfo
  {pages} {035202} (\bibinfo {year} {2012})},\ \Eprint
  {http://arxiv.org/abs/1204.2277} {arXiv:1204.2277 [hep-ph]} \BibitemShut
  {NoStop}%
\bibitem [{\citenamefont {Pavan}\ \emph {et~al.}(2002)\citenamefont {Pavan},
  \citenamefont {Strakovsky}, \citenamefont {Workman},\ and\ \citenamefont
  {Arndt}}]{Pavan:2001wz}%
  \BibitemOpen
  \bibfield  {author} {\bibinfo {author} {\bibfnamefont {M.~M.}\ \bibnamefont
  {Pavan}}, \bibinfo {author} {\bibfnamefont {I.~I.}\ \bibnamefont
  {Strakovsky}}, \bibinfo {author} {\bibfnamefont {R.~L.}\ \bibnamefont
  {Workman}}, \ and\ \bibinfo {author} {\bibfnamefont {R.~A.}\ \bibnamefont
  {Arndt}},\ }\href
  {https://gwdac.phys.gwu.edu/Newsletters/PiN-Newsletter-16.pdf} {\bibfield
  {journal} {\bibinfo  {journal} {PiN Newslett.}\ }\textbf {\bibinfo {volume}
  {16}},\ \bibinfo {pages} {110} (\bibinfo {year} {2002})},\ \Eprint
  {http://arxiv.org/abs/hep-ph/0111066} {arXiv:hep-ph/0111066} \BibitemShut
  {NoStop}%
\bibitem [{\citenamefont {Fettes}\ and\ \citenamefont
  {Mei\ss{}ner}(2000)}]{Fettes:2000xg}%
  \BibitemOpen
  \bibfield  {author} {\bibinfo {author} {\bibfnamefont {N.}~\bibnamefont
  {Fettes}}\ and\ \bibinfo {author} {\bibfnamefont {{\relax
  Ulf-G}.}~\bibnamefont {Mei\ss{}ner}},\ }\href {\doibase
  10.1016/S0375-9474(00)00199-8} {\bibfield  {journal} {\bibinfo  {journal}
  {Nucl. Phys. A}\ }\textbf {\bibinfo {volume} {676}},\ \bibinfo {pages} {311}
  (\bibinfo {year} {2000})},\ \Eprint {http://arxiv.org/abs/hep-ph/0002162}
  {arXiv:hep-ph/0002162} \BibitemShut {NoStop}%
\bibitem [{\citenamefont {Alarc{\'o}n}\ \emph {et~al.}(2012)\citenamefont
  {Alarc{\'o}n}, \citenamefont {Martin~Camalich},\ and\ \citenamefont
  {Oller}}]{Alarcon:2011zs}%
  \BibitemOpen
  \bibfield  {author} {\bibinfo {author} {\bibfnamefont {J.~M.}\ \bibnamefont
  {Alarc{\'o}n}}, \bibinfo {author} {\bibfnamefont {J.}~\bibnamefont
  {Martin~Camalich}}, \ and\ \bibinfo {author} {\bibfnamefont {J.~A.}\
  \bibnamefont {Oller}},\ }\href {\doibase 10.1103/PhysRevD.85.051503}
  {\bibfield  {journal} {\bibinfo  {journal} {Phys. Rev. D}\ }\textbf {\bibinfo
  {volume} {85}},\ \bibinfo {pages} {051503(R)} (\bibinfo {year} {2012})},\
  \Eprint {http://arxiv.org/abs/1110.3797} {arXiv:1110.3797 [hep-ph]}
  \BibitemShut {NoStop}%
\bibitem [{\citenamefont {Koch}(1982)}]{Koch:1982pu}%
  \BibitemOpen
  \bibfield  {author} {\bibinfo {author} {\bibfnamefont {R.}~\bibnamefont
  {Koch}},\ }\href {\doibase 10.1007/BF01571999} {\bibfield  {journal}
  {\bibinfo  {journal} {Z. Phys. C}\ }\textbf {\bibinfo {volume} {15}},\
  \bibinfo {pages} {161} (\bibinfo {year} {1982})}\BibitemShut {NoStop}%
\bibitem [{\citenamefont {Ericson}(1987)}]{Ericson:1987uf}%
  \BibitemOpen
  \bibfield  {author} {\bibinfo {author} {\bibfnamefont {T.~E.~O.}\
  \bibnamefont {Ericson}},\ }\href {\doibase 10.1016/0370-2693(87)91180-4}
  {\bibfield  {journal} {\bibinfo  {journal} {Phys. Lett. B}\ }\textbf
  {\bibinfo {volume} {195}},\ \bibinfo {pages} {116} (\bibinfo {year}
  {1987})}\BibitemShut {NoStop}%
\bibitem [{\citenamefont {H{\"o}hler}(1990)}]{Hohler:1990tz}%
  \BibitemOpen
  \bibfield  {author} {\bibinfo {author} {\bibfnamefont {G.}~\bibnamefont
  {H{\"o}hler}},\ }\href
  {https://gwdac.phys.gwu.edu/Newsletters/PiN-Newsletter-02.pdf} {\bibfield
  {journal} {\bibinfo  {journal} {PiN Newslett.}\ }\textbf {\bibinfo {volume}
  {2}},\ \bibinfo {pages} {1} (\bibinfo {year} {1990})}\BibitemShut {NoStop}%
\bibitem [{\citenamefont {Olsson}(2000)}]{Olsson:1999jt}%
  \BibitemOpen
  \bibfield  {author} {\bibinfo {author} {\bibfnamefont {M.~G.}\ \bibnamefont
  {Olsson}},\ }\href {\doibase 10.1016/S0370-2693(00)00505-0} {\bibfield
  {journal} {\bibinfo  {journal} {Phys. Lett. B}\ }\textbf {\bibinfo {volume}
  {482}},\ \bibinfo {pages} {50} (\bibinfo {year} {2000})},\ \Eprint
  {http://arxiv.org/abs/hep-ph/0001203} {arXiv:hep-ph/0001203} \BibitemShut
  {NoStop}%
\bibitem [{\citenamefont {Hite}\ \emph {et~al.}(2005)\citenamefont {Hite},
  \citenamefont {Kaufmann},\ and\ \citenamefont {Jacob}}]{Hite:2005tg}%
  \BibitemOpen
  \bibfield  {author} {\bibinfo {author} {\bibfnamefont {G.~E.}\ \bibnamefont
  {Hite}}, \bibinfo {author} {\bibfnamefont {W.~B.}\ \bibnamefont {Kaufmann}},
  \ and\ \bibinfo {author} {\bibfnamefont {R.~J.}\ \bibnamefont {Jacob}},\
  }\href {\doibase 10.1103/PhysRevC.71.065201} {\bibfield  {journal} {\bibinfo
  {journal} {Phys. Rev. C}\ }\textbf {\bibinfo {volume} {71}},\ \bibinfo
  {pages} {065201} (\bibinfo {year} {2005})}\BibitemShut {NoStop}%
\bibitem [{\citenamefont {Hadzimehmedovic}\ \emph {et~al.}(2007)\citenamefont
  {Hadzimehmedovic}, \citenamefont {Osmanovic},\ and\ \citenamefont
  {Stahov}}]{Hadzimehmedovic:2007dsb}%
  \BibitemOpen
  \bibfield  {author} {\bibinfo {author} {\bibfnamefont {M.}~\bibnamefont
  {Hadzimehmedovic}}, \bibinfo {author} {\bibfnamefont {H.}~\bibnamefont
  {Osmanovic}}, \ and\ \bibinfo {author} {\bibfnamefont {J.}~\bibnamefont
  {Stahov}},\ }\href {https://www.slac.stanford.edu/econf/C070910/PDF/234.pdf}
  {\bibfield  {journal} {\bibinfo  {journal} {eConf}\ }\textbf {\bibinfo
  {volume} {C070910}},\ \bibinfo {pages} {234} (\bibinfo {year}
  {2007})}\BibitemShut {NoStop}%
\bibitem [{\citenamefont {Stahov}\ \emph {et~al.}(2013)\citenamefont {Stahov},
  \citenamefont {Clement},\ and\ \citenamefont {Wagner}}]{Stahov:2012ca}%
  \BibitemOpen
  \bibfield  {author} {\bibinfo {author} {\bibfnamefont {J.}~\bibnamefont
  {Stahov}}, \bibinfo {author} {\bibfnamefont {H.}~\bibnamefont {Clement}}, \
  and\ \bibinfo {author} {\bibfnamefont {G.~J.}\ \bibnamefont {Wagner}},\
  }\href {\doibase 10.1016/j.physletb.2013.09.018} {\bibfield  {journal}
  {\bibinfo  {journal} {Phys. Lett. B}\ }\textbf {\bibinfo {volume} {726}},\
  \bibinfo {pages} {685} (\bibinfo {year} {2013})},\ \Eprint
  {http://arxiv.org/abs/1211.1148} {arXiv:1211.1148 [nucl-th]} \BibitemShut
  {NoStop}%
\bibitem [{\citenamefont {Matsinos}\ and\ \citenamefont
  {Rasche}(2014)}]{Matsinos:2013era}%
  \BibitemOpen
  \bibfield  {author} {\bibinfo {author} {\bibfnamefont {E.}~\bibnamefont
  {Matsinos}}\ and\ \bibinfo {author} {\bibfnamefont {G.}~\bibnamefont
  {Rasche}},\ }\href {\doibase 10.1016/j.nuclphysa.2014.04.021} {\bibfield
  {journal} {\bibinfo  {journal} {Nucl. Phys. A}\ }\textbf {\bibinfo {volume}
  {927}},\ \bibinfo {pages} {147} (\bibinfo {year} {2014})},\ \Eprint
  {http://arxiv.org/abs/1311.0435} {arXiv:1311.0435 [hep-ph]} \BibitemShut
  {NoStop}%
\bibitem [{\citenamefont {Ditsche}\ \emph {et~al.}(2012)\citenamefont
  {Ditsche}, \citenamefont {Hoferichter}, \citenamefont {Kubis},\ and\
  \citenamefont {Mei{\ss}ner}}]{Ditsche:2012fv}%
  \BibitemOpen
  \bibfield  {author} {\bibinfo {author} {\bibfnamefont {C.}~\bibnamefont
  {Ditsche}}, \bibinfo {author} {\bibfnamefont {M.}~\bibnamefont
  {Hoferichter}}, \bibinfo {author} {\bibfnamefont {B.}~\bibnamefont {Kubis}},
  \ and\ \bibinfo {author} {\bibfnamefont {U.-G.}\ \bibnamefont
  {Mei{\ss}ner}},\ }\href {\doibase 10.1007/JHEP06(2012)043} {\bibfield
  {journal} {\bibinfo  {journal} {JHEP}\ }\textbf {\bibinfo {volume} {06}},\
  \bibinfo {pages} {043} (\bibinfo {year} {2012})},\ \Eprint
  {http://arxiv.org/abs/1203.4758} {arXiv:1203.4758 [hep-ph]} \BibitemShut
  {NoStop}%
\bibitem [{\citenamefont {Hoferichter}\ \emph {et~al.}(2012)\citenamefont
  {Hoferichter}, \citenamefont {Ditsche}, \citenamefont {Kubis},\ and\
  \citenamefont {Mei{\ss}ner}}]{Hoferichter:2012wf}%
  \BibitemOpen
  \bibfield  {author} {\bibinfo {author} {\bibfnamefont {M.}~\bibnamefont
  {Hoferichter}}, \bibinfo {author} {\bibfnamefont {C.}~\bibnamefont
  {Ditsche}}, \bibinfo {author} {\bibfnamefont {B.}~\bibnamefont {Kubis}}, \
  and\ \bibinfo {author} {\bibfnamefont {U.-G.}\ \bibnamefont {Mei{\ss}ner}},\
  }\href {\doibase 10.1007/JHEP06(2012)063} {\bibfield  {journal} {\bibinfo
  {journal} {JHEP}\ }\textbf {\bibinfo {volume} {06}},\ \bibinfo {pages} {063}
  (\bibinfo {year} {2012})},\ \Eprint {http://arxiv.org/abs/1204.6251}
  {arXiv:1204.6251 [hep-ph]} \BibitemShut {NoStop}%
\bibitem [{\citenamefont {Hoferichter}\ \emph
  {et~al.}(2015{\natexlab{a}})\citenamefont {Hoferichter}, \citenamefont
  {Ruiz~de Elvira}, \citenamefont {Kubis},\ and\ \citenamefont
  {Mei\ss{}ner}}]{Hoferichter:2015dsa}%
  \BibitemOpen
  \bibfield  {author} {\bibinfo {author} {\bibfnamefont {M.}~\bibnamefont
  {Hoferichter}}, \bibinfo {author} {\bibfnamefont {J.}~\bibnamefont {Ruiz~de
  Elvira}}, \bibinfo {author} {\bibfnamefont {B.}~\bibnamefont {Kubis}}, \ and\
  \bibinfo {author} {\bibfnamefont {{\relax Ulf-G}.}~\bibnamefont
  {Mei\ss{}ner}},\ }\href {\doibase 10.1103/PhysRevLett.115.092301} {\bibfield
  {journal} {\bibinfo  {journal} {Phys. Rev. Lett.}\ }\textbf {\bibinfo
  {volume} {115}},\ \bibinfo {pages} {092301} (\bibinfo {year}
  {2015}{\natexlab{a}})},\ \Eprint {http://arxiv.org/abs/1506.04142}
  {arXiv:1506.04142 [hep-ph]} \BibitemShut {NoStop}%
\bibitem [{\citenamefont {Hoferichter}\ \emph
  {et~al.}(2015{\natexlab{b}})\citenamefont {Hoferichter}, \citenamefont
  {Ruiz~de Elvira}, \citenamefont {Kubis},\ and\ \citenamefont
  {Mei\ss{}ner}}]{Hoferichter:2015tha}%
  \BibitemOpen
  \bibfield  {author} {\bibinfo {author} {\bibfnamefont {M.}~\bibnamefont
  {Hoferichter}}, \bibinfo {author} {\bibfnamefont {J.}~\bibnamefont {Ruiz~de
  Elvira}}, \bibinfo {author} {\bibfnamefont {B.}~\bibnamefont {Kubis}}, \ and\
  \bibinfo {author} {\bibfnamefont {{\relax Ulf-G}.}~\bibnamefont
  {Mei\ss{}ner}},\ }\href {\doibase 10.1103/PhysRevLett.115.192301} {\bibfield
  {journal} {\bibinfo  {journal} {Phys. Rev. Lett.}\ }\textbf {\bibinfo
  {volume} {115}},\ \bibinfo {pages} {192301} (\bibinfo {year}
  {2015}{\natexlab{b}})},\ \Eprint {http://arxiv.org/abs/1507.07552}
  {arXiv:1507.07552 [nucl-th]} \BibitemShut {NoStop}%
\bibitem [{\citenamefont {Hoferichter}\ \emph
  {et~al.}(2016{\natexlab{a}})\citenamefont {Hoferichter}, \citenamefont
  {Ruiz~de Elvira}, \citenamefont {Kubis},\ and\ \citenamefont
  {Mei\ss{}ner}}]{Hoferichter:2015hva}%
  \BibitemOpen
  \bibfield  {author} {\bibinfo {author} {\bibfnamefont {M.}~\bibnamefont
  {Hoferichter}}, \bibinfo {author} {\bibfnamefont {J.}~\bibnamefont {Ruiz~de
  Elvira}}, \bibinfo {author} {\bibfnamefont {B.}~\bibnamefont {Kubis}}, \ and\
  \bibinfo {author} {\bibfnamefont {{\relax Ulf-G}.}~\bibnamefont
  {Mei\ss{}ner}},\ }\href {\doibase 10.1016/j.physrep.2016.02.002} {\bibfield
  {journal} {\bibinfo  {journal} {Phys. Rept.}\ }\textbf {\bibinfo {volume}
  {625}},\ \bibinfo {pages} {1} (\bibinfo {year} {2016}{\natexlab{a}})},\
  \Eprint {http://arxiv.org/abs/1510.06039} {arXiv:1510.06039 [hep-ph]}
  \BibitemShut {NoStop}%
\bibitem [{\citenamefont {Hoferichter}\ \emph
  {et~al.}(2016{\natexlab{b}})\citenamefont {Hoferichter}, \citenamefont
  {Ruiz~de Elvira}, \citenamefont {Kubis},\ and\ \citenamefont
  {Mei\ss{}ner}}]{Hoferichter:2016ocj}%
  \BibitemOpen
  \bibfield  {author} {\bibinfo {author} {\bibfnamefont {M.}~\bibnamefont
  {Hoferichter}}, \bibinfo {author} {\bibfnamefont {J.}~\bibnamefont {Ruiz~de
  Elvira}}, \bibinfo {author} {\bibfnamefont {B.}~\bibnamefont {Kubis}}, \ and\
  \bibinfo {author} {\bibfnamefont {{\relax Ulf-G}.}~\bibnamefont
  {Mei\ss{}ner}},\ }\href {\doibase 10.1016/j.physletb.2016.06.038} {\bibfield
  {journal} {\bibinfo  {journal} {Phys. Lett. B}\ }\textbf {\bibinfo {volume}
  {760}},\ \bibinfo {pages} {74} (\bibinfo {year} {2016}{\natexlab{b}})},\
  \Eprint {http://arxiv.org/abs/1602.07688} {arXiv:1602.07688 [hep-lat]}
  \BibitemShut {NoStop}%
\bibitem [{\citenamefont {Hoferichter}\ \emph
  {et~al.}(2016{\natexlab{c}})\citenamefont {Hoferichter}, \citenamefont
  {Kubis}, \citenamefont {Ruiz~de Elvira}, \citenamefont {Hammer},\ and\
  \citenamefont {Mei\ss{}ner}}]{Hoferichter:2016duk}%
  \BibitemOpen
  \bibfield  {author} {\bibinfo {author} {\bibfnamefont {M.}~\bibnamefont
  {Hoferichter}}, \bibinfo {author} {\bibfnamefont {B.}~\bibnamefont {Kubis}},
  \bibinfo {author} {\bibfnamefont {J.}~\bibnamefont {Ruiz~de Elvira}},
  \bibinfo {author} {\bibfnamefont {H.-W.}\ \bibnamefont {Hammer}}, \ and\
  \bibinfo {author} {\bibfnamefont {U.-G.}\ \bibnamefont {Mei\ss{}ner}},\
  }\href {\doibase 10.1140/epja/i2016-16331-7} {\bibfield  {journal} {\bibinfo
  {journal} {Eur. Phys. J. A}\ }\textbf {\bibinfo {volume} {52}},\ \bibinfo
  {pages} {331} (\bibinfo {year} {2016}{\natexlab{c}})},\ \Eprint
  {http://arxiv.org/abs/1609.06722} {arXiv:1609.06722 [hep-ph]} \BibitemShut
  {NoStop}%
\bibitem [{\citenamefont {Siemens}\ \emph {et~al.}(2017)\citenamefont
  {Siemens}, \citenamefont {Ruiz~de Elvira}, \citenamefont {Epelbaum},
  \citenamefont {Hoferichter}, \citenamefont {Krebs}, \citenamefont {Kubis},\
  and\ \citenamefont {Mei\ss{}ner}}]{Siemens:2016jwj}%
  \BibitemOpen
  \bibfield  {author} {\bibinfo {author} {\bibfnamefont {D.}~\bibnamefont
  {Siemens}}, \bibinfo {author} {\bibfnamefont {J.}~\bibnamefont {Ruiz~de
  Elvira}}, \bibinfo {author} {\bibfnamefont {E.}~\bibnamefont {Epelbaum}},
  \bibinfo {author} {\bibfnamefont {M.}~\bibnamefont {Hoferichter}}, \bibinfo
  {author} {\bibfnamefont {H.}~\bibnamefont {Krebs}}, \bibinfo {author}
  {\bibfnamefont {B.}~\bibnamefont {Kubis}}, \ and\ \bibinfo {author}
  {\bibfnamefont {U.-G.}\ \bibnamefont {Mei\ss{}ner}},\ }\href {\doibase
  10.1016/j.physletb.2017.04.039} {\bibfield  {journal} {\bibinfo  {journal}
  {Phys. Lett. B}\ }\textbf {\bibinfo {volume} {770}},\ \bibinfo {pages} {27}
  (\bibinfo {year} {2017})},\ \Eprint {http://arxiv.org/abs/1610.08978}
  {arXiv:1610.08978 [nucl-th]} \BibitemShut {NoStop}%
\bibitem [{\citenamefont {Ruiz~de Elvira}\ \emph {et~al.}(2018)\citenamefont
  {Ruiz~de Elvira}, \citenamefont {Hoferichter}, \citenamefont {Kubis},\ and\
  \citenamefont {Mei\ss{}ner}}]{RuizdeElvira:2017stg}%
  \BibitemOpen
  \bibfield  {author} {\bibinfo {author} {\bibfnamefont {J.}~\bibnamefont
  {Ruiz~de Elvira}}, \bibinfo {author} {\bibfnamefont {M.}~\bibnamefont
  {Hoferichter}}, \bibinfo {author} {\bibfnamefont {B.}~\bibnamefont {Kubis}},
  \ and\ \bibinfo {author} {\bibfnamefont {{\relax Ulf-G}.}~\bibnamefont
  {Mei\ss{}ner}},\ }\href {\doibase 10.1088/1361-6471/aa9422} {\bibfield
  {journal} {\bibinfo  {journal} {J. Phys. G}\ }\textbf {\bibinfo {volume}
  {45}},\ \bibinfo {pages} {024001} (\bibinfo {year} {2018})},\ \Eprint
  {http://arxiv.org/abs/1706.01465} {arXiv:1706.01465 [hep-ph]} \BibitemShut
  {NoStop}%
\bibitem [{\citenamefont {Strauch}\ \emph {et~al.}(2011)\citenamefont
  {Strauch}, \citenamefont {Amaro}, \citenamefont {Anagnostopoulos},
  \citenamefont {B{\"u}hler}, \citenamefont {Covita}, \citenamefont {Gorke}
  \emph {et~al.}}]{Strauch:2010vu}%
  \BibitemOpen
  \bibfield  {author} {\bibinfo {author} {\bibfnamefont {{\relax
  Th}.}~\bibnamefont {Strauch}}, \bibinfo {author} {\bibfnamefont {F.~D.}\
  \bibnamefont {Amaro}}, \bibinfo {author} {\bibfnamefont {D.~F.}\ \bibnamefont
  {Anagnostopoulos}}, \bibinfo {author} {\bibfnamefont {P.}~\bibnamefont
  {B{\"u}hler}}, \bibinfo {author} {\bibfnamefont {D.~S.}\ \bibnamefont
  {Covita}}, \bibinfo {author} {\bibfnamefont {H.}~\bibnamefont {Gorke}},
  \emph {et~al.},\ }\href {\doibase 10.1140/epja/i2011-11088-1} {\bibfield
  {journal} {\bibinfo  {journal} {Eur. Phys. J. A}\ }\textbf {\bibinfo {volume}
  {47}},\ \bibinfo {pages} {88} (\bibinfo {year} {2011})},\ \Eprint
  {http://arxiv.org/abs/1011.2415} {arXiv:1011.2415 [nucl-ex]} \BibitemShut
  {NoStop}%
\bibitem [{\citenamefont {Hennebach}\ \emph {et~al.}(2014)\citenamefont
  {Hennebach}, \citenamefont {Anagnostopoulos}, \citenamefont {Dax},
  \citenamefont {Fuhrmann}, \citenamefont {Gotta}, \citenamefont {Gruber} \emph
  {et~al.}}]{Hennebach:2014lsa}%
  \BibitemOpen
  \bibfield  {author} {\bibinfo {author} {\bibfnamefont {M.}~\bibnamefont
  {Hennebach}}, \bibinfo {author} {\bibfnamefont {D.~F.}\ \bibnamefont
  {Anagnostopoulos}}, \bibinfo {author} {\bibfnamefont {A.}~\bibnamefont
  {Dax}}, \bibinfo {author} {\bibfnamefont {H.}~\bibnamefont {Fuhrmann}},
  \bibinfo {author} {\bibfnamefont {D.}~\bibnamefont {Gotta}}, \bibinfo
  {author} {\bibfnamefont {A.}~\bibnamefont {Gruber}},  \emph {et~al.},\ }\href
  {\doibase 10.1140/epja/i2014-14190-x} {\bibfield  {journal} {\bibinfo
  {journal} {Eur. Phys. J. A}\ }\textbf {\bibinfo {volume} {50}},\ \bibinfo
  {pages} {190} (\bibinfo {year} {2014})},\ \bibinfo {note} {[Erratum: Eur.
  Phys. J. A {\bf 55}, 24 (2019)]},\ \Eprint {http://arxiv.org/abs/1406.6525}
  {arXiv:1406.6525 [nucl-ex]} \BibitemShut {NoStop}%
\bibitem [{\citenamefont {Hirtl}\ \emph {et~al.}(2021)\citenamefont {Hirtl},
  \citenamefont {Anagnostopoulos}, \citenamefont {Covita}, \citenamefont
  {Fuhrmann}, \citenamefont {Gorke}, \citenamefont {Gotta} \emph
  {et~al.}}]{Hirtl:2021zqf}%
  \BibitemOpen
  \bibfield  {author} {\bibinfo {author} {\bibfnamefont {A.}~\bibnamefont
  {Hirtl}}, \bibinfo {author} {\bibfnamefont {D.~F.}\ \bibnamefont
  {Anagnostopoulos}}, \bibinfo {author} {\bibfnamefont {D.~S.}\ \bibnamefont
  {Covita}}, \bibinfo {author} {\bibfnamefont {H.}~\bibnamefont {Fuhrmann}},
  \bibinfo {author} {\bibfnamefont {H.}~\bibnamefont {Gorke}}, \bibinfo
  {author} {\bibfnamefont {D.}~\bibnamefont {Gotta}},  \emph {et~al.},\ }\href
  {\doibase 10.1140/epja/s10050-021-00387-x} {\bibfield  {journal} {\bibinfo
  {journal} {Eur. Phys. J. A}\ }\textbf {\bibinfo {volume} {57}},\ \bibinfo
  {pages} {70} (\bibinfo {year} {2021})}\BibitemShut {NoStop}%
\bibitem [{\citenamefont {Baru}\ \emph
  {et~al.}(2011{\natexlab{a}})\citenamefont {Baru}, \citenamefont {Hanhart},
  \citenamefont {Hoferichter}, \citenamefont {Kubis}, \citenamefont {Nogga},\
  and\ \citenamefont {Phillips}}]{Baru:2010xn}%
  \BibitemOpen
  \bibfield  {author} {\bibinfo {author} {\bibfnamefont {V.}~\bibnamefont
  {Baru}}, \bibinfo {author} {\bibfnamefont {C.}~\bibnamefont {Hanhart}},
  \bibinfo {author} {\bibfnamefont {M.}~\bibnamefont {Hoferichter}}, \bibinfo
  {author} {\bibfnamefont {B.}~\bibnamefont {Kubis}}, \bibinfo {author}
  {\bibfnamefont {A.}~\bibnamefont {Nogga}}, \ and\ \bibinfo {author}
  {\bibfnamefont {D.~R.}\ \bibnamefont {Phillips}},\ }\href {\doibase
  10.1016/j.physletb.2010.10.028} {\bibfield  {journal} {\bibinfo  {journal}
  {Phys. Lett. B}\ }\textbf {\bibinfo {volume} {694}},\ \bibinfo {pages} {473}
  (\bibinfo {year} {2011}{\natexlab{a}})},\ \Eprint
  {http://arxiv.org/abs/1003.4444} {arXiv:1003.4444 [nucl-th]} \BibitemShut
  {NoStop}%
\bibitem [{\citenamefont {Baru}\ \emph
  {et~al.}(2011{\natexlab{b}})\citenamefont {Baru}, \citenamefont {Hanhart},
  \citenamefont {Hoferichter}, \citenamefont {Kubis}, \citenamefont {Nogga},\
  and\ \citenamefont {Phillips}}]{Baru:2011bw}%
  \BibitemOpen
  \bibfield  {author} {\bibinfo {author} {\bibfnamefont {V.}~\bibnamefont
  {Baru}}, \bibinfo {author} {\bibfnamefont {C.}~\bibnamefont {Hanhart}},
  \bibinfo {author} {\bibfnamefont {M.}~\bibnamefont {Hoferichter}}, \bibinfo
  {author} {\bibfnamefont {B.}~\bibnamefont {Kubis}}, \bibinfo {author}
  {\bibfnamefont {A.}~\bibnamefont {Nogga}}, \ and\ \bibinfo {author}
  {\bibfnamefont {D.~R.}\ \bibnamefont {Phillips}},\ }\href {\doibase
  10.1016/j.nuclphysa.2011.09.015} {\bibfield  {journal} {\bibinfo  {journal}
  {Nucl. Phys. A}\ }\textbf {\bibinfo {volume} {872}},\ \bibinfo {pages} {69}
  (\bibinfo {year} {2011}{\natexlab{b}})},\ \Eprint
  {http://arxiv.org/abs/1107.5509} {arXiv:1107.5509 [nucl-th]} \BibitemShut
  {NoStop}%
\bibitem [{\citenamefont {Gasser}\ \emph {et~al.}(2002)\citenamefont {Gasser},
  \citenamefont {Ivanov}, \citenamefont {Lipartia}, \citenamefont {Moj{\v
  z}i{\v s}},\ and\ \citenamefont {Rusetsky}}]{Gasser:2002am}%
  \BibitemOpen
  \bibfield  {author} {\bibinfo {author} {\bibfnamefont {J.}~\bibnamefont
  {Gasser}}, \bibinfo {author} {\bibfnamefont {M.~A.}\ \bibnamefont {Ivanov}},
  \bibinfo {author} {\bibfnamefont {E.}~\bibnamefont {Lipartia}}, \bibinfo
  {author} {\bibfnamefont {M.}~\bibnamefont {Moj{\v z}i{\v s}}}, \ and\
  \bibinfo {author} {\bibfnamefont {A.}~\bibnamefont {Rusetsky}},\ }\href
  {\doibase 10.1007/s10052-002-1013-z} {\bibfield  {journal} {\bibinfo
  {journal} {Eur. Phys. J. C}\ }\textbf {\bibinfo {volume} {26}},\ \bibinfo
  {pages} {13} (\bibinfo {year} {2002})},\ \Eprint
  {http://arxiv.org/abs/hep-ph/0206068} {arXiv:hep-ph/0206068} \BibitemShut
  {NoStop}%
\bibitem [{\citenamefont {Hoferichter}\ \emph {et~al.}(2009)\citenamefont
  {Hoferichter}, \citenamefont {Kubis},\ and\ \citenamefont
  {Mei\ss{}ner}}]{Hoferichter:2009ez}%
  \BibitemOpen
  \bibfield  {author} {\bibinfo {author} {\bibfnamefont {M.}~\bibnamefont
  {Hoferichter}}, \bibinfo {author} {\bibfnamefont {B.}~\bibnamefont {Kubis}},
  \ and\ \bibinfo {author} {\bibfnamefont {{\relax Ulf-G}.}~\bibnamefont
  {Mei\ss{}ner}},\ }\href {\doibase 10.1016/j.physletb.2009.05.068} {\bibfield
  {journal} {\bibinfo  {journal} {Phys. Lett.}\ }\textbf {\bibinfo {volume}
  {B678}},\ \bibinfo {pages} {65} (\bibinfo {year} {2009})},\ \Eprint
  {http://arxiv.org/abs/0903.3890} {arXiv:0903.3890 [hep-ph]} \BibitemShut
  {NoStop}%
\bibitem [{\citenamefont {Hoferichter}\ \emph {et~al.}(2010)\citenamefont
  {Hoferichter}, \citenamefont {Kubis},\ and\ \citenamefont
  {Mei\ss{}ner}}]{Hoferichter:2009gn}%
  \BibitemOpen
  \bibfield  {author} {\bibinfo {author} {\bibfnamefont {M.}~\bibnamefont
  {Hoferichter}}, \bibinfo {author} {\bibfnamefont {B.}~\bibnamefont {Kubis}},
  \ and\ \bibinfo {author} {\bibfnamefont {{\relax Ulf-G}.}~\bibnamefont
  {Mei\ss{}ner}},\ }\href {\doibase 10.1016/j.nuclphysa.2009.11.012} {\bibfield
   {journal} {\bibinfo  {journal} {Nucl. Phys. A}\ }\textbf {\bibinfo {volume}
  {833}},\ \bibinfo {pages} {18} (\bibinfo {year} {2010})},\ \Eprint
  {http://arxiv.org/abs/0909.4390} {arXiv:0909.4390 [hep-ph]} \BibitemShut
  {NoStop}%
\bibitem [{\citenamefont {Hoferichter}\ \emph {et~al.}(2013)\citenamefont
  {Hoferichter}, \citenamefont {Baru}, \citenamefont {Hanhart}, \citenamefont
  {Kubis}, \citenamefont {Nogga},\ and\ \citenamefont
  {Phillips}}]{Hoferichter:2012bz}%
  \BibitemOpen
  \bibfield  {author} {\bibinfo {author} {\bibfnamefont {M.}~\bibnamefont
  {Hoferichter}}, \bibinfo {author} {\bibfnamefont {V.}~\bibnamefont {Baru}},
  \bibinfo {author} {\bibfnamefont {C.}~\bibnamefont {Hanhart}}, \bibinfo
  {author} {\bibfnamefont {B.}~\bibnamefont {Kubis}}, \bibinfo {author}
  {\bibfnamefont {A.}~\bibnamefont {Nogga}}, \ and\ \bibinfo {author}
  {\bibfnamefont {D.~R.}\ \bibnamefont {Phillips}},\ }\href {\doibase
  10.22323/1.172.0093} {\bibfield  {journal} {\bibinfo  {journal} {PoS}\
  }\textbf {\bibinfo {volume} {CD12}},\ \bibinfo {pages} {093} (\bibinfo {year}
  {2013})},\ \Eprint {http://arxiv.org/abs/1211.1145} {arXiv:1211.1145
  [nucl-th]} \BibitemShut {NoStop}%
\bibitem [{\citenamefont {Brack}\ \emph {et~al.}(1990)\citenamefont {Brack},
  \citenamefont {Ristinen}, \citenamefont {Kraushaar}, \citenamefont {Loveman},
  \citenamefont {Peterson}, \citenamefont {Smith}, \citenamefont {Gill},
  \citenamefont {Ottewell}, \citenamefont {Sevior}, \citenamefont {Trelle},
  \citenamefont {Mathie}, \citenamefont {Grion},\ and\ \citenamefont
  {Rui}}]{Brack:1989sj}%
  \BibitemOpen
  \bibfield  {author} {\bibinfo {author} {\bibfnamefont {J.~T.}\ \bibnamefont
  {Brack}}, \bibinfo {author} {\bibfnamefont {R.~A.}\ \bibnamefont {Ristinen}},
  \bibinfo {author} {\bibfnamefont {J.~J.}\ \bibnamefont {Kraushaar}}, \bibinfo
  {author} {\bibfnamefont {R.~A.}\ \bibnamefont {Loveman}}, \bibinfo {author}
  {\bibfnamefont {R.~J.}\ \bibnamefont {Peterson}}, \bibinfo {author}
  {\bibfnamefont {G.~R.}\ \bibnamefont {Smith}}, \bibinfo {author}
  {\bibfnamefont {D.~R.}\ \bibnamefont {Gill}}, \bibinfo {author}
  {\bibfnamefont {D.~F.}\ \bibnamefont {Ottewell}}, \bibinfo {author}
  {\bibfnamefont {M.~E.}\ \bibnamefont {Sevior}}, \bibinfo {author}
  {\bibfnamefont {R.~P.}\ \bibnamefont {Trelle}}, \bibinfo {author}
  {\bibfnamefont {E.~L.}\ \bibnamefont {Mathie}}, \bibinfo {author}
  {\bibfnamefont {N.}~\bibnamefont {Grion}}, \ and\ \bibinfo {author}
  {\bibfnamefont {R.}~\bibnamefont {Rui}},\ }\href {\doibase
  10.1103/PhysRevC.41.2202} {\bibfield  {journal} {\bibinfo  {journal} {Phys.
  Rev. C}\ }\textbf {\bibinfo {volume} {41}},\ \bibinfo {pages} {2202}
  (\bibinfo {year} {1990})}\BibitemShut {NoStop}%
\bibitem [{\citenamefont {Joram}\ \emph {et~al.}(1995)\citenamefont {Joram},
  \citenamefont {Metzler}, \citenamefont {Jaki}, \citenamefont {Kluge},
  \citenamefont {Matth\"ay}, \citenamefont {Wieser}, \citenamefont {Barnett},
  \citenamefont {Clement}, \citenamefont {Krell},\ and\ \citenamefont
  {Wagner}}]{Joram:1995gr}%
  \BibitemOpen
  \bibfield  {author} {\bibinfo {author} {\bibfnamefont {{\relax
  Ch}.}~\bibnamefont {Joram}}, \bibinfo {author} {\bibfnamefont
  {M.}~\bibnamefont {Metzler}}, \bibinfo {author} {\bibfnamefont
  {J.}~\bibnamefont {Jaki}}, \bibinfo {author} {\bibfnamefont {W.}~\bibnamefont
  {Kluge}}, \bibinfo {author} {\bibfnamefont {H.}~\bibnamefont {Matth\"ay}},
  \bibinfo {author} {\bibfnamefont {R.}~\bibnamefont {Wieser}}, \bibinfo
  {author} {\bibfnamefont {B.~M.}\ \bibnamefont {Barnett}}, \bibinfo {author}
  {\bibfnamefont {H.}~\bibnamefont {Clement}}, \bibinfo {author} {\bibfnamefont
  {S.}~\bibnamefont {Krell}}, \ and\ \bibinfo {author} {\bibfnamefont {G.~J.}\
  \bibnamefont {Wagner}},\ }\href {\doibase 10.1103/PhysRevC.51.2144}
  {\bibfield  {journal} {\bibinfo  {journal} {Phys. Rev. C}\ }\textbf {\bibinfo
  {volume} {51}},\ \bibinfo {pages} {2144} (\bibinfo {year}
  {1995})}\BibitemShut {NoStop}%
\bibitem [{\citenamefont {Denz}\ \emph {et~al.}(2006)\citenamefont {Denz},
  \citenamefont {Amaudruz}, \citenamefont {Brack}, \citenamefont {Breitschopf},
  \citenamefont {Camerini}, \citenamefont {Clark} \emph
  {et~al.}}]{Denz:2005jq}%
  \BibitemOpen
  \bibfield  {author} {\bibinfo {author} {\bibfnamefont {H.}~\bibnamefont
  {Denz}}, \bibinfo {author} {\bibfnamefont {P.}~\bibnamefont {Amaudruz}},
  \bibinfo {author} {\bibfnamefont {J.~T.}\ \bibnamefont {Brack}}, \bibinfo
  {author} {\bibfnamefont {J.}~\bibnamefont {Breitschopf}}, \bibinfo {author}
  {\bibfnamefont {P.}~\bibnamefont {Camerini}}, \bibinfo {author}
  {\bibfnamefont {J.~L.}\ \bibnamefont {Clark}},  \emph {et~al.},\ }\href
  {\doibase 10.1016/j.physletb.2005.12.017} {\bibfield  {journal} {\bibinfo
  {journal} {Phys. Lett. B}\ }\textbf {\bibinfo {volume} {633}},\ \bibinfo
  {pages} {209} (\bibinfo {year} {2006})},\ \Eprint
  {http://arxiv.org/abs/nucl-ex/0512006} {arXiv:nucl-ex/0512006} \BibitemShut
  {NoStop}%
\bibitem [{\citenamefont {Frle{\v z}}\ \emph {et~al.}(1998)\citenamefont
  {Frle{\v z}}, \citenamefont {Po{\v c}ani\'c}, \citenamefont {Assamagan},
  \citenamefont {Chen}, \citenamefont {Keeter}, \citenamefont {Marshall},
  \citenamefont {Minehart}, \citenamefont {Smith}, \citenamefont {Dodge},
  \citenamefont {Hanna}, \citenamefont {King},\ and\ \citenamefont
  {Knudson}}]{Frlez:1997qu}%
  \BibitemOpen
  \bibfield  {author} {\bibinfo {author} {\bibfnamefont {E.}~\bibnamefont
  {Frle{\v z}}}, \bibinfo {author} {\bibfnamefont {D.}~\bibnamefont {Po{\v
  c}ani\'c}}, \bibinfo {author} {\bibfnamefont {K.~A.}\ \bibnamefont
  {Assamagan}}, \bibinfo {author} {\bibfnamefont {J.~P.}\ \bibnamefont {Chen}},
  \bibinfo {author} {\bibfnamefont {K.~J.}\ \bibnamefont {Keeter}}, \bibinfo
  {author} {\bibfnamefont {R.~M.}\ \bibnamefont {Marshall}}, \bibinfo {author}
  {\bibfnamefont {R.~C.}\ \bibnamefont {Minehart}}, \bibinfo {author}
  {\bibfnamefont {L.~C.}\ \bibnamefont {Smith}}, \bibinfo {author}
  {\bibfnamefont {G.~E.}\ \bibnamefont {Dodge}}, \bibinfo {author}
  {\bibfnamefont {S.~S.}\ \bibnamefont {Hanna}}, \bibinfo {author}
  {\bibfnamefont {B.~H.}\ \bibnamefont {King}}, \ and\ \bibinfo {author}
  {\bibfnamefont {J.~N.}\ \bibnamefont {Knudson}},\ }\href {\doibase
  10.1103/PhysRevC.57.3144} {\bibfield  {journal} {\bibinfo  {journal} {Phys.
  Rev. C}\ }\textbf {\bibinfo {volume} {57}},\ \bibinfo {pages} {3144}
  (\bibinfo {year} {1998})},\ \Eprint {http://arxiv.org/abs/hep-ex/9712024}
  {arXiv:hep-ex/9712024} \BibitemShut {NoStop}%
\bibitem [{\citenamefont {Isenhower}\ \emph {et~al.}(1999)\citenamefont
  {Isenhower}, \citenamefont {Black}, \citenamefont {Brooks}, \citenamefont
  {Brown}, \citenamefont {Graessle} \emph {et~al.}}]{Isenhower:1999aj}%
  \BibitemOpen
  \bibfield  {author} {\bibinfo {author} {\bibfnamefont {L.~D.}\ \bibnamefont
  {Isenhower}}, \bibinfo {author} {\bibfnamefont {T.}~\bibnamefont {Black}},
  \bibinfo {author} {\bibfnamefont {B.~M.}\ \bibnamefont {Brooks}}, \bibinfo
  {author} {\bibfnamefont {A.~D.}\ \bibnamefont {Brown}}, \bibinfo {author}
  {\bibfnamefont {K.}~\bibnamefont {Graessle}},  \emph {et~al.},\ }\href
  {https://gwdac.phys.gwu.edu/Newsletters/PiN-Newsletter-15.pdf} {\bibfield
  {journal} {\bibinfo  {journal} {PiN Newslett.}\ }\textbf {\bibinfo {volume}
  {15}},\ \bibinfo {pages} {292} (\bibinfo {year} {1999})}\BibitemShut
  {NoStop}%
\bibitem [{\citenamefont {Jia}\ \emph {et~al.}(2008)\citenamefont {Jia},
  \citenamefont {Gorringe}, \citenamefont {Hasinoff}, \citenamefont {Kovash},
  \citenamefont {Ojha}, \citenamefont {Pavan}, \citenamefont {Tripathi},\ and\
  \citenamefont {{\.Z}o{\l}nierczuk}}]{Jia:2008rt}%
  \BibitemOpen
  \bibfield  {author} {\bibinfo {author} {\bibfnamefont {Y.}~\bibnamefont
  {Jia}}, \bibinfo {author} {\bibfnamefont {T.~P.}\ \bibnamefont {Gorringe}},
  \bibinfo {author} {\bibfnamefont {M.~D.}\ \bibnamefont {Hasinoff}}, \bibinfo
  {author} {\bibfnamefont {M.~A.}\ \bibnamefont {Kovash}}, \bibinfo {author}
  {\bibfnamefont {M.}~\bibnamefont {Ojha}}, \bibinfo {author} {\bibfnamefont
  {M.~M.}\ \bibnamefont {Pavan}}, \bibinfo {author} {\bibfnamefont
  {S.}~\bibnamefont {Tripathi}}, \ and\ \bibinfo {author} {\bibfnamefont
  {P.~A.}\ \bibnamefont {{\.Z}o{\l}nierczuk}},\ }\href {\doibase
  10.1103/PhysRevLett.101.102301} {\bibfield  {journal} {\bibinfo  {journal}
  {Phys. Rev. Lett.}\ }\textbf {\bibinfo {volume} {101}},\ \bibinfo {pages}
  {102301} (\bibinfo {year} {2008})},\ \Eprint {http://arxiv.org/abs/0804.1531}
  {arXiv:0804.1531 [nucl-ex]} \BibitemShut {NoStop}%
\bibitem [{\citenamefont {Mekterovi{\'c}}\ \emph {et~al.}(2009)\citenamefont
  {Mekterovi{\'c}}, \citenamefont {Supek}, \citenamefont {Abaev}, \citenamefont
  {Bekrenev}, \citenamefont {Bircher}, \citenamefont {Briscoe} \emph
  {et~al.}}]{Mekterovic:2009kw}%
  \BibitemOpen
  \bibfield  {author} {\bibinfo {author} {\bibfnamefont {D.}~\bibnamefont
  {Mekterovi{\'c}}}, \bibinfo {author} {\bibfnamefont {I.}~\bibnamefont
  {Supek}}, \bibinfo {author} {\bibfnamefont {V.}~\bibnamefont {Abaev}},
  \bibinfo {author} {\bibfnamefont {V.}~\bibnamefont {Bekrenev}}, \bibinfo
  {author} {\bibfnamefont {C.}~\bibnamefont {Bircher}}, \bibinfo {author}
  {\bibfnamefont {W.~J.}\ \bibnamefont {Briscoe}},  \emph {et~al.} (\bibinfo
  {collaboration} {Crystal Ball}),\ }\href {\doibase
  10.1103/PhysRevC.80.055207} {\bibfield  {journal} {\bibinfo  {journal} {Phys.
  Rev. C}\ }\textbf {\bibinfo {volume} {80}},\ \bibinfo {pages} {055207}
  (\bibinfo {year} {2009})},\ \Eprint {http://arxiv.org/abs/0908.3845}
  {arXiv:0908.3845 [hep-ex]} \BibitemShut {NoStop}%
\bibitem [{\citenamefont {D{\"u}rr}\ \emph {et~al.}(2012)\citenamefont
  {D{\"u}rr}, \citenamefont {Fodor}, \citenamefont {Hemmert}, \citenamefont
  {Hoelbling}, \citenamefont {Frison}, \citenamefont {Katz}, \citenamefont
  {Krieg}, \citenamefont {Kurth}, \citenamefont {Lellouch}, \citenamefont
  {Lippert}, \citenamefont {Portelli}, \citenamefont {Ramos}, \citenamefont
  {Sch\"afer},\ and\ \citenamefont {Szab\'o}}]{Durr:2011mp}%
  \BibitemOpen
  \bibfield  {author} {\bibinfo {author} {\bibfnamefont {S.}~\bibnamefont
  {D{\"u}rr}}, \bibinfo {author} {\bibfnamefont {Z.}~\bibnamefont {Fodor}},
  \bibinfo {author} {\bibfnamefont {T.}~\bibnamefont {Hemmert}}, \bibinfo
  {author} {\bibfnamefont {C.}~\bibnamefont {Hoelbling}}, \bibinfo {author}
  {\bibfnamefont {J.}~\bibnamefont {Frison}}, \bibinfo {author} {\bibfnamefont
  {S.~D.}\ \bibnamefont {Katz}}, \bibinfo {author} {\bibfnamefont
  {S.}~\bibnamefont {Krieg}}, \bibinfo {author} {\bibfnamefont
  {T.}~\bibnamefont {Kurth}}, \bibinfo {author} {\bibfnamefont
  {L.}~\bibnamefont {Lellouch}}, \bibinfo {author} {\bibfnamefont
  {T.}~\bibnamefont {Lippert}}, \bibinfo {author} {\bibfnamefont
  {A.}~\bibnamefont {Portelli}}, \bibinfo {author} {\bibfnamefont
  {A.}~\bibnamefont {Ramos}}, \bibinfo {author} {\bibfnamefont
  {A.}~\bibnamefont {Sch\"afer}}, \ and\ \bibinfo {author} {\bibfnamefont
  {K.~K.}\ \bibnamefont {Szab\'o}} (\bibinfo {collaboration} {BMWc}),\ }\href
  {\doibase 10.1103/PhysRevD.85.014509} {\bibfield  {journal} {\bibinfo
  {journal} {Phys. Rev. D}\ }\textbf {\bibinfo {volume} {85}},\ \bibinfo
  {pages} {014509} (\bibinfo {year} {2012})},\ \bibinfo {note} {[Erratum: Phys.
  Rev. D {\bf 93}, 039905(E) (2016)]},\ \Eprint
  {http://arxiv.org/abs/1109.4265} {arXiv:1109.4265 [hep-lat]} \BibitemShut
  {NoStop}%
\bibitem [{\citenamefont {Bali}\ \emph {et~al.}(2013)\citenamefont {Bali},
  \citenamefont {Bruns}, \citenamefont {Collins}, \citenamefont {Deka},
  \citenamefont {Gl{\"a}\ss{}le}, \citenamefont {G{\"o}ckeler} \emph
  {et~al.}}]{Bali:2012qs}%
  \BibitemOpen
  \bibfield  {author} {\bibinfo {author} {\bibfnamefont {G.~S.}\ \bibnamefont
  {Bali}}, \bibinfo {author} {\bibfnamefont {P.~C.}\ \bibnamefont {Bruns}},
  \bibinfo {author} {\bibfnamefont {S.}~\bibnamefont {Collins}}, \bibinfo
  {author} {\bibfnamefont {M.}~\bibnamefont {Deka}}, \bibinfo {author}
  {\bibfnamefont {B.}~\bibnamefont {Gl{\"a}\ss{}le}}, \bibinfo {author}
  {\bibfnamefont {M.}~\bibnamefont {G{\"o}ckeler}},  \emph {et~al.} (\bibinfo
  {collaboration} {QCDSF}),\ }\href {\doibase 10.1016/j.nuclphysb.2012.08.009}
  {\bibfield  {journal} {\bibinfo  {journal} {Nucl. Phys. B}\ }\textbf
  {\bibinfo {volume} {866}},\ \bibinfo {pages} {1} (\bibinfo {year} {2013})},\
  \Eprint {http://arxiv.org/abs/1206.7034} {arXiv:1206.7034 [hep-lat]}
  \BibitemShut {NoStop}%
\bibitem [{\citenamefont {D{\"u}rr}\ \emph {et~al.}(2016)\citenamefont
  {D{\"u}rr}, \citenamefont {Fodor}, \citenamefont {Hoelbling}, \citenamefont
  {Katz}, \citenamefont {Krieg}, \citenamefont {Lellouch}, \citenamefont
  {Lippert}, \citenamefont {Metivet}, \citenamefont {Portelli}, \citenamefont
  {Szabo}, \citenamefont {Torrero}, \citenamefont {Toth},\ and\ \citenamefont
  {Varnhorst}}]{Durr:2015dna}%
  \BibitemOpen
  \bibfield  {author} {\bibinfo {author} {\bibfnamefont {S.}~\bibnamefont
  {D{\"u}rr}}, \bibinfo {author} {\bibfnamefont {Z.}~\bibnamefont {Fodor}},
  \bibinfo {author} {\bibfnamefont {C.}~\bibnamefont {Hoelbling}}, \bibinfo
  {author} {\bibfnamefont {S.~D.}\ \bibnamefont {Katz}}, \bibinfo {author}
  {\bibfnamefont {S.}~\bibnamefont {Krieg}}, \bibinfo {author} {\bibfnamefont
  {L.}~\bibnamefont {Lellouch}}, \bibinfo {author} {\bibfnamefont
  {T.}~\bibnamefont {Lippert}}, \bibinfo {author} {\bibfnamefont
  {T.}~\bibnamefont {Metivet}}, \bibinfo {author} {\bibfnamefont
  {A.}~\bibnamefont {Portelli}}, \bibinfo {author} {\bibfnamefont {K.~K.}\
  \bibnamefont {Szabo}}, \bibinfo {author} {\bibfnamefont {C.}~\bibnamefont
  {Torrero}}, \bibinfo {author} {\bibfnamefont {B.~C.}\ \bibnamefont {Toth}}, \
  and\ \bibinfo {author} {\bibfnamefont {L.}~\bibnamefont {Varnhorst}}
  (\bibinfo {collaboration} {Budapest-Marseille-Wuppertal Collaboration}),\
  }\href {\doibase 10.1103/PhysRevLett.116.172001} {\bibfield  {journal}
  {\bibinfo  {journal} {Phys. Rev. Lett.}\ }\textbf {\bibinfo {volume} {116}},\
  \bibinfo {pages} {172001} (\bibinfo {year} {2016})},\ \Eprint
  {http://arxiv.org/abs/1510.08013} {arXiv:1510.08013 [hep-lat]} \BibitemShut
  {NoStop}%
\bibitem [{\citenamefont {Yang}\ \emph {et~al.}(2016)\citenamefont {Yang},
  \citenamefont {Alexandru}, \citenamefont {Draper}, \citenamefont {Liang},\
  and\ \citenamefont {Liu}}]{Yang:2015uis}%
  \BibitemOpen
  \bibfield  {author} {\bibinfo {author} {\bibfnamefont {Y.-B.}\ \bibnamefont
  {Yang}}, \bibinfo {author} {\bibfnamefont {A.}~\bibnamefont {Alexandru}},
  \bibinfo {author} {\bibfnamefont {T.}~\bibnamefont {Draper}}, \bibinfo
  {author} {\bibfnamefont {J.}~\bibnamefont {Liang}}, \ and\ \bibinfo {author}
  {\bibfnamefont {K.-F.}\ \bibnamefont {Liu}} (\bibinfo {collaboration}
  {$\chi$QCD}),\ }\href {\doibase 10.1103/PhysRevD.94.054503} {\bibfield
  {journal} {\bibinfo  {journal} {Phys. Rev. D}\ }\textbf {\bibinfo {volume}
  {94}},\ \bibinfo {pages} {054503} (\bibinfo {year} {2016})},\ \Eprint
  {http://arxiv.org/abs/1511.09089} {arXiv:1511.09089 [hep-lat]} \BibitemShut
  {NoStop}%
\bibitem [{\citenamefont {Abdel-Rehim}\ \emph {et~al.}(2016)\citenamefont
  {Abdel-Rehim}, \citenamefont {Alexandrou}, \citenamefont {Constantinou},
  \citenamefont {Hadjiyiannakou}, \citenamefont {Jansen}, \citenamefont
  {Kallidonis}, \citenamefont {Koutsou},\ and\ \citenamefont {Vaquero
  Avil{\'e}s-Casco}}]{Abdel-Rehim:2016won}%
  \BibitemOpen
  \bibfield  {author} {\bibinfo {author} {\bibfnamefont {A.}~\bibnamefont
  {Abdel-Rehim}}, \bibinfo {author} {\bibfnamefont {C.}~\bibnamefont
  {Alexandrou}}, \bibinfo {author} {\bibfnamefont {M.}~\bibnamefont
  {Constantinou}}, \bibinfo {author} {\bibfnamefont {K.}~\bibnamefont
  {Hadjiyiannakou}}, \bibinfo {author} {\bibfnamefont {K.}~\bibnamefont
  {Jansen}}, \bibinfo {author} {\bibfnamefont {C.}~\bibnamefont {Kallidonis}},
  \bibinfo {author} {\bibfnamefont {G.}~\bibnamefont {Koutsou}}, \ and\
  \bibinfo {author} {\bibfnamefont {A.}~\bibnamefont {Vaquero
  Avil{\'e}s-Casco}} (\bibinfo {collaboration} {ETM}),\ }\href {\doibase
  10.1103/PhysRevLett.116.252001} {\bibfield  {journal} {\bibinfo  {journal}
  {Phys. Rev. Lett.}\ }\textbf {\bibinfo {volume} {116}},\ \bibinfo {pages}
  {252001} (\bibinfo {year} {2016})},\ \Eprint
  {http://arxiv.org/abs/1601.01624} {arXiv:1601.01624 [hep-lat]} \BibitemShut
  {NoStop}%
\bibitem [{\citenamefont {Bali}\ \emph {et~al.}(2016)\citenamefont {Bali},
  \citenamefont {Collins}, \citenamefont {Richtmann}, \citenamefont
  {Sch\"afer}, \citenamefont {S\"oldner},\ and\ \citenamefont
  {Sternbeck}}]{Bali:2016lvx}%
  \BibitemOpen
  \bibfield  {author} {\bibinfo {author} {\bibfnamefont {G.~S.}\ \bibnamefont
  {Bali}}, \bibinfo {author} {\bibfnamefont {S.}~\bibnamefont {Collins}},
  \bibinfo {author} {\bibfnamefont {D.}~\bibnamefont {Richtmann}}, \bibinfo
  {author} {\bibfnamefont {A.}~\bibnamefont {Sch\"afer}}, \bibinfo {author}
  {\bibfnamefont {W.}~\bibnamefont {S\"oldner}}, \ and\ \bibinfo {author}
  {\bibfnamefont {A.}~\bibnamefont {Sternbeck}} (\bibinfo {collaboration}
  {RQCD}),\ }\href {\doibase 10.1103/PhysRevD.93.094504} {\bibfield  {journal}
  {\bibinfo  {journal} {Phys. Rev. D}\ }\textbf {\bibinfo {volume} {93}},\
  \bibinfo {pages} {094504} (\bibinfo {year} {2016})},\ \Eprint
  {http://arxiv.org/abs/1603.00827} {arXiv:1603.00827 [hep-lat]} \BibitemShut
  {NoStop}%
\bibitem [{\citenamefont {Yamanaka}\ \emph {et~al.}(2018)\citenamefont
  {Yamanaka}, \citenamefont {Hashimoto}, \citenamefont {Kaneko},\ and\
  \citenamefont {Ohki}}]{Yamanaka:2018uud}%
  \BibitemOpen
  \bibfield  {author} {\bibinfo {author} {\bibfnamefont {N.}~\bibnamefont
  {Yamanaka}}, \bibinfo {author} {\bibfnamefont {S.}~\bibnamefont {Hashimoto}},
  \bibinfo {author} {\bibfnamefont {T.}~\bibnamefont {Kaneko}}, \ and\ \bibinfo
  {author} {\bibfnamefont {H.}~\bibnamefont {Ohki}} (\bibinfo {collaboration}
  {JLQCD}),\ }\href {\doibase 10.1103/PhysRevD.98.054516} {\bibfield  {journal}
  {\bibinfo  {journal} {Phys. Rev. D}\ }\textbf {\bibinfo {volume} {98}},\
  \bibinfo {pages} {054516} (\bibinfo {year} {2018})},\ \Eprint
  {http://arxiv.org/abs/1805.10507} {arXiv:1805.10507 [hep-lat]} \BibitemShut
  {NoStop}%
\bibitem [{\citenamefont {Alexandrou}\ \emph {et~al.}(2020)\citenamefont
  {Alexandrou}, \citenamefont {Bacchio}, \citenamefont {Constantinou},
  \citenamefont {Finkenrath}, \citenamefont {Hadjiyiannakou}, \citenamefont
  {Jansen}, \citenamefont {Koutsou},\ and\ \citenamefont {Vaquero
  Avil{\'e}s-Casco}}]{Alexandrou:2019brg}%
  \BibitemOpen
  \bibfield  {author} {\bibinfo {author} {\bibfnamefont {C.}~\bibnamefont
  {Alexandrou}}, \bibinfo {author} {\bibfnamefont {S.}~\bibnamefont {Bacchio}},
  \bibinfo {author} {\bibfnamefont {M.}~\bibnamefont {Constantinou}}, \bibinfo
  {author} {\bibfnamefont {J.}~\bibnamefont {Finkenrath}}, \bibinfo {author}
  {\bibfnamefont {K.}~\bibnamefont {Hadjiyiannakou}}, \bibinfo {author}
  {\bibfnamefont {K.}~\bibnamefont {Jansen}}, \bibinfo {author} {\bibfnamefont
  {G.}~\bibnamefont {Koutsou}}, \ and\ \bibinfo {author} {\bibfnamefont
  {A.}~\bibnamefont {Vaquero Avil{\'e}s-Casco}} (\bibinfo {collaboration}
  {ETM}),\ }\href {\doibase 10.1103/PhysRevD.102.054517} {\bibfield  {journal}
  {\bibinfo  {journal} {Phys. Rev. D}\ }\textbf {\bibinfo {volume} {102}},\
  \bibinfo {pages} {054517} (\bibinfo {year} {2020})},\ \Eprint
  {http://arxiv.org/abs/1909.00485} {arXiv:1909.00485 [hep-lat]} \BibitemShut
  {NoStop}%
\bibitem [{\citenamefont {Borsanyi}\ \emph {et~al.}(2020)\citenamefont
  {Borsanyi}, \citenamefont {Fodor}, \citenamefont {Hoelbling}, \citenamefont
  {Lellouch}, \citenamefont {Szabo}, \citenamefont {Torrero},\ and\
  \citenamefont {Varnhorst}}]{Borsanyi:2020bpd}%
  \BibitemOpen
  \bibfield  {author} {\bibinfo {author} {\bibfnamefont {{\relax
  Sz}.}~\bibnamefont {Borsanyi}}, \bibinfo {author} {\bibfnamefont
  {Z.}~\bibnamefont {Fodor}}, \bibinfo {author} {\bibfnamefont
  {C.}~\bibnamefont {Hoelbling}}, \bibinfo {author} {\bibfnamefont
  {L.}~\bibnamefont {Lellouch}}, \bibinfo {author} {\bibfnamefont {K.~K.}\
  \bibnamefont {Szabo}}, \bibinfo {author} {\bibfnamefont {C.}~\bibnamefont
  {Torrero}}, \ and\ \bibinfo {author} {\bibfnamefont {L.}~\bibnamefont
  {Varnhorst}} (\bibinfo {collaboration} {BMWc}),\ }\href@noop {} {\  (\bibinfo
  {year} {2020})},\ \Eprint {http://arxiv.org/abs/2007.03319} {arXiv:2007.03319
  [hep-lat]} \BibitemShut {NoStop}%
\bibitem [{\citenamefont {Alexandrou}\ \emph {et~al.}(2014)\citenamefont
  {Alexandrou}, \citenamefont {Drach}, \citenamefont {Jansen}, \citenamefont
  {Kallidonis},\ and\ \citenamefont {Koutsou}}]{Alexandrou:2014sha}%
  \BibitemOpen
  \bibfield  {author} {\bibinfo {author} {\bibfnamefont {C.}~\bibnamefont
  {Alexandrou}}, \bibinfo {author} {\bibfnamefont {V.}~\bibnamefont {Drach}},
  \bibinfo {author} {\bibfnamefont {K.}~\bibnamefont {Jansen}}, \bibinfo
  {author} {\bibfnamefont {C.}~\bibnamefont {Kallidonis}}, \ and\ \bibinfo
  {author} {\bibfnamefont {G.}~\bibnamefont {Koutsou}},\ }\href {\doibase
  10.1103/PhysRevD.90.074501} {\bibfield  {journal} {\bibinfo  {journal} {Phys.
  Rev. D}\ }\textbf {\bibinfo {volume} {90}},\ \bibinfo {pages} {074501}
  (\bibinfo {year} {2014})},\ \Eprint {http://arxiv.org/abs/1406.4310}
  {arXiv:1406.4310 [hep-lat]} \BibitemShut {NoStop}%
\bibitem [{\citenamefont {Aoki}\ \emph {et~al.}(2020)\citenamefont {Aoki},
  \citenamefont {Aoki}, \citenamefont {Be{\v c}irevi\'c}, \citenamefont {Blum},
  \citenamefont {Colangelo}, \citenamefont {Collins} \emph
  {et~al.}}]{Aoki:2019cca}%
  \BibitemOpen
  \bibfield  {author} {\bibinfo {author} {\bibfnamefont {S.}~\bibnamefont
  {Aoki}}, \bibinfo {author} {\bibfnamefont {Y.}~\bibnamefont {Aoki}}, \bibinfo
  {author} {\bibfnamefont {D.}~\bibnamefont {Be{\v c}irevi\'c}}, \bibinfo
  {author} {\bibfnamefont {T.}~\bibnamefont {Blum}}, \bibinfo {author}
  {\bibfnamefont {G.}~\bibnamefont {Colangelo}}, \bibinfo {author}
  {\bibfnamefont {S.}~\bibnamefont {Collins}},  \emph {et~al.} (\bibinfo
  {collaboration} {Flavour Lattice Averaging Group}),\ }\href {\doibase
  10.1140/epjc/s10052-019-7354-7} {\bibfield  {journal} {\bibinfo  {journal}
  {Eur. Phys. J. C}\ }\textbf {\bibinfo {volume} {80}},\ \bibinfo {pages} {113}
  (\bibinfo {year} {2020})},\ \Eprint {http://arxiv.org/abs/1902.08191}
  {arXiv:1902.08191 [hep-lat]} \BibitemShut {NoStop}%
\bibitem [{\citenamefont {Jenkins}\ and\ \citenamefont
  {Manohar}(1991)}]{Jenkins:1990jv}%
  \BibitemOpen
  \bibfield  {author} {\bibinfo {author} {\bibfnamefont {E.~E.}\ \bibnamefont
  {Jenkins}}\ and\ \bibinfo {author} {\bibfnamefont {A.~V.}\ \bibnamefont
  {Manohar}},\ }\href {\doibase 10.1016/0370-2693(91)90266-S} {\bibfield
  {journal} {\bibinfo  {journal} {Phys. Lett. B}\ }\textbf {\bibinfo {volume}
  {255}},\ \bibinfo {pages} {558} (\bibinfo {year} {1991})}\BibitemShut
  {NoStop}%
\bibitem [{\citenamefont {Bernard}\ \emph {et~al.}(1992)\citenamefont
  {Bernard}, \citenamefont {Kaiser}, \citenamefont {Kambor},\ and\
  \citenamefont {Mei{\ss}ner}}]{Bernard:1992qa}%
  \BibitemOpen
  \bibfield  {author} {\bibinfo {author} {\bibfnamefont {V.}~\bibnamefont
  {Bernard}}, \bibinfo {author} {\bibfnamefont {N.}~\bibnamefont {Kaiser}},
  \bibinfo {author} {\bibfnamefont {J.}~\bibnamefont {Kambor}}, \ and\ \bibinfo
  {author} {\bibfnamefont {{\relax Ulf-G}.}~\bibnamefont {Mei{\ss}ner}},\
  }\href {\doibase 10.1016/0550-3213(92)90615-I} {\bibfield  {journal}
  {\bibinfo  {journal} {Nucl. Phys. B}\ }\textbf {\bibinfo {volume} {388}},\
  \bibinfo {pages} {315} (\bibinfo {year} {1992})}\BibitemShut {NoStop}%
\bibitem [{\citenamefont {Follana}\ \emph {et~al.}(2007)\citenamefont
  {Follana}, \citenamefont {Mason}, \citenamefont {Davies}, \citenamefont
  {Hornbostel}, \citenamefont {Lepage}, \citenamefont {Shigemitsu},
  \citenamefont {Trottier},\ and\ \citenamefont {Wong}}]{Follana:2006rc}%
  \BibitemOpen
  \bibfield  {author} {\bibinfo {author} {\bibfnamefont {E.}~\bibnamefont
  {Follana}}, \bibinfo {author} {\bibfnamefont {Q.}~\bibnamefont {Mason}},
  \bibinfo {author} {\bibfnamefont {C.}~\bibnamefont {Davies}}, \bibinfo
  {author} {\bibfnamefont {K.}~\bibnamefont {Hornbostel}}, \bibinfo {author}
  {\bibfnamefont {G.~P.}\ \bibnamefont {Lepage}}, \bibinfo {author}
  {\bibfnamefont {J.}~\bibnamefont {Shigemitsu}}, \bibinfo {author}
  {\bibfnamefont {H.}~\bibnamefont {Trottier}}, \ and\ \bibinfo {author}
  {\bibfnamefont {K.}~\bibnamefont {Wong}} (\bibinfo {collaboration} {HPQCD,
  UKQCD}),\ }\href {\doibase 10.1103/PhysRevD.75.054502} {\bibfield  {journal}
  {\bibinfo  {journal} {Phys. Rev. D}\ }\textbf {\bibinfo {volume} {75}},\
  \bibinfo {pages} {054502} (\bibinfo {year} {2007})},\ \Eprint
  {http://arxiv.org/abs/hep-lat/0610092} {arXiv:hep-lat/0610092 [hep-lat]}
  \BibitemShut {NoStop}%
\bibitem [{\citenamefont {Bazavov}\ \emph {et~al.}(2013)\citenamefont
  {Bazavov}, \citenamefont {Bernard}, \citenamefont {Komijani}, \citenamefont
  {De{T}ar}, \citenamefont {Levkova}, \citenamefont {Freeman}, \citenamefont
  {Gottlieb}, \citenamefont {Zhou}, \citenamefont {Heller}, \citenamefont
  {Hetrick}, \citenamefont {Laiho}, \citenamefont {Osborn}, \citenamefont
  {Sugar}, \citenamefont {Toussaint},\ and\ \citenamefont {Van~de
  Water}}]{Bazavov:2012xda}%
  \BibitemOpen
  \bibfield  {author} {\bibinfo {author} {\bibfnamefont {A.}~\bibnamefont
  {Bazavov}}, \bibinfo {author} {\bibfnamefont {C.}~\bibnamefont {Bernard}},
  \bibinfo {author} {\bibfnamefont {J.}~\bibnamefont {Komijani}}, \bibinfo
  {author} {\bibfnamefont {C.}~\bibnamefont {De{T}ar}}, \bibinfo {author}
  {\bibfnamefont {L.}~\bibnamefont {Levkova}}, \bibinfo {author} {\bibfnamefont
  {W.}~\bibnamefont {Freeman}}, \bibinfo {author} {\bibfnamefont
  {S.}~\bibnamefont {Gottlieb}}, \bibinfo {author} {\bibfnamefont
  {R.}~\bibnamefont {Zhou}}, \bibinfo {author} {\bibfnamefont {U.~M.}\
  \bibnamefont {Heller}}, \bibinfo {author} {\bibfnamefont {J.~E.}\
  \bibnamefont {Hetrick}}, \bibinfo {author} {\bibfnamefont {J.}~\bibnamefont
  {Laiho}}, \bibinfo {author} {\bibfnamefont {J.}~\bibnamefont {Osborn}},
  \bibinfo {author} {\bibfnamefont {R.~L.}\ \bibnamefont {Sugar}}, \bibinfo
  {author} {\bibfnamefont {D.}~\bibnamefont {Toussaint}}, \ and\ \bibinfo
  {author} {\bibfnamefont {R.~S.}\ \bibnamefont {Van~de Water}} (\bibinfo
  {collaboration} {MILC}),\ }\href {\doibase 10.1103/PhysRevD.87.054505}
  {\bibfield  {journal} {\bibinfo  {journal} {Phys. Rev. D}\ }\textbf {\bibinfo
  {volume} {87}},\ \bibinfo {pages} {054505} (\bibinfo {year} {2013})},\
  \Eprint {http://arxiv.org/abs/1212.4768} {arXiv:1212.4768 [hep-lat]}
  \BibitemShut {NoStop}%
\bibitem [{\citenamefont {Bhattacharya}\ \emph {et~al.}(2015)\citenamefont
  {Bhattacharya}, \citenamefont {Cirigliano}, \citenamefont {Cohen},
  \citenamefont {Gupta}, \citenamefont {Joseph}, \citenamefont {Lin},\ and\
  \citenamefont {Yoon}}]{Bhattacharya:2015wna}%
  \BibitemOpen
  \bibfield  {author} {\bibinfo {author} {\bibfnamefont {T.}~\bibnamefont
  {Bhattacharya}}, \bibinfo {author} {\bibfnamefont {V.}~\bibnamefont
  {Cirigliano}}, \bibinfo {author} {\bibfnamefont {S.~D.}\ \bibnamefont
  {Cohen}}, \bibinfo {author} {\bibfnamefont {R.}~\bibnamefont {Gupta}},
  \bibinfo {author} {\bibfnamefont {A.}~\bibnamefont {Joseph}}, \bibinfo
  {author} {\bibfnamefont {H.-W.}\ \bibnamefont {Lin}}, \ and\ \bibinfo
  {author} {\bibfnamefont {B.}~\bibnamefont {Yoon}} (\bibinfo {collaboration}
  {PNDME}),\ }\href {\doibase 10.1103/PhysRevD.92.094511} {\bibfield  {journal}
  {\bibinfo  {journal} {Phys. Rev. D}\ }\textbf {\bibinfo {volume} {92}},\
  \bibinfo {pages} {094511} (\bibinfo {year} {2015})},\ \Eprint
  {http://arxiv.org/abs/1506.06411} {arXiv:1506.06411 [hep-lat]} \BibitemShut
  {NoStop}%
\bibitem [{\citenamefont {Gupta}\ \emph {et~al.}(2018)\citenamefont {Gupta},
  \citenamefont {Jang}, \citenamefont {Yoon}, \citenamefont {Lin},
  \citenamefont {Cirigliano},\ and\ \citenamefont
  {Bhattacharya}}]{Gupta:2018qil}%
  \BibitemOpen
  \bibfield  {author} {\bibinfo {author} {\bibfnamefont {R.}~\bibnamefont
  {Gupta}}, \bibinfo {author} {\bibfnamefont {Y.-C.}\ \bibnamefont {Jang}},
  \bibinfo {author} {\bibfnamefont {B.}~\bibnamefont {Yoon}}, \bibinfo {author}
  {\bibfnamefont {H.-W.}\ \bibnamefont {Lin}}, \bibinfo {author} {\bibfnamefont
  {V.}~\bibnamefont {Cirigliano}}, \ and\ \bibinfo {author} {\bibfnamefont
  {T.}~\bibnamefont {Bhattacharya}},\ }\href {\doibase
  10.1103/PhysRevD.98.034503} {\bibfield  {journal} {\bibinfo  {journal} {Phys.
  Rev. D}\ }\textbf {\bibinfo {volume} {98}},\ \bibinfo {pages} {034503}
  (\bibinfo {year} {2018})},\ \Eprint {http://arxiv.org/abs/1806.09006}
  {arXiv:1806.09006 [hep-lat]} \BibitemShut {NoStop}%
\bibitem [{\citenamefont {Jang}\ \emph {et~al.}(2020)\citenamefont {Jang},
  \citenamefont {Gupta}, \citenamefont {Yoon},\ and\ \citenamefont
  {Bhattacharya}}]{Jang:2019vkm}%
  \BibitemOpen
  \bibfield  {author} {\bibinfo {author} {\bibfnamefont {Y.-C.}\ \bibnamefont
  {Jang}}, \bibinfo {author} {\bibfnamefont {R.}~\bibnamefont {Gupta}},
  \bibinfo {author} {\bibfnamefont {B.}~\bibnamefont {Yoon}}, \ and\ \bibinfo
  {author} {\bibfnamefont {T.}~\bibnamefont {Bhattacharya}},\ }\href {\doibase
  10.1103/PhysRevLett.124.072002} {\bibfield  {journal} {\bibinfo  {journal}
  {Phys. Rev. Lett.}\ }\textbf {\bibinfo {volume} {124}},\ \bibinfo {pages}
  {072002} (\bibinfo {year} {2020})},\ \Eprint
  {http://arxiv.org/abs/1905.06470} {arXiv:1905.06470 [hep-lat]} \BibitemShut
  {NoStop}%
\bibitem [{\citenamefont {Park}\ \emph {et~al.}(2021)\citenamefont {Park},
  \citenamefont {Gupta}, \citenamefont {Yoon}, \citenamefont {Mondal},
  \citenamefont {Bhattacharya}, \citenamefont {Jang}, \citenamefont {Jo\'o},\
  and\ \citenamefont {Winter}}]{Park:2021ypf}%
  \BibitemOpen
  \bibfield  {author} {\bibinfo {author} {\bibfnamefont {S.}~\bibnamefont
  {Park}}, \bibinfo {author} {\bibfnamefont {R.}~\bibnamefont {Gupta}},
  \bibinfo {author} {\bibfnamefont {B.}~\bibnamefont {Yoon}}, \bibinfo {author}
  {\bibfnamefont {S.}~\bibnamefont {Mondal}}, \bibinfo {author} {\bibfnamefont
  {T.}~\bibnamefont {Bhattacharya}}, \bibinfo {author} {\bibfnamefont {Y.-C.}\
  \bibnamefont {Jang}}, \bibinfo {author} {\bibfnamefont {B.}~\bibnamefont
  {Jo\'o}}, \ and\ \bibinfo {author} {\bibfnamefont {F.}~\bibnamefont {Winter}}
  (\bibinfo {collaboration} {Nucleon Matrix Elements}),\ }\href@noop {} {\
  (\bibinfo {year} {2021})},\ \Eprint {http://arxiv.org/abs/2103.05599}
  {arXiv:2103.05599 [hep-lat]} \BibitemShut {NoStop}%
\bibitem [{\citenamefont {B{\"a}r}(2015)}]{Bar:2015zwa}%
  \BibitemOpen
  \bibfield  {author} {\bibinfo {author} {\bibfnamefont {O.}~\bibnamefont
  {B{\"a}r}},\ }\href {\doibase 10.1103/PhysRevD.92.074504} {\bibfield
  {journal} {\bibinfo  {journal} {Phys. Rev. D}\ }\textbf {\bibinfo {volume}
  {92}},\ \bibinfo {pages} {074504} (\bibinfo {year} {2015})},\ \Eprint
  {http://arxiv.org/abs/1503.03649} {arXiv:1503.03649 [hep-lat]} \BibitemShut
  {NoStop}%
\bibitem [{\citenamefont {Tiburzi}(2015)}]{Tiburzi:2015tta}%
  \BibitemOpen
  \bibfield  {author} {\bibinfo {author} {\bibfnamefont {B.~C.}\ \bibnamefont
  {Tiburzi}},\ }\href {\doibase 10.1103/PhysRevD.91.094510} {\bibfield
  {journal} {\bibinfo  {journal} {Phys. Rev. D}\ }\textbf {\bibinfo {volume}
  {91}},\ \bibinfo {pages} {094510} (\bibinfo {year} {2015})},\ \Eprint
  {http://arxiv.org/abs/1503.06329} {arXiv:1503.06329 [hep-lat]} \BibitemShut
  {NoStop}%
\bibitem [{\citenamefont {B\"ar}(2016)}]{Bar:2016uoj}%
  \BibitemOpen
  \bibfield  {author} {\bibinfo {author} {\bibfnamefont {O.}~\bibnamefont
  {B\"ar}},\ }\href {\doibase 10.1103/PhysRevD.94.054505} {\bibfield  {journal}
  {\bibinfo  {journal} {Phys. Rev. D}\ }\textbf {\bibinfo {volume} {94}},\
  \bibinfo {pages} {054505} (\bibinfo {year} {2016})},\ \Eprint
  {http://arxiv.org/abs/1606.09385} {arXiv:1606.09385 [hep-lat]} \BibitemShut
  {NoStop}%
\bibitem [{\citenamefont {B{\"a}r}(2017)}]{Bar:2016jof}%
  \BibitemOpen
  \bibfield  {author} {\bibinfo {author} {\bibfnamefont {O.}~\bibnamefont
  {B{\"a}r}},\ }\href {\doibase 10.1103/PhysRevD.95.034506} {\bibfield
  {journal} {\bibinfo  {journal} {Phys. Rev. D}\ }\textbf {\bibinfo {volume}
  {95}},\ \bibinfo {pages} {034506} (\bibinfo {year} {2017})},\ \Eprint
  {http://arxiv.org/abs/1612.08336} {arXiv:1612.08336 [hep-lat]} \BibitemShut
  {NoStop}%
\bibitem [{\citenamefont {B{\"a}r}(2018)}]{Bar:2018wco}%
  \BibitemOpen
  \bibfield  {author} {\bibinfo {author} {\bibfnamefont {O.}~\bibnamefont
  {B{\"a}r}},\ }\href {\doibase 10.1103/PhysRevD.97.094507} {\bibfield
  {journal} {\bibinfo  {journal} {Phys. Rev. D}\ }\textbf {\bibinfo {volume}
  {97}},\ \bibinfo {pages} {094507} (\bibinfo {year} {2018})},\ \Eprint
  {http://arxiv.org/abs/1802.10442} {arXiv:1802.10442 [hep-lat]} \BibitemShut
  {NoStop}%
\bibitem [{\citenamefont {B{\"a}r}(2019{\natexlab{a}})}]{Bar:2018xyi}%
  \BibitemOpen
  \bibfield  {author} {\bibinfo {author} {\bibfnamefont {O.}~\bibnamefont
  {B{\"a}r}},\ }\href {\doibase 10.1103/PhysRevD.99.054506} {\bibfield
  {journal} {\bibinfo  {journal} {Phys. Rev. D}\ }\textbf {\bibinfo {volume}
  {99}},\ \bibinfo {pages} {054506} (\bibinfo {year} {2019}{\natexlab{a}})},\
  \Eprint {http://arxiv.org/abs/1812.09191} {arXiv:1812.09191 [hep-lat]}
  \BibitemShut {NoStop}%
\bibitem [{\citenamefont {B{\"a}r}(2019{\natexlab{b}})}]{Bar:2019gfx}%
  \BibitemOpen
  \bibfield  {author} {\bibinfo {author} {\bibfnamefont {O.}~\bibnamefont
  {B{\"a}r}},\ }\href {\doibase 10.1103/PhysRevD.100.054507} {\bibfield
  {journal} {\bibinfo  {journal} {Phys. Rev. D}\ }\textbf {\bibinfo {volume}
  {100}},\ \bibinfo {pages} {054507} (\bibinfo {year} {2019}{\natexlab{b}})},\
  \Eprint {http://arxiv.org/abs/1906.03652} {arXiv:1906.03652 [hep-lat]}
  \BibitemShut {NoStop}%
\bibitem [{\citenamefont {B{\" a}r}\ and\ \citenamefont {{\v
  C}oli{\'c}}(2021)}]{Bar:2021crj}%
  \BibitemOpen
  \bibfield  {author} {\bibinfo {author} {\bibfnamefont {O.}~\bibnamefont {B{\"
  a}r}}\ and\ \bibinfo {author} {\bibfnamefont {H.}~\bibnamefont {{\v
  C}oli{\'c}}},\ }\href {\doibase 10.1103/PhysRevD.103.114514} {\bibfield
  {journal} {\bibinfo  {journal} {Phys. Rev. D}\ }\textbf {\bibinfo {volume}
  {103}},\ \bibinfo {pages} {114514} (\bibinfo {year} {2021})},\ \Eprint
  {http://arxiv.org/abs/2104.00329} {arXiv:2104.00329 [hep-lat]} \BibitemShut
  {NoStop}%
\bibitem [{\citenamefont {Dmitra{\v s}inovi{\'c}}\ \emph
  {et~al.}(2010)\citenamefont {Dmitra{\v s}inovi{\'c}}, \citenamefont
  {Hosaka},\ and\ \citenamefont {Nagata}}]{Dmitrasinovic:2009vp}%
  \BibitemOpen
  \bibfield  {author} {\bibinfo {author} {\bibfnamefont {V.}~\bibnamefont
  {Dmitra{\v s}inovi{\'c}}}, \bibinfo {author} {\bibfnamefont {A.}~\bibnamefont
  {Hosaka}}, \ and\ \bibinfo {author} {\bibfnamefont {K.}~\bibnamefont
  {Nagata}},\ }\href {\doibase 10.1142/S0217732310032494} {\bibfield  {journal}
  {\bibinfo  {journal} {Mod. Phys. Lett. A}\ }\textbf {\bibinfo {volume}
  {25}},\ \bibinfo {pages} {233} (\bibinfo {year} {2010})},\ \Eprint
  {http://arxiv.org/abs/0912.2372} {arXiv:0912.2372 [hep-ph]} \BibitemShut
  {NoStop}%
\bibitem [{\citenamefont {Lin}\ \emph {et~al.}(2018)\citenamefont {Lin},
  \citenamefont {Nocera}, \citenamefont {Olness}, \citenamefont {Orginos},
  \citenamefont {Rojo} \emph {et~al.}}]{Lin:2017snn}%
  \BibitemOpen
  \bibfield  {author} {\bibinfo {author} {\bibfnamefont {H.-W.}\ \bibnamefont
  {Lin}}, \bibinfo {author} {\bibfnamefont {E.~R.}\ \bibnamefont {Nocera}},
  \bibinfo {author} {\bibfnamefont {F.}~\bibnamefont {Olness}}, \bibinfo
  {author} {\bibfnamefont {K.}~\bibnamefont {Orginos}}, \bibinfo {author}
  {\bibfnamefont {J.}~\bibnamefont {Rojo}},  \emph {et~al.},\ }\href {\doibase
  10.1016/j.ppnp.2018.01.007} {\bibfield  {journal} {\bibinfo  {journal} {Prog.
  Part. Nucl. Phys.}\ }\textbf {\bibinfo {volume} {100}},\ \bibinfo {pages}
  {107} (\bibinfo {year} {2018})},\ \Eprint {http://arxiv.org/abs/1711.07916}
  {arXiv:1711.07916 [hep-ph]} \BibitemShut {NoStop}%
\bibitem [{\citenamefont {Edwards}\ and\ \citenamefont
  {Jo{\'o}}(2005)}]{Edwards:2004sx}%
  \BibitemOpen
  \bibfield  {author} {\bibinfo {author} {\bibfnamefont {R.~G.}\ \bibnamefont
  {Edwards}}\ and\ \bibinfo {author} {\bibfnamefont {B.}~\bibnamefont
  {Jo{\'o}}} (\bibinfo {collaboration} {SciDAC, LHPC, UKQCD}),\ }\href
  {\doibase 10.1016/j.nuclphysbps.2004.11.254} {\bibfield  {journal} {\bibinfo
  {journal} {Nucl. Phys. Proc. Suppl.}\ }\textbf {\bibinfo {volume} {140}},\
  \bibinfo {pages} {832} (\bibinfo {year} {2005})},\ \Eprint
  {http://arxiv.org/abs/hep-lat/0409003} {arXiv:hep-lat/0409003 [hep-lat]}
  \BibitemShut {NoStop}%
\bibitem [{\citenamefont {Gusken}\ \emph {et~al.}(1989)\citenamefont {Gusken},
  \citenamefont {Low}, \citenamefont {Mutter}, \citenamefont {Sommer},
  \citenamefont {Patel},\ and\ \citenamefont {Schilling}}]{Gusken:1989ad}%
  \BibitemOpen
  \bibfield  {author} {\bibinfo {author} {\bibfnamefont {S.}~\bibnamefont
  {Gusken}}, \bibinfo {author} {\bibfnamefont {U.}~\bibnamefont {Low}},
  \bibinfo {author} {\bibfnamefont {K.~H.}\ \bibnamefont {Mutter}}, \bibinfo
  {author} {\bibfnamefont {R.}~\bibnamefont {Sommer}}, \bibinfo {author}
  {\bibfnamefont {A.}~\bibnamefont {Patel}}, \ and\ \bibinfo {author}
  {\bibfnamefont {K.}~\bibnamefont {Schilling}},\ }\href {\doibase
  10.1016/S0370-2693(89)80034-6} {\bibfield  {journal} {\bibinfo  {journal}
  {Phys. Lett.}\ }\textbf {\bibinfo {volume} {B227}},\ \bibinfo {pages} {266}
  (\bibinfo {year} {1989})}\BibitemShut {NoStop}%
\bibitem [{\citenamefont {Bali}\ \emph {et~al.}(2010)\citenamefont {Bali},
  \citenamefont {Collins},\ and\ \citenamefont {Sch{\"a}fer}}]{Bali:2009hu}%
  \BibitemOpen
  \bibfield  {author} {\bibinfo {author} {\bibfnamefont {G.~S.}\ \bibnamefont
  {Bali}}, \bibinfo {author} {\bibfnamefont {S.}~\bibnamefont {Collins}}, \
  and\ \bibinfo {author} {\bibfnamefont {A.}~\bibnamefont {Sch{\"a}fer}},\
  }\href {\doibase 10.1016/j.cpc.2010.05.008} {\bibfield  {journal} {\bibinfo
  {journal} {Comput. Phys. Commun.}\ }\textbf {\bibinfo {volume} {181}},\
  \bibinfo {pages} {1570} (\bibinfo {year} {2010})},\ \Eprint
  {http://arxiv.org/abs/0910.3970} {arXiv:0910.3970 [hep-lat]} \BibitemShut
  {NoStop}%
\bibitem [{\citenamefont {Blum}\ \emph {et~al.}(2013)\citenamefont {Blum},
  \citenamefont {Izubuchi},\ and\ \citenamefont {Shintani}}]{Blum:2012uh}%
  \BibitemOpen
  \bibfield  {author} {\bibinfo {author} {\bibfnamefont {T.}~\bibnamefont
  {Blum}}, \bibinfo {author} {\bibfnamefont {T.}~\bibnamefont {Izubuchi}}, \
  and\ \bibinfo {author} {\bibfnamefont {E.}~\bibnamefont {Shintani}},\ }\href
  {\doibase 10.1103/PhysRevD.88.094503} {\bibfield  {journal} {\bibinfo
  {journal} {Phys. Rev. D}\ }\textbf {\bibinfo {volume} {88}},\ \bibinfo
  {pages} {094503} (\bibinfo {year} {2013})},\ \Eprint
  {http://arxiv.org/abs/1208.4349} {arXiv:1208.4349 [hep-lat]} \BibitemShut
  {NoStop}%
\bibitem [{\citenamefont {Gasser}\ \emph
  {et~al.}(1988{\natexlab{b}})\citenamefont {Gasser}, \citenamefont {Sainio},\
  and\ \citenamefont {{\v S}varc}}]{Gasser:1987rb}%
  \BibitemOpen
  \bibfield  {author} {\bibinfo {author} {\bibfnamefont {J.}~\bibnamefont
  {Gasser}}, \bibinfo {author} {\bibfnamefont {M.~E.}\ \bibnamefont {Sainio}},
  \ and\ \bibinfo {author} {\bibfnamefont {A.}~\bibnamefont {{\v S}varc}},\
  }\href {\doibase 10.1016/0550-3213(88)90108-3} {\bibfield  {journal}
  {\bibinfo  {journal} {Nucl. Phys. B}\ }\textbf {\bibinfo {volume} {307}},\
  \bibinfo {pages} {779} (\bibinfo {year} {1988}{\natexlab{b}})}\BibitemShut
  {NoStop}%
\bibitem [{\citenamefont {Bernard}\ \emph {et~al.}(1995)\citenamefont
  {Bernard}, \citenamefont {Kaiser},\ and\ \citenamefont
  {Mei{\ss}ner}}]{Bernard:1995dp}%
  \BibitemOpen
  \bibfield  {author} {\bibinfo {author} {\bibfnamefont {V.}~\bibnamefont
  {Bernard}}, \bibinfo {author} {\bibfnamefont {N.}~\bibnamefont {Kaiser}}, \
  and\ \bibinfo {author} {\bibfnamefont {{\relax Ulf-G}.}~\bibnamefont
  {Mei{\ss}ner}},\ }\href {\doibase 10.1142/S0218301395000092} {\bibfield
  {journal} {\bibinfo  {journal} {Int. J. Mod. Phys.}\ }\textbf {\bibinfo
  {volume} {E04}},\ \bibinfo {pages} {193} (\bibinfo {year} {1995})},\ \Eprint
  {http://arxiv.org/abs/hep-ph/9501384} {arXiv:hep-ph/9501384 [hep-ph]}
  \BibitemShut {NoStop}%
\bibitem [{\citenamefont {Borasoy}\ and\ \citenamefont
  {Mei{\ss}ner}(1997)}]{Borasoy:1996bx}%
  \BibitemOpen
  \bibfield  {author} {\bibinfo {author} {\bibfnamefont {B.}~\bibnamefont
  {Borasoy}}\ and\ \bibinfo {author} {\bibfnamefont {{\relax
  Ulf-G}.}~\bibnamefont {Mei{\ss}ner}},\ }\href {\doibase
  10.1006/aphy.1996.5630} {\bibfield  {journal} {\bibinfo  {journal} {Annals
  Phys.}\ }\textbf {\bibinfo {volume} {254}},\ \bibinfo {pages} {192} (\bibinfo
  {year} {1997})},\ \Eprint {http://arxiv.org/abs/hep-ph/9607432}
  {arXiv:hep-ph/9607432} \BibitemShut {NoStop}%
\bibitem [{\citenamefont {Mei{\ss}ner}\ and\ \citenamefont
  {Steininger}(1998)}]{Meissner:1997ii}%
  \BibitemOpen
  \bibfield  {author} {\bibinfo {author} {\bibfnamefont {{\relax
  Ulf-G}.}~\bibnamefont {Mei{\ss}ner}}\ and\ \bibinfo {author} {\bibfnamefont
  {S.}~\bibnamefont {Steininger}},\ }\href {\doibase
  10.1016/S0370-2693(97)01418-4} {\bibfield  {journal} {\bibinfo  {journal}
  {Phys. Lett. B}\ }\textbf {\bibinfo {volume} {419}},\ \bibinfo {pages} {403}
  (\bibinfo {year} {1998})},\ \Eprint {http://arxiv.org/abs/hep-ph/9709453}
  {arXiv:hep-ph/9709453} \BibitemShut {NoStop}%
\bibitem [{\citenamefont {Steininger}\ \emph {et~al.}(1998)\citenamefont
  {Steininger}, \citenamefont {Mei{\ss}ner},\ and\ \citenamefont
  {Fettes}}]{Steininger:1998ya}%
  \BibitemOpen
  \bibfield  {author} {\bibinfo {author} {\bibfnamefont {S.}~\bibnamefont
  {Steininger}}, \bibinfo {author} {\bibfnamefont {{\relax
  Ulf-G}.}~\bibnamefont {Mei{\ss}ner}}, \ and\ \bibinfo {author} {\bibfnamefont
  {N.}~\bibnamefont {Fettes}},\ }\href {\doibase 10.1088/1126-6708/1998/09/008}
  {\bibfield  {journal} {\bibinfo  {journal} {JHEP}\ }\textbf {\bibinfo
  {volume} {09}},\ \bibinfo {pages} {008} (\bibinfo {year} {1998})},\ \Eprint
  {http://arxiv.org/abs/hep-ph/9808280} {arXiv:hep-ph/9808280} \BibitemShut
  {NoStop}%
\bibitem [{\citenamefont {Kambor}\ and\ \citenamefont {Moj{\v z}i{\v
  s}}(1999)}]{Kambor:1998pi}%
  \BibitemOpen
  \bibfield  {author} {\bibinfo {author} {\bibfnamefont {J.}~\bibnamefont
  {Kambor}}\ and\ \bibinfo {author} {\bibfnamefont {M.}~\bibnamefont {Moj{\v
  z}i{\v s}}},\ }\href {\doibase 10.1088/1126-6708/1999/04/031} {\bibfield
  {journal} {\bibinfo  {journal} {JHEP}\ }\textbf {\bibinfo {volume} {04}},\
  \bibinfo {pages} {031} (\bibinfo {year} {1999})},\ \Eprint
  {http://arxiv.org/abs/hep-ph/9901235} {arXiv:hep-ph/9901235} \BibitemShut
  {NoStop}%
\bibitem [{\citenamefont {McGovern}\ and\ \citenamefont
  {Birse}(1999)}]{McGovern:1998tm}%
  \BibitemOpen
  \bibfield  {author} {\bibinfo {author} {\bibfnamefont {J.~A.}\ \bibnamefont
  {McGovern}}\ and\ \bibinfo {author} {\bibfnamefont {M.~C.}\ \bibnamefont
  {Birse}},\ }\href {\doibase 10.1016/S0370-2693(98)01550-0} {\bibfield
  {journal} {\bibinfo  {journal} {Phys. Lett. B}\ }\textbf {\bibinfo {volume}
  {446}},\ \bibinfo {pages} {300} (\bibinfo {year} {1999})},\ \Eprint
  {http://arxiv.org/abs/hep-ph/9807384} {arXiv:hep-ph/9807384} \BibitemShut
  {NoStop}%
\bibitem [{\citenamefont {Becher}\ and\ \citenamefont
  {Leutwyler}(1999)}]{Becher:1999he}%
  \BibitemOpen
  \bibfield  {author} {\bibinfo {author} {\bibfnamefont {T.}~\bibnamefont
  {Becher}}\ and\ \bibinfo {author} {\bibfnamefont {H.}~\bibnamefont
  {Leutwyler}},\ }\href {\doibase 10.1007/PL00021673} {\bibfield  {journal}
  {\bibinfo  {journal} {Eur. Phys. J. C}\ }\textbf {\bibinfo {volume} {9}},\
  \bibinfo {pages} {643} (\bibinfo {year} {1999})},\ \Eprint
  {http://arxiv.org/abs/hep-ph/9901384} {arXiv:hep-ph/9901384} \BibitemShut
  {NoStop}%
\bibitem [{\citenamefont {McGovern}\ and\ \citenamefont
  {Birse}(2006)}]{McGovern:2006fm}%
  \BibitemOpen
  \bibfield  {author} {\bibinfo {author} {\bibfnamefont {J.~A.}\ \bibnamefont
  {McGovern}}\ and\ \bibinfo {author} {\bibfnamefont {M.~C.}\ \bibnamefont
  {Birse}},\ }\href {\doibase 10.1103/PhysRevD.74.097501} {\bibfield  {journal}
  {\bibinfo  {journal} {Phys. Rev. D}\ }\textbf {\bibinfo {volume} {74}},\
  \bibinfo {pages} {097501} (\bibinfo {year} {2006})},\ \Eprint
  {http://arxiv.org/abs/hep-lat/0608002} {arXiv:hep-lat/0608002} \BibitemShut
  {NoStop}%
\bibitem [{\citenamefont {Schindler}\ \emph {et~al.}(2007)\citenamefont
  {Schindler}, \citenamefont {Djukanovic}, \citenamefont {Gegelia},\ and\
  \citenamefont {Scherer}}]{Schindler:2006ha}%
  \BibitemOpen
  \bibfield  {author} {\bibinfo {author} {\bibfnamefont {M.~R.}\ \bibnamefont
  {Schindler}}, \bibinfo {author} {\bibfnamefont {D.}~\bibnamefont
  {Djukanovic}}, \bibinfo {author} {\bibfnamefont {J.}~\bibnamefont {Gegelia}},
  \ and\ \bibinfo {author} {\bibfnamefont {S.}~\bibnamefont {Scherer}},\ }\href
  {\doibase 10.1016/j.physletb.2007.04.034} {\bibfield  {journal} {\bibinfo
  {journal} {Phys. Lett. B}\ }\textbf {\bibinfo {volume} {649}},\ \bibinfo
  {pages} {390} (\bibinfo {year} {2007})},\ \Eprint
  {http://arxiv.org/abs/hep-ph/0612164} {arXiv:hep-ph/0612164} \BibitemShut
  {NoStop}%
\bibitem [{\citenamefont {Schindler}\ \emph {et~al.}(2008)\citenamefont
  {Schindler}, \citenamefont {Djukanovic}, \citenamefont {Gegelia},\ and\
  \citenamefont {Scherer}}]{Schindler:2007dr}%
  \BibitemOpen
  \bibfield  {author} {\bibinfo {author} {\bibfnamefont {M.~R.}\ \bibnamefont
  {Schindler}}, \bibinfo {author} {\bibfnamefont {D.}~\bibnamefont
  {Djukanovic}}, \bibinfo {author} {\bibfnamefont {J.}~\bibnamefont {Gegelia}},
  \ and\ \bibinfo {author} {\bibfnamefont {S.}~\bibnamefont {Scherer}},\ }\href
  {\doibase 10.1016/j.nuclphysa.2008.01.023} {\bibfield  {journal} {\bibinfo
  {journal} {Nucl. Phys. A}\ }\textbf {\bibinfo {volume} {803}},\ \bibinfo
  {pages} {68} (\bibinfo {year} {2008})},\ \bibinfo {note} {[Erratum: Nucl.
  Phys. A {\bf 1010}, 122175 (2021)]},\ \Eprint
  {http://arxiv.org/abs/0707.4296} {arXiv:0707.4296 [hep-ph]} \BibitemShut
  {NoStop}%
\bibitem [{\citenamefont {Colangelo}\ and\ \citenamefont
  {D{\"u}rr}(2004)}]{Colangelo:2003hf}%
  \BibitemOpen
  \bibfield  {author} {\bibinfo {author} {\bibfnamefont {G.}~\bibnamefont
  {Colangelo}}\ and\ \bibinfo {author} {\bibfnamefont {S.}~\bibnamefont
  {D{\"u}rr}},\ }\href {\doibase 10.1140/epjc/s2004-01593-y} {\bibfield
  {journal} {\bibinfo  {journal} {Eur. Phys. J. C}\ }\textbf {\bibinfo {volume}
  {33}},\ \bibinfo {pages} {543} (\bibinfo {year} {2004})},\ \Eprint
  {http://arxiv.org/abs/hep-lat/0311023} {arXiv:hep-lat/0311023} \BibitemShut
  {NoStop}%
\bibitem [{\citenamefont {Zyla}\ \emph {et~al.}(2020)\citenamefont {Zyla},
  \citenamefont {Barnett}, \citenamefont {Beringer}, \citenamefont {Dahl},
  \citenamefont {Dwyer}, \citenamefont {Groom} \emph {et~al.}}]{Zyla:2020zbs}%
  \BibitemOpen
  \bibfield  {author} {\bibinfo {author} {\bibfnamefont {P.~A.}\ \bibnamefont
  {Zyla}}, \bibinfo {author} {\bibfnamefont {R.~M.}\ \bibnamefont {Barnett}},
  \bibinfo {author} {\bibfnamefont {J.}~\bibnamefont {Beringer}}, \bibinfo
  {author} {\bibfnamefont {O.}~\bibnamefont {Dahl}}, \bibinfo {author}
  {\bibfnamefont {D.~A.}\ \bibnamefont {Dwyer}}, \bibinfo {author}
  {\bibfnamefont {D.~E.}\ \bibnamefont {Groom}},  \emph {et~al.} (\bibinfo
  {collaboration} {Particle Data Group}),\ }\href {\doibase
  10.1093/ptep/ptaa104} {\bibfield  {journal} {\bibinfo  {journal} {PTEP}\
  }\textbf {\bibinfo {volume} {2020}},\ \bibinfo {pages} {083C01} (\bibinfo
  {year} {2020})}\BibitemShut {NoStop}%
\bibitem [{\citenamefont {Beane}(2004)}]{Beane:2004tw}%
  \BibitemOpen
  \bibfield  {author} {\bibinfo {author} {\bibfnamefont {S.~R.}\ \bibnamefont
  {Beane}},\ }\href {\doibase 10.1103/PhysRevD.70.034507} {\bibfield  {journal}
  {\bibinfo  {journal} {Phys. Rev. D}\ }\textbf {\bibinfo {volume} {70}},\
  \bibinfo {pages} {034507} (\bibinfo {year} {2004})},\ \Eprint
  {http://arxiv.org/abs/hep-lat/0403015} {arXiv:hep-lat/0403015} \BibitemShut
  {NoStop}%
\bibitem [{\citenamefont {Bali}\ \emph {et~al.}(2020)\citenamefont {Bali},
  \citenamefont {Barca}, \citenamefont {Collins}, \citenamefont {Gruber},
  \citenamefont {L\"offler}, \citenamefont {Sch\"afer} \emph
  {et~al.}}]{Bali:2019yiy}%
  \BibitemOpen
  \bibfield  {author} {\bibinfo {author} {\bibfnamefont {G.~S.}\ \bibnamefont
  {Bali}}, \bibinfo {author} {\bibfnamefont {L.}~\bibnamefont {Barca}},
  \bibinfo {author} {\bibfnamefont {S.}~\bibnamefont {Collins}}, \bibinfo
  {author} {\bibfnamefont {M.}~\bibnamefont {Gruber}}, \bibinfo {author}
  {\bibfnamefont {M.}~\bibnamefont {L\"offler}}, \bibinfo {author}
  {\bibfnamefont {A.}~\bibnamefont {Sch\"afer}},  \emph {et~al.} (\bibinfo
  {collaboration} {RQCD}),\ }\href {\doibase 10.1007/JHEP05(2020)126}
  {\bibfield  {journal} {\bibinfo  {journal} {JHEP}\ }\textbf {\bibinfo
  {volume} {05}},\ \bibinfo {pages} {126} (\bibinfo {year} {2020})},\ \Eprint
  {http://arxiv.org/abs/1911.13150} {arXiv:1911.13150 [hep-lat]} \BibitemShut
  {NoStop}%
\bibitem [{\citenamefont {Bhattacharya}\ \emph {et~al.}(2006)\citenamefont
  {Bhattacharya}, \citenamefont {Gupta}, \citenamefont {Lee}, \citenamefont
  {Sharpe},\ and\ \citenamefont {Wu}}]{Bhattacharya:2005rb}%
  \BibitemOpen
  \bibfield  {author} {\bibinfo {author} {\bibfnamefont {T.}~\bibnamefont
  {Bhattacharya}}, \bibinfo {author} {\bibfnamefont {R.}~\bibnamefont {Gupta}},
  \bibinfo {author} {\bibfnamefont {W.}~\bibnamefont {Lee}}, \bibinfo {author}
  {\bibfnamefont {S.~R.}\ \bibnamefont {Sharpe}}, \ and\ \bibinfo {author}
  {\bibfnamefont {J.~M.~S.}\ \bibnamefont {Wu}},\ }\href {\doibase
  10.1103/PhysRevD.73.034504} {\bibfield  {journal} {\bibinfo  {journal} {Phys.
  Rev.}\ }\textbf {\bibinfo {volume} {D73}},\ \bibinfo {pages} {034504}
  (\bibinfo {year} {2006})},\ \Eprint {http://arxiv.org/abs/hep-lat/0511014}
  {arXiv:hep-lat/0511014 [hep-lat]} \BibitemShut {NoStop}%
\bibitem [{\citenamefont {Ohki}\ \emph {et~al.}(2008)\citenamefont {Ohki},
  \citenamefont {Fukaya}, \citenamefont {Hashimoto}, \citenamefont {Kaneko},
  \citenamefont {Matsufuru}, \citenamefont {Noaki}, \citenamefont {Onogi},
  \citenamefont {Shintani},\ and\ \citenamefont {Yamada}}]{Ohki:2008ff}%
  \BibitemOpen
  \bibfield  {author} {\bibinfo {author} {\bibfnamefont {H.}~\bibnamefont
  {Ohki}}, \bibinfo {author} {\bibfnamefont {H.}~\bibnamefont {Fukaya}},
  \bibinfo {author} {\bibfnamefont {S.}~\bibnamefont {Hashimoto}}, \bibinfo
  {author} {\bibfnamefont {T.}~\bibnamefont {Kaneko}}, \bibinfo {author}
  {\bibfnamefont {H.}~\bibnamefont {Matsufuru}}, \bibinfo {author}
  {\bibfnamefont {J.}~\bibnamefont {Noaki}}, \bibinfo {author} {\bibfnamefont
  {T.}~\bibnamefont {Onogi}}, \bibinfo {author} {\bibfnamefont
  {E.}~\bibnamefont {Shintani}}, \ and\ \bibinfo {author} {\bibfnamefont
  {N.}~\bibnamefont {Yamada}},\ }\href {\doibase 10.1103/PhysRevD.78.054502}
  {\bibfield  {journal} {\bibinfo  {journal} {Phys. Rev. D}\ }\textbf {\bibinfo
  {volume} {78}},\ \bibinfo {pages} {054502} (\bibinfo {year} {2008})},\
  \Eprint {http://arxiv.org/abs/0806.4744} {arXiv:0806.4744 [hep-lat]}
  \BibitemShut {NoStop}%
\bibitem [{\citenamefont {Ishikawa}\ \emph {et~al.}(2009)\citenamefont
  {Ishikawa}, \citenamefont {Ishizuka}, \citenamefont {Izubuchi}, \citenamefont
  {Kadoh}, \citenamefont {Kanaya}, \citenamefont {Kuramashi} \emph
  {et~al.}}]{Ishikawa:2009vc}%
  \BibitemOpen
  \bibfield  {author} {\bibinfo {author} {\bibfnamefont {{\relax
  K.-I}.}~\bibnamefont {Ishikawa}}, \bibinfo {author} {\bibfnamefont
  {N.}~\bibnamefont {Ishizuka}}, \bibinfo {author} {\bibfnamefont
  {T.}~\bibnamefont {Izubuchi}}, \bibinfo {author} {\bibfnamefont
  {D.}~\bibnamefont {Kadoh}}, \bibinfo {author} {\bibfnamefont
  {K.}~\bibnamefont {Kanaya}}, \bibinfo {author} {\bibfnamefont
  {Y.}~\bibnamefont {Kuramashi}},  \emph {et~al.} (\bibinfo {collaboration}
  {PACS-CS}),\ }\href {\doibase 10.1103/PhysRevD.80.054502} {\bibfield
  {journal} {\bibinfo  {journal} {Phys. Rev. D}\ }\textbf {\bibinfo {volume}
  {80}},\ \bibinfo {pages} {054502} (\bibinfo {year} {2009})},\ \Eprint
  {http://arxiv.org/abs/0905.0962} {arXiv:0905.0962 [hep-lat]} \BibitemShut
  {NoStop}%
\bibitem [{\citenamefont {Horsley}\ \emph {et~al.}(2012)\citenamefont
  {Horsley}, \citenamefont {Nakamura}, \citenamefont {Perlt}, \citenamefont
  {Pleiter}, \citenamefont {Rakow}, \citenamefont {Schierholz}, \citenamefont
  {Schiller}, \citenamefont {Stuben}, \citenamefont {Winter},\ and\
  \citenamefont {Zanotti}}]{Horsley:2011wr}%
  \BibitemOpen
  \bibfield  {author} {\bibinfo {author} {\bibfnamefont {R.}~\bibnamefont
  {Horsley}}, \bibinfo {author} {\bibfnamefont {Y.}~\bibnamefont {Nakamura}},
  \bibinfo {author} {\bibfnamefont {H.}~\bibnamefont {Perlt}}, \bibinfo
  {author} {\bibfnamefont {D.}~\bibnamefont {Pleiter}}, \bibinfo {author}
  {\bibfnamefont {P.~E.~L.}\ \bibnamefont {Rakow}}, \bibinfo {author}
  {\bibfnamefont {G.}~\bibnamefont {Schierholz}}, \bibinfo {author}
  {\bibfnamefont {A.}~\bibnamefont {Schiller}}, \bibinfo {author}
  {\bibfnamefont {H.}~\bibnamefont {Stuben}}, \bibinfo {author} {\bibfnamefont
  {F.}~\bibnamefont {Winter}}, \ and\ \bibinfo {author} {\bibfnamefont {J.~M.}\
  \bibnamefont {Zanotti}} (\bibinfo {collaboration} {QCDSF-UKQCD}),\ }\href
  {\doibase 10.1103/PhysRevD.85.034506} {\bibfield  {journal} {\bibinfo
  {journal} {Phys. Rev. D}\ }\textbf {\bibinfo {volume} {85}},\ \bibinfo
  {pages} {034506} (\bibinfo {year} {2012})},\ \Eprint
  {http://arxiv.org/abs/1110.4971} {arXiv:1110.4971 [hep-lat]} \BibitemShut
  {NoStop}%
\bibitem [{\citenamefont {Martin~Camalich}\ \emph {et~al.}(2010)\citenamefont
  {Martin~Camalich}, \citenamefont {Geng},\ and\ \citenamefont
  {Vicente-Vacas}}]{MartinCamalich:2010fp}%
  \BibitemOpen
  \bibfield  {author} {\bibinfo {author} {\bibfnamefont {J.}~\bibnamefont
  {Martin~Camalich}}, \bibinfo {author} {\bibfnamefont {L.~S.}\ \bibnamefont
  {Geng}}, \ and\ \bibinfo {author} {\bibfnamefont {M.~J.}\ \bibnamefont
  {Vicente-Vacas}},\ }\href {\doibase 10.1103/PhysRevD.82.074504} {\bibfield
  {journal} {\bibinfo  {journal} {Phys. Rev. D}\ }\textbf {\bibinfo {volume}
  {82}},\ \bibinfo {pages} {074504} (\bibinfo {year} {2010})},\ \Eprint
  {http://arxiv.org/abs/1003.1929} {arXiv:1003.1929 [hep-lat]} \BibitemShut
  {NoStop}%
\bibitem [{\citenamefont {Shanahan}\ \emph {et~al.}(2013)\citenamefont
  {Shanahan}, \citenamefont {Thomas},\ and\ \citenamefont
  {Young}}]{Shanahan:2012wh}%
  \BibitemOpen
  \bibfield  {author} {\bibinfo {author} {\bibfnamefont {P.~E.}\ \bibnamefont
  {Shanahan}}, \bibinfo {author} {\bibfnamefont {A.~W.}\ \bibnamefont
  {Thomas}}, \ and\ \bibinfo {author} {\bibfnamefont {R.~D.}\ \bibnamefont
  {Young}},\ }\href {\doibase 10.1103/PhysRevD.87.074503} {\bibfield  {journal}
  {\bibinfo  {journal} {Phys. Rev. D}\ }\textbf {\bibinfo {volume} {87}},\
  \bibinfo {pages} {074503} (\bibinfo {year} {2013})},\ \Eprint
  {http://arxiv.org/abs/1205.5365} {arXiv:1205.5365 [nucl-th]} \BibitemShut
  {NoStop}%
\bibitem [{\citenamefont {Ohki}\ \emph {et~al.}(2013)\citenamefont {Ohki},
  \citenamefont {Takeda}, \citenamefont {Aoki}, \citenamefont {Hashimoto},
  \citenamefont {Kaneko}, \citenamefont {Matsufuru}, \citenamefont {Noaki},\
  and\ \citenamefont {Onogi}}]{Oksuzian:2012rzb}%
  \BibitemOpen
  \bibfield  {author} {\bibinfo {author} {\bibfnamefont {H.}~\bibnamefont
  {Ohki}}, \bibinfo {author} {\bibfnamefont {K.}~\bibnamefont {Takeda}},
  \bibinfo {author} {\bibfnamefont {S.}~\bibnamefont {Aoki}}, \bibinfo {author}
  {\bibfnamefont {S.}~\bibnamefont {Hashimoto}}, \bibinfo {author}
  {\bibfnamefont {T.}~\bibnamefont {Kaneko}}, \bibinfo {author} {\bibfnamefont
  {H.}~\bibnamefont {Matsufuru}}, \bibinfo {author} {\bibfnamefont
  {J.}~\bibnamefont {Noaki}}, \ and\ \bibinfo {author} {\bibfnamefont
  {T.}~\bibnamefont {Onogi}} (\bibinfo {collaboration} {JLQCD}),\ }\href
  {\doibase 10.1103/PhysRevD.87.034509} {\bibfield  {journal} {\bibinfo
  {journal} {Phys. Rev. D}\ }\textbf {\bibinfo {volume} {87}},\ \bibinfo
  {pages} {034509} (\bibinfo {year} {2013})},\ \Eprint
  {http://arxiv.org/abs/1208.4185} {arXiv:1208.4185 [hep-lat]} \BibitemShut
  {NoStop}%
\bibitem [{\citenamefont {Junnarkar}\ and\ \citenamefont
  {Walker-Loud}(2013)}]{Junnarkar:2013ac}%
  \BibitemOpen
  \bibfield  {author} {\bibinfo {author} {\bibfnamefont {P.~M.}\ \bibnamefont
  {Junnarkar}}\ and\ \bibinfo {author} {\bibfnamefont {A.}~\bibnamefont
  {Walker-Loud}},\ }\href {\doibase 10.1103/PhysRevD.87.114510} {\bibfield
  {journal} {\bibinfo  {journal} {Phys. Rev. D}\ }\textbf {\bibinfo {volume}
  {87}},\ \bibinfo {pages} {114510} (\bibinfo {year} {2013})},\ \Eprint
  {http://arxiv.org/abs/1301.1114} {arXiv:1301.1114 [hep-lat]} \BibitemShut
  {NoStop}%
\bibitem [{\citenamefont {Procura}\ \emph {et~al.}(2006)\citenamefont
  {Procura}, \citenamefont {Musch}, \citenamefont {Wollenweber}, \citenamefont
  {Hemmert},\ and\ \citenamefont {Weise}}]{Procura:2006bj}%
  \BibitemOpen
  \bibfield  {author} {\bibinfo {author} {\bibfnamefont {M.}~\bibnamefont
  {Procura}}, \bibinfo {author} {\bibfnamefont {B.~U.}\ \bibnamefont {Musch}},
  \bibinfo {author} {\bibfnamefont {T.}~\bibnamefont {Wollenweber}}, \bibinfo
  {author} {\bibfnamefont {T.~R.}\ \bibnamefont {Hemmert}}, \ and\ \bibinfo
  {author} {\bibfnamefont {W.}~\bibnamefont {Weise}},\ }\href {\doibase
  10.1103/PhysRevD.73.114510} {\bibfield  {journal} {\bibinfo  {journal} {Phys.
  Rev. D}\ }\textbf {\bibinfo {volume} {73}},\ \bibinfo {pages} {114510}
  (\bibinfo {year} {2006})},\ \Eprint {http://arxiv.org/abs/hep-lat/0603001}
  {arXiv:hep-lat/0603001} \BibitemShut {NoStop}%
\bibitem [{\citenamefont {Walker-Loud}\ \emph {et~al.}(2009)\citenamefont
  {Walker-Loud}, \citenamefont {Lin}, \citenamefont {Richards}, \citenamefont
  {Edwards}, \citenamefont {Engelhardt}, \citenamefont {Fleming} \emph
  {et~al.}}]{WalkerLoud:2008bp}%
  \BibitemOpen
  \bibfield  {author} {\bibinfo {author} {\bibfnamefont {A.}~\bibnamefont
  {Walker-Loud}}, \bibinfo {author} {\bibfnamefont {{\relax
  H.-W.}.}~\bibnamefont {Lin}}, \bibinfo {author} {\bibfnamefont {D.~G.}\
  \bibnamefont {Richards}}, \bibinfo {author} {\bibfnamefont {R.~G.}\
  \bibnamefont {Edwards}}, \bibinfo {author} {\bibfnamefont {M.}~\bibnamefont
  {Engelhardt}}, \bibinfo {author} {\bibfnamefont {G.~T.}\ \bibnamefont
  {Fleming}},  \emph {et~al.},\ }\href {\doibase 10.1103/PhysRevD.79.054502}
  {\bibfield  {journal} {\bibinfo  {journal} {Phys. Rev. D}\ }\textbf {\bibinfo
  {volume} {79}},\ \bibinfo {pages} {054502} (\bibinfo {year} {2009})},\
  \Eprint {http://arxiv.org/abs/0806.4549} {arXiv:0806.4549 [hep-lat]}
  \BibitemShut {NoStop}%
\bibitem [{\citenamefont {Walker-Loud}(2008)}]{WalkerLoud:2008pj}%
  \BibitemOpen
  \bibfield  {author} {\bibinfo {author} {\bibfnamefont {A.}~\bibnamefont
  {Walker-Loud}},\ }\href {\doibase 10.22323/1.066.0005} {\bibfield  {journal}
  {\bibinfo  {journal} {PoS}\ }\textbf {\bibinfo {volume} {LATTICE2008}},\
  \bibinfo {pages} {005} (\bibinfo {year} {2008})},\ \Eprint
  {http://arxiv.org/abs/0810.0663} {arXiv:0810.0663 [hep-lat]} \BibitemShut
  {NoStop}%
\bibitem [{\citenamefont {Young}\ and\ \citenamefont
  {Thomas}(2010)}]{Young:2009zb}%
  \BibitemOpen
  \bibfield  {author} {\bibinfo {author} {\bibfnamefont {R.~D.}\ \bibnamefont
  {Young}}\ and\ \bibinfo {author} {\bibfnamefont {A.~W.}\ \bibnamefont
  {Thomas}},\ }\href {\doibase 10.1103/PhysRevD.81.014503} {\bibfield
  {journal} {\bibinfo  {journal} {Phys. Rev. D}\ }\textbf {\bibinfo {volume}
  {81}},\ \bibinfo {pages} {014503} (\bibinfo {year} {2010})},\ \Eprint
  {http://arxiv.org/abs/0901.3310} {arXiv:0901.3310 [hep-lat]} \BibitemShut
  {NoStop}%
\bibitem [{\citenamefont {Ren}\ \emph {et~al.}(2012)\citenamefont {Ren},
  \citenamefont {Geng}, \citenamefont {Martin~Camalich}, \citenamefont {Meng},\
  and\ \citenamefont {Toki}}]{Ren:2012aj}%
  \BibitemOpen
  \bibfield  {author} {\bibinfo {author} {\bibfnamefont {{\relax
  X.-L}.}~\bibnamefont {Ren}}, \bibinfo {author} {\bibfnamefont {L.~S.}\
  \bibnamefont {Geng}}, \bibinfo {author} {\bibfnamefont {J.}~\bibnamefont
  {Martin~Camalich}}, \bibinfo {author} {\bibfnamefont {J.}~\bibnamefont
  {Meng}}, \ and\ \bibinfo {author} {\bibfnamefont {H.}~\bibnamefont {Toki}},\
  }\href {\doibase 10.1007/JHEP12(2012)073} {\bibfield  {journal} {\bibinfo
  {journal} {JHEP}\ }\textbf {\bibinfo {volume} {12}},\ \bibinfo {pages} {073}
  (\bibinfo {year} {2012})},\ \Eprint {http://arxiv.org/abs/1209.3641}
  {arXiv:1209.3641 [nucl-th]} \BibitemShut {NoStop}%
\bibitem [{\citenamefont {Walker-Loud}(2013)}]{Walker-Loud:2013yua}%
  \BibitemOpen
  \bibfield  {author} {\bibinfo {author} {\bibfnamefont {A.}~\bibnamefont
  {Walker-Loud}},\ }\href {\doibase 10.22323/1.172.0017} {\bibfield  {journal}
  {\bibinfo  {journal} {PoS}\ }\textbf {\bibinfo {volume} {CD12}},\ \bibinfo
  {pages} {017} (\bibinfo {year} {2013})},\ \Eprint
  {http://arxiv.org/abs/1304.6341} {arXiv:1304.6341 [hep-lat]} \BibitemShut
  {NoStop}%
\bibitem [{\citenamefont {Alvarez-Ruso}\ \emph {et~al.}(2013)\citenamefont
  {Alvarez-Ruso}, \citenamefont {Ledwig}, \citenamefont {Martin~Camalich},\
  and\ \citenamefont {Vicente-Vacas}}]{Alvarez-Ruso:2013fza}%
  \BibitemOpen
  \bibfield  {author} {\bibinfo {author} {\bibfnamefont {L.}~\bibnamefont
  {Alvarez-Ruso}}, \bibinfo {author} {\bibfnamefont {T.}~\bibnamefont
  {Ledwig}}, \bibinfo {author} {\bibfnamefont {J.}~\bibnamefont
  {Martin~Camalich}}, \ and\ \bibinfo {author} {\bibfnamefont {M.~J.}\
  \bibnamefont {Vicente-Vacas}},\ }\href {\doibase 10.1103/PhysRevD.88.054507}
  {\bibfield  {journal} {\bibinfo  {journal} {Phys. Rev. D}\ }\textbf {\bibinfo
  {volume} {88}},\ \bibinfo {pages} {054507} (\bibinfo {year} {2013})},\
  \Eprint {http://arxiv.org/abs/1304.0483} {arXiv:1304.0483 [hep-ph]}
  \BibitemShut {NoStop}%
\bibitem [{\citenamefont {Lutz}\ \emph {et~al.}(2014)\citenamefont {Lutz},
  \citenamefont {Bavontaweepanya}, \citenamefont {Kobdaj},\ and\ \citenamefont
  {Schwarz}}]{Lutz:2014oxa}%
  \BibitemOpen
  \bibfield  {author} {\bibinfo {author} {\bibfnamefont {M.~F.~M.}\
  \bibnamefont {Lutz}}, \bibinfo {author} {\bibfnamefont {R.}~\bibnamefont
  {Bavontaweepanya}}, \bibinfo {author} {\bibfnamefont {C.}~\bibnamefont
  {Kobdaj}}, \ and\ \bibinfo {author} {\bibfnamefont {K.}~\bibnamefont
  {Schwarz}},\ }\href {\doibase 10.1103/PhysRevD.90.054505} {\bibfield
  {journal} {\bibinfo  {journal} {Phys. Rev. D}\ }\textbf {\bibinfo {volume}
  {90}},\ \bibinfo {pages} {054505} (\bibinfo {year} {2014})},\ \Eprint
  {http://arxiv.org/abs/1401.7805} {arXiv:1401.7805 [hep-lat]} \BibitemShut
  {NoStop}%
\bibitem [{\citenamefont {Ren}\ \emph {et~al.}(2015)\citenamefont {Ren},
  \citenamefont {Geng},\ and\ \citenamefont {Meng}}]{Ren:2014vea}%
  \BibitemOpen
  \bibfield  {author} {\bibinfo {author} {\bibfnamefont {X.-L.}\ \bibnamefont
  {Ren}}, \bibinfo {author} {\bibfnamefont {L.-S.}\ \bibnamefont {Geng}}, \
  and\ \bibinfo {author} {\bibfnamefont {J.}~\bibnamefont {Meng}},\ }\href
  {\doibase 10.1103/PhysRevD.91.051502} {\bibfield  {journal} {\bibinfo
  {journal} {Phys. Rev. D}\ }\textbf {\bibinfo {volume} {91}},\ \bibinfo
  {pages} {051502} (\bibinfo {year} {2015})},\ \Eprint
  {http://arxiv.org/abs/1404.4799} {arXiv:1404.4799 [hep-ph]} \BibitemShut
  {NoStop}%
\bibitem [{\citenamefont {Ren}\ \emph {et~al.}(2017)\citenamefont {Ren},
  \citenamefont {Alvarez-Ruso}, \citenamefont {Geng}, \citenamefont {Ledwig},
  \citenamefont {Meng},\ and\ \citenamefont {Vicente-Vacas}}]{Ren:2016aeo}%
  \BibitemOpen
  \bibfield  {author} {\bibinfo {author} {\bibfnamefont {X.-L.}\ \bibnamefont
  {Ren}}, \bibinfo {author} {\bibfnamefont {L.}~\bibnamefont {Alvarez-Ruso}},
  \bibinfo {author} {\bibfnamefont {L.-S.}\ \bibnamefont {Geng}}, \bibinfo
  {author} {\bibfnamefont {T.}~\bibnamefont {Ledwig}}, \bibinfo {author}
  {\bibfnamefont {J.}~\bibnamefont {Meng}}, \ and\ \bibinfo {author}
  {\bibfnamefont {M.~J.}\ \bibnamefont {Vicente-Vacas}},\ }\href {\doibase
  10.1016/j.physletb.2017.01.024} {\bibfield  {journal} {\bibinfo  {journal}
  {Phys. Lett. B}\ }\textbf {\bibinfo {volume} {766}},\ \bibinfo {pages} {325}
  (\bibinfo {year} {2017})},\ \Eprint {http://arxiv.org/abs/1606.03820}
  {arXiv:1606.03820 [nucl-th]} \BibitemShut {NoStop}%
\bibitem [{\citenamefont {Alexandrou}\ and\ \citenamefont
  {Kallidonis}(2017)}]{Alexandrou:2017xwd}%
  \BibitemOpen
  \bibfield  {author} {\bibinfo {author} {\bibfnamefont {C.}~\bibnamefont
  {Alexandrou}}\ and\ \bibinfo {author} {\bibfnamefont {C.}~\bibnamefont
  {Kallidonis}},\ }\href {\doibase 10.1103/PhysRevD.96.034511} {\bibfield
  {journal} {\bibinfo  {journal} {Phys. Rev. D}\ }\textbf {\bibinfo {volume}
  {96}},\ \bibinfo {pages} {034511} (\bibinfo {year} {2017})},\ \Eprint
  {http://arxiv.org/abs/1704.02647} {arXiv:1704.02647 [hep-lat]} \BibitemShut
  {NoStop}%
\bibitem [{\citenamefont {Ren}\ \emph {et~al.}(2018)\citenamefont {Ren},
  \citenamefont {Ling},\ and\ \citenamefont {Geng}}]{Ling:2017jyz}%
  \BibitemOpen
  \bibfield  {author} {\bibinfo {author} {\bibfnamefont {X.-L.}\ \bibnamefont
  {Ren}}, \bibinfo {author} {\bibfnamefont {X.-Z.}\ \bibnamefont {Ling}}, \
  and\ \bibinfo {author} {\bibfnamefont {L.-S.}\ \bibnamefont {Geng}},\ }\href
  {\doibase 10.1016/j.physletb.2018.05.063} {\bibfield  {journal} {\bibinfo
  {journal} {Phys. Lett. B}\ }\textbf {\bibinfo {volume} {783}},\ \bibinfo
  {pages} {7} (\bibinfo {year} {2018})},\ \Eprint
  {http://arxiv.org/abs/1710.07164} {arXiv:1710.07164 [hep-ph]} \BibitemShut
  {NoStop}%
\bibitem [{\citenamefont {Lutz}\ \emph {et~al.}(2018)\citenamefont {Lutz},
  \citenamefont {Heo},\ and\ \citenamefont {Guo}}]{Lutz:2018cqo}%
  \BibitemOpen
  \bibfield  {author} {\bibinfo {author} {\bibfnamefont {M.~F.~M.}\
  \bibnamefont {Lutz}}, \bibinfo {author} {\bibfnamefont {Y.}~\bibnamefont
  {Heo}}, \ and\ \bibinfo {author} {\bibfnamefont {X.-Y.}\ \bibnamefont
  {Guo}},\ }\href {\doibase 10.1016/j.nuclphysa.2018.05.007} {\bibfield
  {journal} {\bibinfo  {journal} {Nucl. Phys. A}\ }\textbf {\bibinfo {volume}
  {977}},\ \bibinfo {pages} {146} (\bibinfo {year} {2018})},\ \Eprint
  {http://arxiv.org/abs/1801.06417} {arXiv:1801.06417 [hep-lat]} \BibitemShut
  {NoStop}%
\bibitem [{\citenamefont {Chen}\ \emph {et~al.}(2013)\citenamefont {Chen},
  \citenamefont {Yao},\ and\ \citenamefont {Zheng}}]{Chen:2012nx}%
  \BibitemOpen
  \bibfield  {author} {\bibinfo {author} {\bibfnamefont {Y.-H.}\ \bibnamefont
  {Chen}}, \bibinfo {author} {\bibfnamefont {D.-L.}\ \bibnamefont {Yao}}, \
  and\ \bibinfo {author} {\bibfnamefont {H.~Q.}\ \bibnamefont {Zheng}},\ }\href
  {\doibase 10.1103/PhysRevD.87.054019} {\bibfield  {journal} {\bibinfo
  {journal} {Phys. Rev. D}\ }\textbf {\bibinfo {volume} {87}},\ \bibinfo
  {pages} {054019} (\bibinfo {year} {2013})},\ \Eprint
  {http://arxiv.org/abs/1212.1893} {arXiv:1212.1893 [hep-ph]} \BibitemShut
  {NoStop}%
\end{thebibliography}%

\end{document}